\pdfoutput=1
\documentclass[12pt,a4paper]{article}

\usepackage{ifthen} 
\newboolean{pdflatex}
\setboolean{pdflatex}{true} 

\newboolean{articletitles}
\setboolean{articletitles}{true} 

\newboolean{uprightparticles}
\setboolean{uprightparticles}{false} 

\def\paperauthors{LHCb collaboration} 
\def\paperasciititle{Determination of short- and long-distance contributions in B0->Kstar mu mu decays} 
\def\papertitle{Determination of short- and long-distance contributions in $B^{0}\to K^{*0}\mu^+\mu^-$ decays} 
\def\paperkeywords{{High Energy Physics}, {LHCb}} 
\def\papercopyright{\the\year\ CERN for the benefit of the LHCb collaboration} 
\def\paperlicence{CC BY 4.0 licence}
\def\paperlicenceurl{https://creativecommons.org/licenses/by/4.0/}

\def\flavio {\mbox{\textsc{Flavio}}\xspace}
\def\eos {\mbox{\textsc{Eos}}\xspace}
\def\im { \thinspace \mathrm{Im}}
\def\re { \thinspace \mathrm{Re}}
\def\thetal {\ensuremath{\theta_{\ell}}\xspace}                              
\def\thetak {\ensuremath{\theta_{K}}\xspace}

\usepackage[top=1in, bottom=1.25in, left=1in, right=1in]{geometry}

\usepackage{microtype}
\usepackage{lineno}  
\usepackage{xspace} 
\usepackage{caption} 

\usepackage{graphicx}  
\usepackage{color}
\usepackage{colortbl}
\graphicspath{{./figs/}} 

\usepackage{amsmath} 
\usepackage{amssymb}
\usepackage{amsfonts}
\usepackage{upgreek} 

\newcommand*\patchAmsMathEnvironmentForLineno[1]{%
\expandafter\let\csname old#1\expandafter\endcsname\csname #1\endcsname
\expandafter\let\csname oldend#1\expandafter\endcsname\csname
end#1\endcsname
 \renewenvironment{#1}%
   {\linenomath\csname old#1\endcsname}%
   {\csname oldend#1\endcsname\endlinenomath}%
}
\newcommand*\patchBothAmsMathEnvironmentsForLineno[1]{%
  \patchAmsMathEnvironmentForLineno{#1}%
  \patchAmsMathEnvironmentForLineno{#1*}%
}
\AtBeginDocument{%
\patchBothAmsMathEnvironmentsForLineno{equation}%
\patchBothAmsMathEnvironmentsForLineno{align}%
\patchBothAmsMathEnvironmentsForLineno{flalign}%
\patchBothAmsMathEnvironmentsForLineno{alignat}%
\patchBothAmsMathEnvironmentsForLineno{gather}%
\patchBothAmsMathEnvironmentsForLineno{multline}%
\patchBothAmsMathEnvironmentsForLineno{eqnarray}%
}

\usepackage{hyperxmp}

\usepackage[pdftex,
            pdfauthor={\paperauthors},
            pdftitle={\paperasciititle},
            pdfkeywords={\paperkeywords},
            pdfcopyright={Copyright (C) \papercopyright},
            pdflicenseurl={\paperlicenceurl}]{hyperref}

\usepackage[colorinlistoftodos,textsize=scriptsize]{todonotes}

\usepackage[bottom,flushmargin,hang,multiple]{footmisc}

\usepackage[all]{hypcap} 

\usepackage{xspace} 
\usepackage{upgreek}

\def\lhcb   {\mbox{LHCb}\xspace}

\def\MagUp {\mbox{\em Mag\kern -0.05em Up}\xspace}

\ifthenelse{\boolean{uprightparticles}}%
{

 \def\Pmu         {\ensuremath{\upmu}\xspace}

 \def\Ppsi        {\ensuremath{\uppsi}\xspace}

 \def\PDelta      {\ensuremath{\Delta}\xspace}                 
 \def\PXi         {\ensuremath{\Xi}\xspace}                 
 \def\PLambda     {\ensuremath{\Lambda}\xspace}                 
 \def\PSigma      {\ensuremath{\Sigma}\xspace}                 
 \def\POmega      {\ensuremath{\Omega}\xspace}                 
 \def\PUpsilon    {\ensuremath{\Upsilon}\xspace}
 \let\oldPi\Pi
 \def\PPi         {\ensuremath{\oldPi}\xspace}

 \def\PB      {\ensuremath{\mathrm{B}}\xspace}                 
                  
 \def\PD      {\ensuremath{\mathrm{D}}\xspace}

 \def\PJ      {\ensuremath{\mathrm{J}}\xspace}                 
 \def\PK      {\ensuremath{\mathrm{K}}\xspace}

 \def\Pb      {\ensuremath{\mathrm{b}}\xspace}                 
 \def\Pc      {\ensuremath{\mathrm{c}}\xspace}

 \def\Pi      {\ensuremath{\mathrm{i}}\xspace}

 \def\Ps      {\ensuremath{\mathrm{s}}\xspace}

 \def\thebaroffset{0.0em}
}
{

 \def\Pmu         {\ensuremath{\mu}\xspace}

 \def\Ppsi        {\ensuremath{\psi}\xspace}                 
                  
 \mathchardef\PDelta="7101
 \mathchardef\PXi="7104
 \mathchardef\PLambda="7103
 \mathchardef\PSigma="7106
 \mathchardef\POmega="710A
 \mathchardef\PUpsilon="7107
 \mathchardef\PPi="7105
                  
 \def\PB      {\ensuremath{B}\xspace}                 
                  
 \def\PD      {\ensuremath{D}\xspace}

 \def\PJ      {\ensuremath{J}\xspace}                 
 \def\PK      {\ensuremath{K}\xspace}

 \def\Pb      {\ensuremath{b}\xspace}                 
 \def\Pc      {\ensuremath{c}\xspace}

 \def\Pi      {\ensuremath{i}\xspace}

 \def\Ps      {\ensuremath{s}\xspace}

 \def\thebaroffset{0.18em}
}
\newcommand{\offsetoverline}[2][\thebaroffset]{\kern #1\overline{\kern -#1 #2}}%

\makeatletter
\ifcase \@ptsize \relax
  \newcommand{\miniscule}{\@setfontsize\miniscule{4}{5}}
\or
  \newcommand{\miniscule}{\@setfontsize\miniscule{5}{6}}
\or
  \newcommand{\miniscule}{\@setfontsize\miniscule{5}{6}}
\fi
\makeatother

\DeclareRobustCommand{\optbar}[1]{\shortstack{{\miniscule (\rule[.5ex]{1.25em}{.18mm})}
  \\ [-.7ex] $#1$}}

\def\mup        {{\ensuremath{\Pmu^+}}\xspace}
\def\mun        {{\ensuremath{\Pmu^-}}\xspace} 

\def\mumu       {{\ensuremath{\Pmu^+\Pmu^-}}\xspace}

\def\squark    {{\ensuremath{\Ps}}\xspace}

\def\cquark    {{\ensuremath{\Pc}}\xspace}

\def\bquark    {{\ensuremath{\Pb}}\xspace}

\def\kaon    {{\ensuremath{\PK}}\xspace}
\def\Kbar    {{\ensuremath{\offsetoverline{\PK}}}\xspace}

\def\KorKbar {\kern \thebaroffset\optbar{\kern -\thebaroffset \PK}{}\xspace}

\def\Km      {{\ensuremath{\kaon^-}}\xspace}

\def\Kstarz  {{\ensuremath{\kaon^{*0}}}\xspace}
\def\Kstarzb {{\ensuremath{\Kbar{}^{*0}}}\xspace}

\def\D       {{\ensuremath{\PD}}\xspace}

\def\DorDbar {\kern \thebaroffset\optbar{\kern -\thebaroffset \PD}\xspace}

\def\Dp      {{\ensuremath{\D^+}}\xspace}
\def\Dm      {{\ensuremath{\D^-}}\xspace}

\def\DpDm    {\ensuremath{\Dp {\kern -0.16em \Dm}}\xspace}

\def\B       {{\ensuremath{\PB}}\xspace}
\def\Bbar    {{\ensuremath{\offsetoverline{\PB}}}\xspace}

\def\BorBbar {\kern \thebaroffset\optbar{\kern -\thebaroffset \PB}\xspace}
\def\Bz      {{\ensuremath{\B^0}}\xspace}
\def\Bzb     {{\ensuremath{\Bbar{}^0}}\xspace}
\def\Bd      {{\ensuremath{\B^0}}\xspace}

\def\BdorBdbar {\kern \thebaroffset\optbar{\kern -\thebaroffset \Bd}\xspace}

\def\Bs      {{\ensuremath{\B^0_\squark}}\xspace}
\def\Bsb     {{\ensuremath{\Bbar{}^0_\squark}}\xspace}
\def\BsorBsbar {\kern \thebaroffset\optbar{\kern -\thebaroffset \Bs}\xspace}

\def\jpsi     {{\ensuremath{{\PJ\mskip -3mu/\mskip -2mu\Ppsi}}}\xspace}
\def\psitwos  {{\ensuremath{\Ppsi{(2S)}}}\xspace}

\def\Y#1S{\ensuremath{\PUpsilon{(#1S)}}\xspace}

\def\Lz          {{\ensuremath{\PLambda}}\xspace}

\def\LorLbar     {\kern \thebaroffset\optbar{\kern -\thebaroffset \PLambda}\xspace}

\def\Lb           {{\ensuremath{\Lz^0_\bquark}}\xspace}

\newcommand{\decay}[2]{\ensuremath{#1\!\to #2}\xspace} 

\def\to                 {\ensuremath{\rightarrow}\xspace}

\def\qsq       {{\ensuremath{q^2}}\xspace}

\def\AT#1     {\ensuremath{A_{\mathrm{T}}^{#1}}\xspace}           

\def\C#1      {\ensuremath{\mathcal{C}_{#1}}\xspace}                       
\def\Cp#1     {\ensuremath{\mathcal{C}_{#1}^{'}}\xspace}                    
\def\Ceff#1   {\ensuremath{\mathcal{C}_{#1}^{\mathrm{(eff)}}}\xspace}        
\def\Cpeff#1  {\ensuremath{\mathcal{C}_{#1}^{'\mathrm{(eff)}}}\xspace}       
\def\Ope#1    {\ensuremath{\mathcal{O}_{#1}}\xspace}                       
\def\Opep#1   {\ensuremath{\mathcal{O}_{#1}^{'}}\xspace}                    

\newcommand{\nospaceunit}[1]{\ensuremath{\text{#1}}}       
\newcommand{\aunit}[1]{\ensuremath{\text{\,#1}}}       

\newcommand{\tev}{\aunit{Te\kern -0.1em V}\xspace}
\newcommand{\gev}{\aunit{Ge\kern -0.1em V}\xspace}
\newcommand{\mev}{\aunit{Me\kern -0.1em V}\xspace}
\newcommand{\kev}{\aunit{ke\kern -0.1em V}\xspace}
\newcommand{\ev}{\aunit{e\kern -0.1em V}\xspace}
 
\newcommand{\mevc}{\ensuremath{\aunit{Me\kern -0.1em V\!/}c}\xspace}
\newcommand{\gevc}{\ensuremath{\aunit{Ge\kern -0.1em V\!/}c}\xspace}
\newcommand{\mevcc}{\ensuremath{\aunit{Me\kern -0.1em V\!/}c^2}\xspace}
\newcommand{\gevcc}{\ensuremath{\aunit{Ge\kern -0.1em V\!/}c^2}\xspace}
\newcommand{\gevgevcc}{\ensuremath{\gev^2\!/c^2}\xspace} 
\newcommand{\gevgevcccc}{\ensuremath{\gev^2\!/c^4}\xspace} 

\def\mum  {\ensuremath{\,\upmu\nospaceunit{m}}\xspace}

\def\fb   {\ensuremath{\aunit{fb}}\xspace}
\def\invfb   {\ensuremath{\fb^{-1}}\xspace}

\def\gsim{{~\raise.15em\hbox{$>$}\kern-.85em
          \lower.35em\hbox{$\sim$}~}\xspace}
\def\lsim{{~\raise.15em\hbox{$<$}\kern-.85em
          \lower.35em\hbox{$\sim$}~}\xspace}

\def\pt         {\ensuremath{p_{\mathrm{T}}}\xspace}

\def\ptot       {\ensuremath{p}\xspace}

\def\evtgen     {\mbox{\textsc{EvtGen}}\xspace}

\def\geant      {\mbox{\textsc{Geant4}}\xspace}

\def\photos     {\mbox{\textsc{Photos}}\xspace}

\def\pythia     {\mbox{\textsc{Pythia}}\xspace}

\def\tell1  {TELL1\xspace}
\def\ukl1   {UKL1\xspace}

\newcommand{\phz}{\phantom{0}}

\newcommand{\lhcborcid}[1]{\href{https://orcid.org/#1}{\hspace*{0.1em}\raisebox{-0.45ex}{\includegraphics[width=1em]{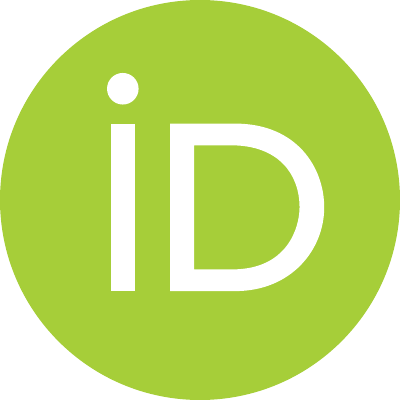}}}}

\usepackage{cite} 
\usepackage{mciteplus}

\usepackage{longtable} 

\usepackage{multirow}
\usepackage{siunitx}
\usepackage{makecell}
\usepackage{booktabs}
\usepackage{cleveref}

\begin{document}

\renewcommand{\thefootnote}{\fnsymbol{footnote}}
\setcounter{footnote}{1}


\begin{titlepage}
\pagenumbering{roman}

\vspace*{-1.5cm}
\centerline{\large EUROPEAN ORGANIZATION FOR NUCLEAR RESEARCH (CERN)}
\vspace*{1.5cm}
\noindent
\begin{tabular*}{\linewidth}{lc@{\extracolsep{\fill}}r@{\extracolsep{0pt}}}
\ifthenelse{\boolean{pdflatex}}
{\vspace*{-1.5cm}\mbox{\!\!\!\includegraphics[width=.14\textwidth]{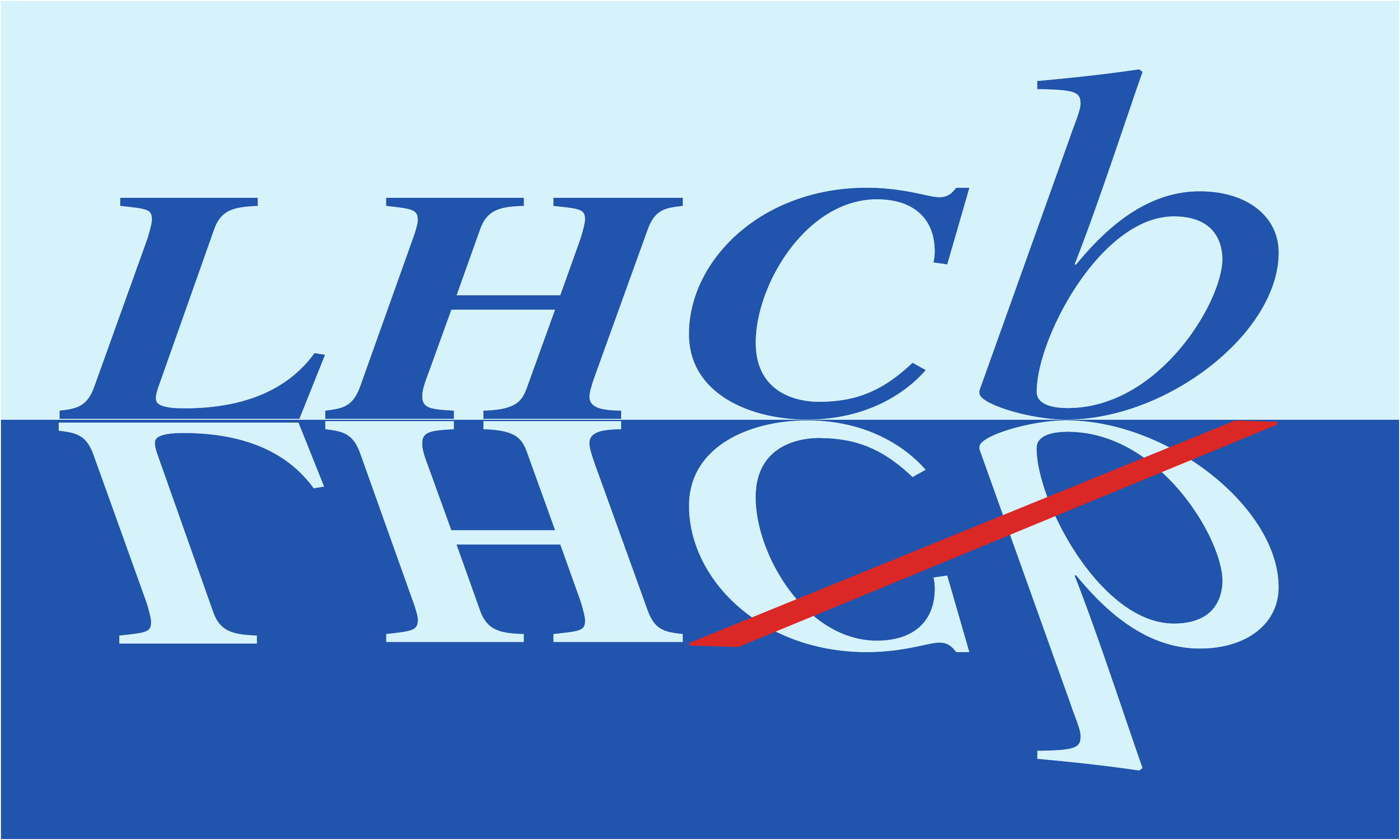}} & &}%
{\vspace*{-1.2cm}\mbox{\!\!\!\includegraphics[width=.12\textwidth]{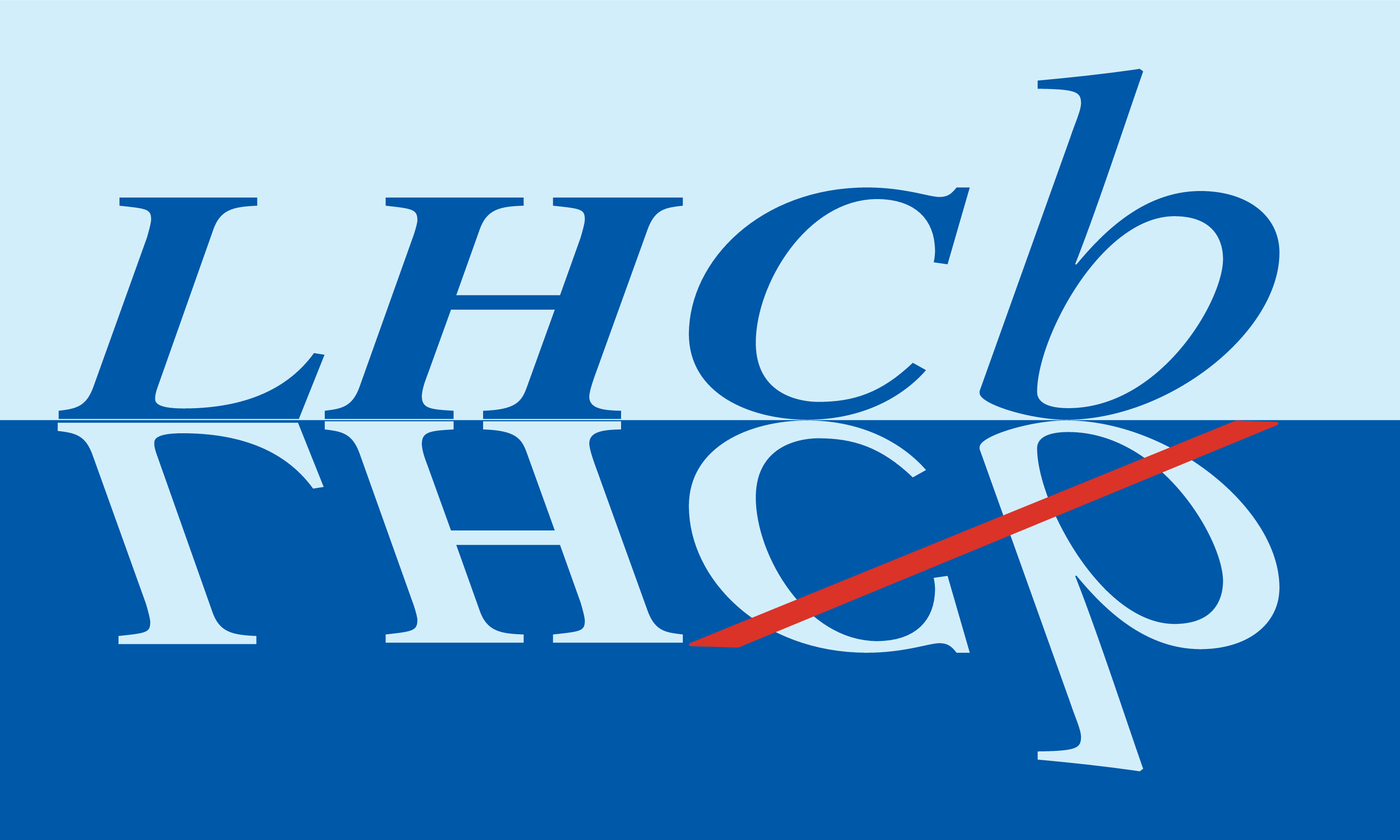}} & &}%
\\
 & & CERN-EP-2023-237 \\  
 & & LHCb-PAPER-2023-032 \\  
 & & \today \\ 
 & & \\
\end{tabular*}

\vspace*{3.5cm}

{\normalfont\bfseries\boldmath\huge
\begin{center}
  \papertitle 
\end{center}
}

\vspace*{1.5cm}

\begin{center}
\paperauthors\footnote{Authors are listed at the end of this paper.}
\end{center}

\vspace{\fill}

\begin{abstract}
  \noindent
An amplitude analysis of the $B^0 \to K^{*0} \mumu$ decay is presented. The analysis is based on data collected by the LHCb experiment from proton-proton collisions at $\sqrt{s} = 7,\,8$ and $13\tev$, corresponding to an integrated luminosity of $4.7\invfb$.
For the first time, Wilson coefficients and non-local hadronic contributions are accessed directly from the unbinned data, where the latter are parameterised as a function of $q^2$ with a polynomial expansion.
Wilson coefficients and non-local hadronic parameters are determined under two alternative hypotheses: the first relies on experimental information alone, while the second one includes information from theoretical predictions for the non-local contributions.
Both models obtain similar results for the parameters of interest.
The overall level of compatibility with the Standard Model is evaluated to be 
between 1.8 and 1.9 standard deviations 
when looking at the $\mathcal{C}_9$ Wilson coefficient alone, and between 1.3 and 1.4 standard deviations when considering the full set of $\mathcal{C}_9, \, \mathcal{C}_{10}, \, \mathcal{C}_9^\prime$ and $\mathcal{C}_{10}^\prime$ Wilson coefficients. The ranges reflect the theoretical assumptions made in the analysis.

\end{abstract}

\vspace*{2.0cm}

\begin{center}
  Published in Phys. Rev. D 109 (2024) 052009
\end{center}

\vspace{\fill}

{\footnotesize 
\centerline{\copyright~\papercopyright. \href{\paperlicenceurl}{\paperlicence}.}}
\vspace*{2mm}

\end{titlepage}


\newpage
\setcounter{page}{2}
\mbox{~}
%
%
%
%


\renewcommand{\thefootnote}{\arabic{footnote}}
\setcounter{footnote}{0}

\cleardoublepage


\pagestyle{plain} 
\setcounter{page}{1}
\pagenumbering{arabic}


\section{Introduction}
\label{sec:Introduction}
Flavour-changing neutral current decays of beauty hadrons, mediated through electroweak loop diagrams,
constitute powerful indirect probes of the Standard Model (SM).
Several studies of these processes have shown intriguing discrepancies with respect to predictions from the SM,
notably in final states with a pair of muons~\cite{LHCb-PAPER-2016-012,LHCb-PAPER-2014-006,LHCb-PAPER-2013-017,LHCb-PAPER-2015-009,LHCb-PAPER-2022-050,LHCb-PAPER-2020-041}.
In particular, angular analyses of $B^0 \to K^{*0} \mumu$ decays\footnote{Throughout this article, $K^{*0}$ is used to refer to the $K^*(892)^0$ resonance and the inclusion of charge-conjugated processes is implied.}~\cite{LHCb-PAPER-2013-019,LHCb-PAPER-2015-051,LHCb-PAPER-2020-002,Belle:2016fev,Aaboud:2018krd,Sirunyan:2017dhj} reported a tension in the measured value of the angular observable $P_5^\prime$~\cite{Descotes-Genon:2012isb} with respect to its calculations based on the SM~\cite{Khodjamirian:2010vf,Descotes-Genon:2014uoa,Gubernari:2022hxn}.

In order to further investigate these anomalies, it is convenient to describe rare $b \to s$ decays within an effective field theory in terms of local (short-distance) operators, $\mathcal{O}_i$, that only contain light SM fields, and corresponding effective couplings, $\mathcal{C}_i$, known as Wilson coefficients.
Only a subset of dimension-six operators is relevant for $B^0\to K^{*0}\mu^+\mu^-$ decays~\cite{Bobeth:2012vn}: 
the radiative operator $\mathcal{O}_7$, the semileptonic operators $\mathcal{O}_9$ and $\mathcal{O}_{10}$ that drive the vector and axial leptonic currents, respectively, and their corresponding right-handed counterparts $\mathcal{O}_{7,9,10}^\prime$.
Within this approach, all non-perturbative QCD effects are contained in the local (form factors) and non-local hadronic matrix elements, while physics beyond the SM would appear as a shift in the values of one or more Wilson coefficients.
The lack of knowledge on these hadronic matrix elements prevents a clean determination of the Wilson coefficients.
Non-local contributions originating from $b \to s c \bar{c}$ virtual loops are particularly difficult to assess reliably from first principles as they receive non-perturbative corrections from soft-gluon emissions~\cite{Khodjamirian:2010vf}.
Over the last two decades, different approaches have been proposed to determine the impact of these contributions, from phenomenological~\cite{Lyon:2014hpa,Ciuchini:2015qxb,Ciuchini:2017mik} and data~\cite{LHCb-PAPER-2016-045,Hurth:2017sqw} analyses to the use of analytic~\cite{Bobeth:2017vxj,Gubernari:2020eft,Gubernari:2022hxn,Chrzaszcz:2018yza} and dispersion~\cite{Cornella:2020aoq} relations,
where all these methods rely on the perturbative treatment of four-quark operators in the regime of $q^2 \ll 4 m_c^2$, with $q^2$ the dimuon invariant mass squared  and $m_c$ the mass of the charm quark.
While the study and development of all these techniques enabled a clear progress in the understanding of such contributions, a definitive consensus on their impact has not been reached yet~\cite{Ciuchini:2022wbq}.
A deeper comprehension of these non-local hadronic effects is therefore crucial for the understanding of the anomalies seen in rare $b \to s$ hadron decays.

The analysis described in this paper aims at the first determination of such non-local hadronic contributions in  $B^0\to K^{*0}\mu^+\mu^-$ decays from data by performing a $q^2$-unbinned amplitude analysis of the decay.
Non-local hadronic contributions are determined following two alternative strategies, the first relies purely   on experimental data and the second includes state-of-the-art theoretical inputs.
The four Wilson coefficients $\mathcal{C}_9, \, \mathcal{C}_{10}, \, \mathcal{C}_9^\prime $ and $\mathcal{C}_{10}^\prime$
are determined simultaneously from data together with the local and non-local hadronic matrix elements, while the coefficients $\mathcal{C}_7$ and $\mathcal{C}^{\prime}_7$ are already precisely known from radiative $B$ meson decays~\cite{Paul:2016urs} and are fixed to their SM values~\cite{Bobeth:1999mk,Gorbahn:2004my}.
This measurement therefore provides a complementary and more in-depth set of information with respect to previous binned analyses of the same decay~\cite{LHCb-PAPER-2016-012,LHCb-PAPER-2020-002}.
A more concise description of this work is reported in a companion article~\cite{LHCb-PAPER-2023-033}.
The employed dataset corresponds to an integrated luminosity of $4.7\invfb$ of $pp$ collisions collected with the LHCb experiment during the years 2011, 2012 and 2016 at centre-of-mass energies of 7, 8 and $13\tev$, respectively.

This paper is structured as follows: the formalism employed to describe $B^0 \to K^{*0}\mumu$ decays is presented in Sec.~\ref{sec:formalism}; the LHCb detector and the procedure used to generate simulated samples is introduced in Sec.~\ref{sec:detector}; the selection of signal candidates is described in Sec.~\ref{sec:selection}; the amplitude fit method is discussed in Sec.~\ref{sec:amplitude_fit}; the fit to data is presented in Sec.~\ref{sec:data}; sources of systematic uncertainties are discussed in Sec.~\ref{sec:systematics}; Sec.~\ref{sec:results} shows the obtained results; Sec.~\ref{sec:discussion} is dedicated to a final discussion and conclusions are presented in Sec.~\ref{sec:conclusions}.


\section{Formalism}
\label{sec:formalism}

The $B^0 \to K^{*0}  \mumu$ decay, where the $K^{*0}$ meson is reconstructed through the decay $K^{*0} \to K^+ \pi^-$, can be described by five kinematic variables: 
the dimuon  invariant mass squared, $q^2$, the $K^{+}\pi^{-}$  invariant mass squared, $k^2$, and the three decay angles ${\vec{\Omega}} = ( \theta_{\ell}, \theta_{K}, \phi)$. 
Here, $\theta_{\ell}$ is the angle between the $\mu^+\,(\mu^-)$ and the direction opposite to that of the $\Bz \,(\Bzb)$ in the rest frame of the dimuon system,
$\theta_{K}$ is the angle between the direction of the $K^+\,(K^-)$ and the $\Bz \,(\Bzb)$ in the rest frame of the $\Kstarz \,(\Kstarzb)$ system,
and $\phi$ is the angle between the plane defined by the dimuon pair and the plane defined by the kaon and pion in the $\Bz \,(\Bzb)$ rest frame.
A full description of the angular basis is provided in Ref.~\cite{LHCb-PAPER-2013-019}.  
The differential decay rate of $B^0 \to K^{*0} (\to K^+ \pi^-) \mumu$ decays,
where the $K^+ \pi^-$ pair comes from an intermediate resonance of spin-1 (P-wave), 
can be written as~\cite{Altmannshofer:2008dz,LHCb-PAPER-2015-051}
\begin{equation}
\begin{split}
  \frac{\textrm{d}^4\Gamma[B^0 \to K^{*0}\mumu]}{\textrm{d} q^2  \, \textrm{d}  \vec{\Omega}} =  \frac{9}{32\pi} &  \sum_{i} I_{i}(q^{2})f_{i}(\vec{\Omega}) \\
 = \frac{9}{32\pi} & 
\left[ \frac{}{} {I_{1s}} \sin^{2} \theta_{K} + 
{I_{1c}}  \cos^{2}\theta_{K} 
+ {I_{2s}} \sin^{2} \theta_{K} \cos 2\theta_{\ell} ~+ \right.  \\ 
&   \left. ~\frac{}{}  {I_{2c}} \cos^{2} \theta_{K} \cos 2\theta_{\ell} 
+ {I_{3}}  \sin^{2}\theta_{K} \sin^{2} \theta_{\ell} \cos 2\phi ~ + \right. \\ 
&   \left. ~\frac{}{}  {{ I_{4} \sin 2\theta_{K} \sin 2\theta_{\ell} \cos\phi}} 
+ {{{I_{5}} \sin 2\theta_{K} \sin\theta_{\ell}\cos\phi}} ~+ \right. \\ 
& ~\frac{}{} \left.  I_{6} \sin^{2}\theta_{K}  \cos\theta_{\ell} 
+ {{{I_{7}}  \sin 2\theta_{K} \sin\theta_{\ell} \sin\phi}} ~+ \right. \\ 
& ~\frac{}{} \left. {{{I_{8}} \sin 2\theta_{K} \sin 2\theta_{\ell}\sin\phi}} 
+ I_{9} \sin^{2} \theta_{K} \sin^{2}\theta_{\ell} \sin 2\phi \frac{}{}\right],
\end{split}
\label{eq:d4Gamma_P}
\end{equation}
where the $I_i$ are the $q^2$-dependent angular coefficients defined in Appendix~\ref{app:angular_coefficients},   
which can be conveniently expressed in terms of the amplitudes  $\mathcal{A}_\lambda^{L,R}$ and $\mathcal{A}_t$.
The former are transversity amplitudes where the symbol $\lambda = 0,\perp,\parallel$ refers to the transversity state of the $K^{*0}$ meson and the indices $L$ and $R$ denote the chirality of the lepton current. 
The amplitude $\mathcal{A}_t$ corresponds to a time-like polarisation of the virtual vector boson decaying into a dimuon pair and longitudinal polarisation of the $K^{*0}$ meson~\cite{Altmannshofer:2008dz}.
These amplitudes describe the decay process 
and can be expressed as~\cite{Bobeth:2012vn,Bobeth:2017vxj}
\begin{eqnarray}
\label{eq:Ampl}
\mathcal{A}_\perp^{L,R} & = & \mathcal{N}\Big\{  \Big[ (\mathcal{C}_9 + \mathcal{C}^\prime_{9})  
										\mp (\mathcal{C}_{10} +   \mathcal{C}^\prime_{10}) \Big]     \mathcal{F}_\perp(q^2, k^2)      \nonumber   \\   
							&&  \quad\;\;\;\,  +  \frac{2 m_b M_B}{q^2}    \Big[   (\mathcal{C}_7 + \mathcal{C}^\prime_{7})  
										 \mathcal{F}_\perp^T(q^2, k^2)   -  16 \pi^2  \frac{M_B}{m_b} \mathcal{H}_\perp(q^2) \Big]  \Big\}  \, ,   \nonumber   \\
\mathcal{A}_\parallel^{L,R} & = & - \mathcal{N} \Big\{  \Big[ (\mathcal{C}_9 - \mathcal{C}^\prime_{9})  
											\mp (\mathcal{C}_{10} -   \mathcal{C}^\prime_{10}) \Big]     \mathcal{F}_\parallel(q^2, k^2)      \\   
								&&  \qquad \; +  \frac{2 m_b M_B}{q^2}  \Big[    (\mathcal{C}_7 - \mathcal{C}^\prime_{7})  
											\mathcal{F}_\parallel^T(q^2, k^2)  -  16 \pi^2  \frac{M_B}{m_b} \mathcal{H}_\parallel(q^2) \Big]  \Big\}  \, ,   \nonumber 
\end{eqnarray}
\begin{eqnarray}
\mathcal{A}_0^{L,R} & = & - \mathcal{N} \frac{M_B}{\sqrt{q^2}} \Big\{  \Big[ (\mathcal{C}_9 - \mathcal{C}^\prime_{9})  
								\mp (\mathcal{C}_{10} -   \mathcal{C}^\prime_{10}) \Big]     \mathcal{F}_0(q^2, k^2)   \nonumber  \\   
					   &&  \qquad \;   +  \frac{2 m_b M_B}{q^2}   \Big[   (\mathcal{C}_7 - \mathcal{C}^\prime_{7})   
								\mathcal{F}_0^T(q^2, k^2)  -  16 \pi^2  \frac{M_B}{m_b} \mathcal{H}_0(q^2) \Big]   \Big\}  \, , \nonumber  \\
\mathcal{A}_t  &  = &  - 2 \mathcal{N} \,  (\mathcal{C}_{10} -  \mathcal{C}^\prime_{10})  \mathcal{F}_t(q^2, k^2) \, ,  \nonumber
\end{eqnarray}
where the functions $\mathcal{F}_\lambda^{(T)}$ (form factors) and $\mathcal{H}_\lambda$ encode the local  and non-local hadronic matrix 
elements, respectively, and $m_b$ and $M_B$ correspond to the $b$ quark and $B^0$ meson masses. 
The coefficient $\mathcal{N}$ is a normalisation factor given by 
\begin{equation}
\label{eq:ampl_norm}
\mathcal{N} = G_F \alpha_e V_{tb} V^*_{ts}  \sqrt{ \frac{ q^2 \beta_\mu  \sqrt{\lambda(M_B^2, q^2, k^2)} }{ 3 \cdot 2^{10} \pi^5 M_B}} \,,
\end{equation}
where  $\lambda(M_B^2, q^2, k^2)$ is a kinematical factor, with $\lambda(a,b,c) = a^2 + b^2 +c^2 -2ab -2ac -2bc$,
$\beta_\mu = \sqrt{1-4m_{\mu}^2/q^2}$,
$V_{tb}$ and $V^*_{ts}$ are elements of the Cabibbo-Kobayashi-Maskawa (CKM) quark-mixing matrix, $G_F$ is the Fermi constant and $\alpha_e$ is the fine structure constant.
The exact definition of the form factors $\mathcal{F}_\lambda^{(T)}(q^2,k^2)$
is given in Appendix~\ref{app:form_factors}, while
the definition of the non-local functions $\mathcal{H}_\lambda(q^2)$ follows what has been proposed in Refs.~\cite{Bobeth:2017vxj,Gubernari:2020eft,Gubernari:2022hxn,Chrzaszcz:2018yza} where the analytic properties of the hadronic matrix elements are exploited through the mapping~\cite{Boyd:1995cf,Bourrely:2008za} 
\begin{equation}\label{eq:z_maps_q2}
q^{2} \mapsto z(q^{2}) \equiv \frac{\sqrt{t_+ - q^2} - \sqrt{t_+ - t_0}}{\sqrt{t_+ - q^2} + \sqrt{t_+ - t_0}} \, ,
\end{equation}
where $t_+ = 4 M_D^2$, with $M_D$ the $D^0$ meson mass, and 
$t_0$ 
can be arbitrarily chosen such that \mbox{$z(q^2=t_0) = 0$}.
After this transformation,\footnote{The functional form of $\mathcal{H}_\lambda$ defined in Eq.~\ref{eq:H} is defined as function of the variable $z$. Throughout this article, the expression $\mathcal{H}_\lambda(q^2)$ implies $\mathcal{H}_\lambda(z(q^2))$, where the contracted form is used to improve legibility.} the non-local hadronic functions can be expressed as
\begin{equation}\label{eq:H}
\mathcal{H}_\lambda(z) = \frac{1 - z z_{\jpsi}}{z - z_{\jpsi}} \frac{1 - z z_{\psi(2S)}}{z - z_{\psi(2S)}} \,  \hat{\mathcal{H}}_\lambda(z)\,,
\end{equation}
where the first and second terms remove the singularities due to the 
$\jpsi$ and $\psi(2S)$ poles. The
$\hat{\mathcal{H}}_\lambda(z)$ are analytic functions which can be further decomposed as
\begin{equation}\label{eq:Hhat_definition}
\hat{\mathcal{H}}_\lambda(z) = \phi_\lambda^{-1}(z)   \sum_{k} \alpha_{\lambda,k} z^k \,,
\end{equation}
where $\phi_\lambda^{-1}(z)$ are so-called outer functions that ensure the correct kinematic dependence~\cite{Gubernari:2020eft}, 
\textit{e.g.} $\mathcal{H}_0(q^2=0) = 0$,
and $\alpha_{\lambda,k}$ are the coefficients of a polynomial expansion.

The $K^+\pi^-$ system in the final state can also appear in a scalar (S-wave) configuration,
which introduces two additional amplitudes~\cite{Becirevic:2012dp},
\begin{eqnarray}
\label{eq:Ampl_Swave}
\mathcal{A}_{S\,0}^{L,R} & = & - \mathcal{N}  \frac{\sqrt{\lambda(M_B^2, q^2, k^2)}}{ M_B  \sqrt{q^2}   }  \Big\{  \Big[ (\mathcal{C}_9 - \mathcal{C}^\prime_{9})  
								\mp (\mathcal{C}_{10} -   \mathcal{C}^\prime_{10}) \Big]   f_+(q^2,k^2)  \nonumber \\ 
					&& \qquad \qquad \qquad \qquad \qquad    +  \frac{2 m_b M_B}{q^2}    (\mathcal{C}_7 - \mathcal{C}^\prime_{7})  f_T(q^2,k^2)    \Big\}  ,   \\
\mathcal{A}_{S\,t}  &  = &  - 2 \mathcal{N}  \frac{M_B^2 - k^2}{  M_B \sqrt{q^2}} \,  (\mathcal{C}_{10} -  \mathcal{C}^\prime_{10}) f_0(q^2,k^2)  ,  \nonumber
\end{eqnarray}
with three new form factors, $f_+, f_T$ and $f_0$ whose definitions can be found in Appendix~\ref{app:form_factors}.
In the following, contributions from non-local hadronic matrix elements to the scalar amplitudes are ignored.
This assumption is studied as a source of systematic uncertainty in Sec.~\ref{sec:systematics}.

In order to better separate the contribution of the S-wave amplitudes from those of the P-wave,
the $k^2$ dependence is included in the model.
This is achieved by multiplying the decay amplitudes of Eq.~\ref{eq:Ampl} 
by a relativistic Breit-Wigner function for the resonant P-wave~\cite{Descotes-Genon:2019bud} and 
the scalar amplitudes of Eq. \ref{eq:Ampl_Swave} by the LASS parameterisation~\cite{Aston:1987ir} for the slowly varying S-wave,
\textit{i.e.} 
\begin{equation}\label{eq:4Dto5D}
\begin{aligned}
  \mathcal{A}^{L,R}_{0,\perp,\parallel,t}   &  \mapsto    \mathcal{A}^{L,R}_{0,\perp,\parallel,t}   \times  f_P(k^2)   \, , \\ 
  \mathcal{A}^{L,R}_{S\,0,\,S\,t}   &  \mapsto    \mathcal{A}^{L,R}_{S\,0,\,S\,t}   \times  f_S(k^2)    \, ,   
\end{aligned}
\end{equation}
where $f_P$ and $f_S$ encode the P- and S-wave $k^2$ dependence, respectively.
In principle, the  proportion between P- and S-waves can be determined from the  normalisation of the decay amplitudes, however, an accurate prediction of the two relative contributions involves a complete analysis of the $B^0 \to K^+ \pi^-$ form factors in the full $K^+ \pi^-$ spectrum~\cite{Descotes-Genon:2019bud,Descotes-Genon:2023ukb}. 
Given the limited $k^2$ range analysed around the $K^{*0}$ mass, it is therefore practical to decouple the normalisation of the P- and S-wave amplitudes and introduce an additional complex coefficient to control the relative magnitude and phase between the two.
A detailed definition of $f_P$ and $f_S$ is given in Appendix~\ref{app:mKpi_lineshapes}.

Finally, when taking into account the full set of P-wave and S-wave amplitudes, the total $B^0 \to  K^+ \pi^- \mumu$ 
differential decay rate reads as
\begin{eqnarray}
\label{eq:d5Gamma_SP}
\frac{32\pi}{9} \frac{\textrm{d}^5\Gamma}{\textrm{d} q^2 \,\textrm{d} k^2 \,\textrm{d} \vec{\Omega} } 
	&  =  &    \frac{32\pi}{9} \frac{\textrm{d}^5\Gamma^{\rm P}}{\textrm{d} q^2 \,\textrm{d} k^2 \,\textrm{d} \vec{\Omega}}  
 \nonumber \\ 
	&  +  &    \big( I^S_{1c} + I^S_{2c}  \cos 2 \theta_\ell \big)							  \nonumber \\ 
	&  +  &	  \big( \tilde{I}_{1c} + \tilde{I}_{2c}  \cos 2 \theta_\ell \big)   \cos \theta_K										  \\ 
	&  +  &	  \big( \tilde{I}_4  \sin 2 \theta_\ell  + \tilde{I}_5  \sin \theta_\ell \big)   \sin \theta_K \cos \phi				  \nonumber \\ 
	&  +  &	  \big( \tilde{I}_7  \sin  \theta_\ell  + \tilde{I}_8  \sin 2 \theta_\ell \big)   \sin \theta_K \sin \phi	 \,,			  \nonumber 
\end{eqnarray}	
where the first term corresponds to the P-wave differential decay rate of Eq.~\ref{eq:d4Gamma_P} extended to the $k^2$ dimension via Eq.~\ref{eq:4Dto5D},
$I_i^S$ are pure S-wave angular coefficients and 
$\tilde{I}_i$ denote interference terms between the S- and P-wave amplitudes, as defined in Appendix~\ref{app:angular_coefficients}.

\section{LHCb detector and simulation}
\label{sec:detector}

The \lhcb detector~\cite{LHCb-DP-2008-001,LHCb-DP-2014-002} is a single-arm forward
spectrometer covering the \mbox{pseudorapidity} range $2<\eta <5$,
designed for the study of particles containing \bquark or \cquark
quarks. The detector includes a high-precision tracking system
consisting of a silicon-strip vertex detector surrounding the $pp$
interaction region, a large-area silicon-strip detector located
upstream of a dipole magnet with a bending power of about
$4{\mathrm{\,Tm}}$, and three stations of silicon-strip detectors and straw
drift tubes
placed downstream of the magnet.
The tracking system provides a measurement of the momentum, \ptot, of charged particles with
a relative uncertainty that varies from 0.5\% at low momentum to 1.0\% at 200\gevc.
The minimum distance of a track to a primary $pp$ collision vertex (PV), the impact parameter (IP), 
is measured with a resolution of $(15+29/\pt)\mum$,
where \pt is the component of the momentum transverse to the beam, in\,\gevc.
Different types of charged hadrons are distinguished using information
from two ring-imaging Cherenkov detectors. 
Photons, electrons and hadrons are identified by a calorimeter system consisting of
scintillating-pad and preshower detectors, an electromagnetic
and a hadronic calorimeter. Muons are identified by a
system composed of alternating layers of iron and multiwire
proportional chambers.
The online event selection is performed by a trigger, 
which consists of a hardware stage, based on information from the calorimeter and muon
systems, followed by a software stage, which applies a full event
reconstruction.
In the hardware stage, depending on the data-taking conditions signal candidates are required to have at least one muon with a \pt greater than 1 to 2\gevc or a pair of muons with the product of their \pt above 1 to 4\gevgevcc. The software trigger requires a two-, three- or four-track secondary vertex with a significant displacement from any PV. At least one charged particle must have \pt greater than 1\gevc and be inconsistent with originating from a PV. A multivariate algorithm~\cite{BBDT} is used for the identification of secondary vertices consistent with the decay of a $b$-hadron.

Simulation is required to model the effects of the detector acceptance and the
imposed selection requirements.
In the simulation, $pp$ collisions are generated using
\pythia~\cite{Sjostrand:2007gs,*Sjostrand:2006za} 
with a specific \lhcb configuration~\cite{LHCb-PROC-2010-056}.
Decays of unstable particles
are described by \evtgen~\cite{Lange:2001uf}, in which final-state
radiation is generated using \photos~\cite{davidson2015photos}.
The interaction of the generated particles with the detector, and its response,
are implemented using the \geant
toolkit~\cite{Allison:2006ve, *Agostinelli:2002hh} as described in
Ref.~\cite{LHCb-PROC-2011-006}.

Corrections derived from the data are applied to the simulation to account for mismodelling 
of the charge multiplicity of the event, the $B^0$ momentum spectrum and the $B^0$ vertex quality.
Similarly, the simulated particle identification (PID) performance is corrected to match that 
determined from control samples selected from the data~\cite{Anderlini:2202412,Aaij:2018vrk}.


\section{Selection of signal candidates}
\label{sec:selection}

The same dataset analysed in Ref.~\cite{LHCb-PAPER-2020-002} is considered for this measurement, as well as the same simulated samples and data-simulation corrections.
Signal candidates are formed from a pair of oppositely charged tracks that are identified as muons, combined with a $K^{*0}$ candidate,  
formed from two  tracks identified as a kaon and a pion, respectively. 
Signal candidates are required to have a reconstructed $K^+\pi^-\mumu$ invariant mass, $m_{K\pi\mu\mu}$, in the range $[5170, \, 5700]\mevcc$ and a reconstructed mass of the $K^+ \pi^-$ system, $m_{K\pi}$, in a window of $\pm 100\mevcc$ around the known mass of the $K^{*0}$ meson~\cite{PDG2022}.
In addition, the following requirements are imposed, following the strategy in Ref.~\cite{LHCb-PAPER-2020-002}.
The tracks of the four final-state particles must have a significant impact parameter with respect to all PVs in the event and must form a good-quality  vertex. 
The $B^0$ candidate must be compatible with being produced in one of the PVs, \textit{i.e.} the IP is small, and its decay vertex must be significantly displaced from the identified PV.
The angle between the reconstructed $B^0$ momentum and the vector connecting the identified PV to the reconstructed $B^0$ decay vertex is also required to be small.

Two types of backgrounds are considered: combinatorial background, where the selected  final-state particles do not originate from a single $b$-hadron decay; and peaking backgrounds, where specific background processes can mimic the signal if their final-state particles are misidentified or misreconstructed.
Combinatorial background events are further reduced using a boosted decision tree (BDT) algorithm~\cite{Breiman,AdaBoost}. 
The BDT classifier used for the present analysis is identical to that described in Ref.~\cite{LHCb-PAPER-2020-002} and the same working point is used. 
The BDT selection rejects more than 97\% of the remaining combinatorial background, while retaining more than 85\% of the signal. 

Peaking backgrounds arise from decays of $B_s^0 \to \phi(1020)(\to K^+K^-)\mumu$ and $\Lb \to pK^- \mumu$ where a kaon or proton is misidentified as a pion, and
$\Bzb \to \Kstarzb \mumu$ 
in which the kaon and pion from the $\Kstarzb$ decay are both misidentified.
These contributions are rejected by applying dedicated vetoes combining PID information with the invariant mass of the candidates
recomputed with the relevant change in the particle mass hypotheses.
Cabibbo-suppressed decays of $\Bsb \to K^{*0} \mumu$, which have the correct final state, contribute at the level of 1\% of the signal~\cite{LHCb-PAPER-2018-004} and are neglected when considering the $B^0\to K^{*0}\mu^+\mu^-$ decay.
Decays of $B^+ \to K^+ \mumu$ can pass the above selection if a low-momentum pion from the rest of the event 
is added to form a four-particle final state. 
The resulting invariant mass $m_{K\pi\mu\mu}$ will be larger than the known $B^0$ mass 
but can fall into the upper-mass sideband. 
Such decays can therefore distort the estimate of the residual combinatorial background distributions which are assessed from this sideband
and are suppressed by removing candidates with $m_{K\mu\mu}$ within $60\mevcc$ (corresponding to approximately three times the detector resolution) around the known $B^+$ mass~\cite{PDG2022}.
The residual contamination from these peaking backgrounds is studied using simulated events
and is estimated to be at the level of 1\% or less.
Therefore, the peaking backgrounds are sufficiently small to be neglected in the analysis and are considered only as sources of systematic uncertainty.


\section{Amplitude fit method}
\label{sec:amplitude_fit}

\begin{figure}[t]
\centering
\includegraphics[width=0.65\textwidth]{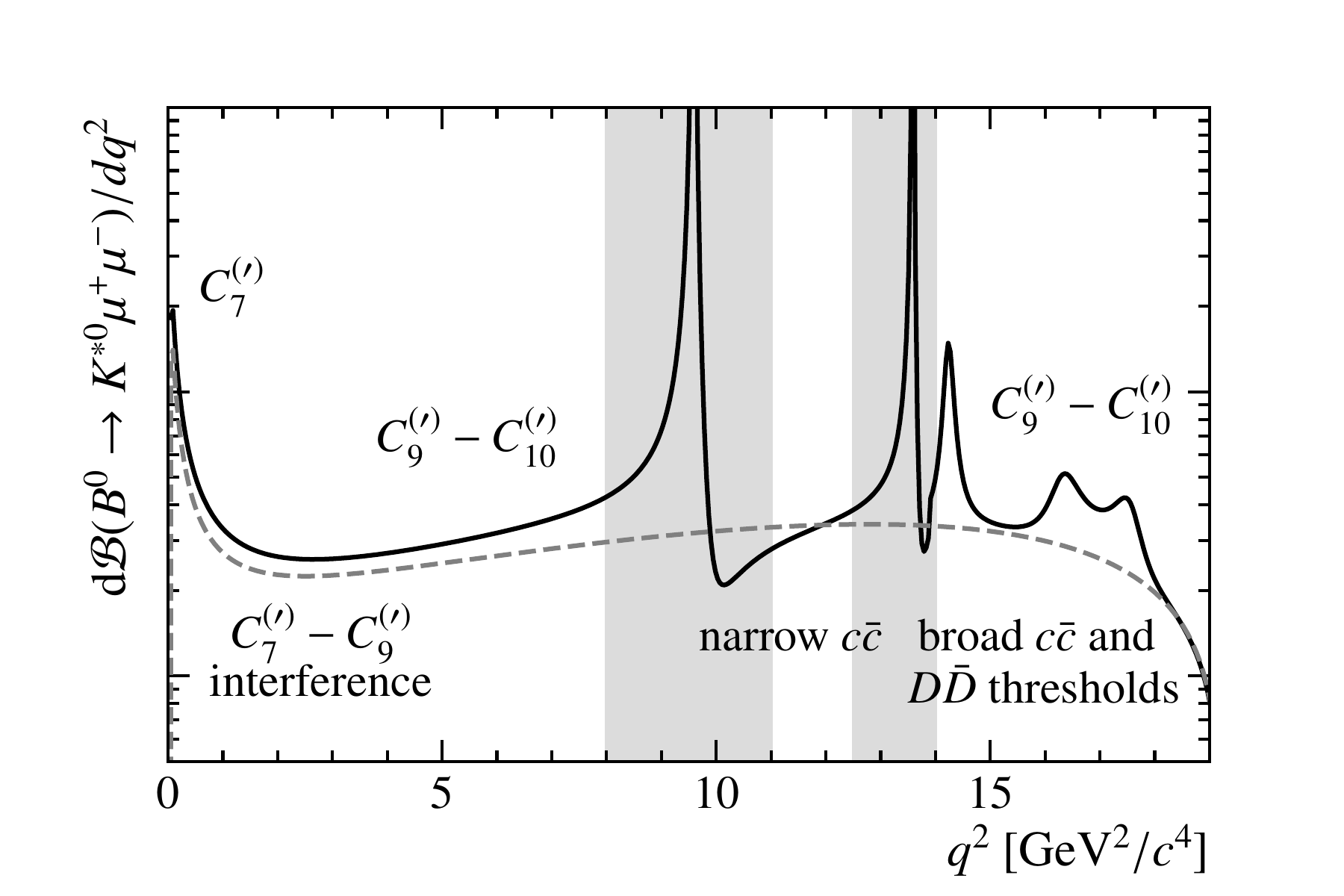}
\caption{Illustration of the $q^2$ spectrum in $B^0\to K^{*0}\mumu$ decays. The dashed line corresponds to the pure rare semileptonic decay, while the solid line includes the impact of different charmonium resonances. The gray bands correspond to the regions dominated by $B^0\to \jpsi K^{*0}$ and $B^0\to \psitwos K^{*0}$ tree-level decays, which are selected as control channels. Magnitudes and phases of $c\bar{c}$ resonant components have been arbitrarily chosen for illustrative purposes. The dominant Wilson coefficients in each region of the spectrum are also highlighted for reference.
 \label{fig:q2}}
\end{figure}

The $q^2$ spectrum of $\Bz \to \Kstarz \mumu$ decays is shaped by a variety of contributions, as illustrated in Fig.~\ref{fig:q2}. 
In this analysis, signal candidates are only considered in two $q^2$ regions, \textit{i.e.} $[1.1, \, 8.0]$ and $[11,\, 12.5]\gevgevcccc$.
The region below $1.1\gevgevcccc$ is not investigated due to the presence of light-quark resonances, 
\textit{e.g.} $\eta, \, \rho^0(770), \, \omega(792)$ and $\phi(1020)$, that would need to be explicitly modelled in the fit.
Similarly, the $q^2$ region above the $\psitwos$ resonance is populated by  broad charmonium states,
\textit{e.g.} $\psi(3770), \, \psi(4040)$, that lie beyond the region of validity of the
$\mathcal{H}_\lambda(q^2)$ parameterisation described in Sec.~\ref{sec:formalism}. 
Future theoretical developments on the treatment of non-local hadronic matrix elements are necessary
in order to incorporate such contributions in the analysis.
On the contrary, narrow charmonium resonances appearing in 
$B^0 \to \psi_n K^{*0}$ tree-level decays, with $\psi_n = \jpsi$ or $\psi(2S)$ and $\psi_n \to \mumu$,
are embedded in the described formalism  and appear as poles of the non-local hadronic 
parameterisation as shown in Eq.~\ref{eq:H}.
Regions with $q^2 \in [8.0, \, 11.0]$ and $[12.5, \, 14.5]\gevgevcccc$, which are dominated by 
these resonant decays, are removed from the signal selection 
and considered only as control regions for the analysis.

Since \Bz and \Bzb decays are treated together in this analysis, 
no sensitivity to the imaginary part of the Wilson coefficients can be achieved, and thus 
all these coefficients are assumed to be real without loss of generality. 
The  parameters of interest are determined using an extended unbinned maximum-likelihood fit. The fit is  performed simultaneously on the 2011 and 2012 (referred to as Run~1) and 2016 datasets and on the two 
$q^2$ regions of the analysis, with all the physics signal parameters shared in the fit.

\subsection{Efficiency correction}
\label{sec:efficiency}

The reconstruction and selection of signal candidates distort the distributions of the decay angles and $q^2$.
To account for this acceptance effect, a four-dimensional acceptance function is introduced according to
\begin{equation}
\label{eq:acceptance}
 \sum_{k,l,m,n} c_{klmn} L_k(\cos\theta_K) L_l(\cos\theta_{\ell})  L_m(\phi) L_n(\qsq),
\end{equation}
where $L_m(x)$ are Legendre polynomials in $x$ of order $m$, and $q^2$ is parameterised in the range between 1 and $14\gevgevcccc$,
slightly beyond the $q^2$ domain of this analysis in order to avoid border effects.  The $\phi$ angle and $q^2$
are each rescaled to the range $[-1, 1]$ when evaluating the Legendre polynomial.
The acceptance is modelled using polynomials of the lowest order that show
a good description of the detector response;
Legendre polynomials up to order three are used for $q^{2}$;
for the decay angles, polynomials up to order five, four and six are used for $\cos\thetak$, 
 $\cos\thetal$ 
and the angle $\phi$, respectively.
Only even orders are used for the angle $\phi$.
The coefficients $c_{klmn}$ are determined separately for the Run~1 and 2016 datasets by exploiting the orthogonality of the Legendre polynomials on large samples of simulated  three-body $B^0 \to K^{*0}\mu^+\mu^-$ phase space decays. 
Resolution effects are effectively taken into account by evaluating the acceptance function on the reconstructed angles and $q^2$ after final-state radiation effects.
The impact of the  resolution on $k^2,\theta_\ell, \theta_K$ and $\phi$  is minimal.
The acceptance parameterisation is also found to be independent of
$k^2$ in the range considered in this analysis.
In addition, as the acceptance is parameterised in terms of all the kinematic variables needed to describe the decay, it does not depend on the model used in the simulation.

The validity of the obtained acceptance parameterisation is tested on a large set of simulated signal events
generated based on the physics model provided by \evtgen~\cite{Ali:1999mm,Lange:2001uf}.
The  differential decay rate of Eq.~\ref{eq:d4Gamma_P} is multiplied by the acceptance function of Eq.~\ref{eq:acceptance}  and the extracted signal parameters are found to be in good agreement with the generated values.

\subsection{Signal modelling}
\label{sec:sig_model}

The parameterisation of the signal is built from the five-dimensional differential decay rate of Eq.~\ref{eq:d5Gamma_SP} multiplied by the acceptance function of Eq.~\ref{eq:acceptance}.
The signal yields are directly extracted from the $m_{K\pi\mu\mu}$ distribution, which is included as an additional dimension in the fit to separate signal from background. 
The probability density function (p.d.f.) of $m_{K\pi\mu\mu}$ for the signal 
is parameterised with the sum of two Gaussian functions with power-law tails on both sides of the peak. 
The two Gaussian functions share the same mean and tail parameters but are allowed to have different widths. 
All  the parameters of the signal $m_{K\pi\mu\mu}$ model are fixed from the simulation with the exception of the mean and widths,
which receive a correction 
from a fit to the $m_{K\pi\mu\mu}$ invariant mass distribution of the $B^0 \to \jpsi K^{*0}$ control channel.

The resulting six-dimensional signal p.d.f. depends on the four Wilson coefficients $\mathcal{C}_9^{(\prime)} $ and $\mathcal{C}_{10}^{(\prime)}$, which are varied freely in the fit, and on a large number of parameters that are constrained to different external inputs as described in the next subsections.
These are the 4 CKM parameters, 19 (9) local P-wave (S-wave) form factor parameters and between 18 and 30 non-local hadronic parameters depending of the truncation order of $\mathcal{H}_\lambda(z)$.
External constraints are applied in the fit by multiplying the likelihood function by multi-dimensional Gaussian functions.
Finally, the two parameters describing the relative magnitude and phase between the P- and S-waves, as introduced in Appendix~\ref{app:mKpi_lineshapes}, are also free to vary in fit.

\subsection{External inputs}
\label{sec:external_inputs}

Several external inputs are used to constrain different parts of the signal decay amplitudes.
The CKM elements $V_{tb}V^*_{ts}$ are expressed through the Wolfenstein parameterisation, whose parameters $\lbrace \lambda, A, \bar{\rho}, \bar{\eta}\rbrace$ are constrained to the values obtained from an SM fit of the unitarity triangle~\cite{Charles:2004jd}, as summarised in Table~\ref{tab:ckm}.

\begin{table}[t]
\caption{Central values and uncertainties used to constrain the CKM parameters in the fit.
All values are taken from the CKMfitter group as of Summer 2019~\cite{Charles:2004jd}. }
    \centering
    \renewcommand{\arraystretch}{1.2}
    \begin{tabular}{ c    c   }
    \toprule
    CKM parameter	&	 Value	\\
    \midrule
		$A$		&	$0.8235 \pm 0.0145$ 	\\
		$\lambda$		&	$ 0.224837 \pm 0.000251$ 	\\
		$\bar{\eta}$		&	$0.3499 \pm  0.0079$ 	\\
		$\bar{\rho}$		&	$0.1569 \pm 0.0102$ 	\\
    \bottomrule  	      
    \end{tabular}
\label{tab:ckm}
\end{table}

Local form factors can be assessed either from first principles through lattice QCD (LQCD) simulations~\cite{Becirevic:2006nm,Horgan:2013hoa}, or from quark-hadron-duality arguments through QCD light-cone sum rules (LCSR)~\cite{Ball:1998kk,Khodjamirian:2006st}. 
In this analysis, $B^0 \to K^{*0}$ form factors $\mathcal{F}_\lambda(q^2)$ are constrained to the combination of LCSR~\cite{Gubernari:2018wyi,Gubernari:2022hxn}
and LQCD~\cite{Horgan:2015vla} results 
provided by the \eos software package~\cite{EOSAuthors:2021xpv}, with  the correlations amongst the form factor parameters fully taken into account.
The current knowledge on the $B^0 \to [K \pi]_{J=0}$ scalar form factors is very limited and
only recently  these effects have been investigated in the context of $B^0 \to K^+ \pi^- \ell^+ \ell^-$ decays~\cite{Descotes-Genon:2023ukb}.
The resulting $q^2$ dependence is found to be almost entirely driven by the kinematic factors.
In the following, the S-wave form factors $f_+, \,  f_T$ and $f_0$ are constrained to those obtained for $B^+ \to K^+$ transitions from LQCD~\cite{Bailey:2015dka}, with uncertainties inflated by a factor of three to compensate for the different meson species.
This assumption is further studied as a source of systematic uncertainty with alternative parametric expressions~\cite{Doring:2013wka} and it is found to be consistent with the numerical analysis of Ref.~\cite{Descotes-Genon:2023ukb}.

While the  determination of the non-local hadronic contributions $\mathcal{H}_\lambda(q^2)$
is one of the key aspects of this analysis, additional external inputs can be used to constrain these elements.
First, experimental measurements of the decays $B^0 \to \psi_{n}K^{*0}$ 
can provide information on the magnitude and phase of the non-local matrix elements evaluated at the pole of the corresponding resonance. 
The amplitude of these decays can be expressed in terms of the residues of the functions ${\mathcal{H}}_\lambda(q^{2})$ at the $\psi_n$ poles as~\cite{Bobeth:2017vxj}
\begin{equation}
\label{eq:Residue_H_lambda}
\operatorname*{Res}_{q^2\to M^2_{\psi_n}} \frac{\mathcal{H}_\lambda(q^2)}{\mathcal{F}_\lambda(q^2)}  = 
  \frac{M_{\psi_n} f^*_{\psi_n} {\cal{A}}^{\psi_n}_\lambda}{M_B^2\, {\mathcal{F}}_\lambda(M_{\psi_n}^2)}\,, 
\end{equation}
where the $M_{\psi_n}$ and $f^*_{\psi_n}$ are masses and decay constants for the $\psi_n$ resonances~\cite{PDG2022,Khodjamirian:2010vf,Beneke:2006hg}, respectively, 
and ${\cal{A}}^{\psi_n}_\lambda$ are the transversity amplitudes of the tree-level decays to charmonia. 
These amplitudes are experimentally related to the measured polarisation fractions and phase differences~\cite{Aubert:2007hz,Chilikin:2013tch,Chilikin:2014bkk,LHCb-PAPER-2013-023}
\begin{eqnarray} 
\label{eq:psi_frac_and_phase_diff}
f^{\psi_{n}}_\lambda=\frac{|\mathcal{A}^{\psi_{n}}_{\lambda}|^2}{|\mathcal{A}^{\psi_{n}}_0|^2+|\mathcal{A}^{\psi_{n}}_\parallel|^2+
|\mathcal{A}^{\psi_{n}}_\perp|^2}\,,~~
  \delta^{\psi_{n}}_{\parallel,\perp} = \textrm{arg}\frac{\mathcal{A}^{\psi_{n}}_{\parallel,\perp}}{\mathcal{A}^{\psi_{n}}_{0}}\,,
\end{eqnarray}
and to branching fraction measurements~\cite{Chilikin:2013tch,Chilikin:2014bkk,LHCb-PAPER-2012-010}
\begin{equation}
\label{eq:BRJpsiKst}
\mathcal{B}(B^0 \to \psi_{n} K^{*0}) = \frac{\tau_B}{\hslash} \frac{ \sqrt{\lambda(M_B^2,M^2_{\psi_n},M_{K^{*0}}^2)}}{2 \pi M_B} G_F^2 |V_{cb} V^*_{cs}|^2 \sum_\lambda | \mathcal{A}^{\psi_{n}}_\lambda|^2 \, ,
\end{equation}
where $M_{K^{*0}}$ is the pole mass of the $K^{*0}$ meson and $V_{cb}$ and $V^*_{cs}$ are elements of the CKM matrix.
Table~\ref{tab:inputs_B2KstPsi} collects all the external inputs from $B^0 \to \psi_{n}K^{*0}$ measurements 
used to constrain the values of $\mathcal{H}_\lambda(q^2)$ in the analysis.

\begin{table}[t]
\caption{Summary of external inputs from $B^0 \to \psi_{n}K^{*0}$ measurements used in the analysis.
When measurements from different experiments are available, these are averaged taking into account correlations.
A shift of $+\pi$ is introduced for the phase differences $\delta^{\psi_{n}}_{\perp,\parallel}$ to account for the different convention 
between this analysis and Refs.~\cite{Aubert:2007hz,Chilikin:2014bkk,LHCb-PAPER-2013-023}.  }
\centering
\renewcommand{\arraystretch}{1.2}
\begin{tabular}{
  c 
  r 
  @{ $\pm$ }  
  l 
  @{\hspace{-5ex}}  
  r
  r 
  @{ $\pm$ }  
  l 
  @{\hspace{0ex}}  
  l
  }
\toprule
						&  \multicolumn{3}{c}{$B^0 \to \jpsi K^{*0}$}	&  \multicolumn{3}{c}{$B^0 \to \psi(2S) K^{*0}$} 	\\
\midrule	     
   $f_0$				&			\multicolumn{2}{c}{-}	&			&		0.455 & 0.057 & \cite{Chilikin:2013tch}		\\
	$f_\parallel$		&		0.227 & 0.006 & \cite{Aubert:2007hz,Chilikin:2014bkk,LHCb-PAPER-2013-023}	 &		0.22 & 0.06 & \cite{Aubert:2007hz}			\\
	$f_\perp$			&		0.209 & 0.005 & \cite{Aubert:2007hz,Chilikin:2014bkk,LHCb-PAPER-2013-023}	 &		0.30  & 0.06 & \cite{Aubert:2007hz}		\\
	$\delta_\parallel$	&		0.20  & 0.03 & \cite{Aubert:2007hz,Chilikin:2014bkk,LHCb-PAPER-2013-023}	 &		0.34  & 0.4 & \cite{Aubert:2007hz}		\\
	$\delta_\perp$		&		$-0.21$  & 0.03 & \cite{Aubert:2007hz,Chilikin:2014bkk,LHCb-PAPER-2013-023}	 &		$-0.34$  & 0.3 & \cite{Aubert:2007hz}			\\
    \midrule  	     
	$\mathcal{B}(B^0 \to \psi_n K^{*0})$		&	$(1.19$ &  $0.08)\times 10^{-3}$ & \cite{Chilikin:2014bkk}  &  $(5.55$ & $0.87)\times 10^{-4}$ & \cite{Chilikin:2013tch}	\\
	$\frac{\mathcal{B}(B^0 \to \psi(2S) K^{*0})}{\mathcal{B}(B^0 \to \jpsi K^{*0})}$		&	\multicolumn{6}{c}{$0.487\, \pm \, 0.021 $~\cite{LHCb-PAPER-2012-010}	}	\\
    \bottomrule  	      
\end{tabular}
\label{tab:inputs_B2KstPsi}
\end{table}

Theoretical predictions of the ratio $\mathcal{H}_\lambda / \mathcal{F}_\lambda$ 
can be obtained at $q^2 <0 \gevgevcccc$.
In this regime, 
the effects of four-quark operators can be treated perturbatively
and SM theory predictions can be accessed by a light-cone operator-product 
expansion~\cite{Khodjamirian:2010vf}.
Precise predictions of the real and imaginary parts of the ratio $\mathcal{H}_\lambda / \mathcal{F}_\lambda$ are provided by Refs.~\cite{Gubernari:2022hxn,Gubernari:2020eft}
at three $q^2$ points, \textit{i.e.} $q^2 \in \{-7,-5,-3\}\gevgevcccc$.
While information at $q^2 = -1 \gevgevcccc$ is also available,
in this measurement it is used instead as 
a test point to check the consistency of the result of the fit.
In order to explore the impact and the consistency of these theory predictions, two fit configurations are considered in the analysis: 
the first includes external constraints to the above-mentioned theoretical prediction on the values of $\mathcal{H}_\lambda/ \mathcal{F}_\lambda$ at negative $q^2$, hereafter referred to as ``$q^{2} < 0$ constraints'';
while the second ignores those constraints and provides a pure data-driven determination of the non-local hadronic matrix elements, referred to as ``$q^{2} > 0$ only''.

\subsection{Constraints from branching fraction determination}
\label{sec:branching_fraction}

\begin{figure}[t]
\centering
  \includegraphics[width=0.49\textwidth]{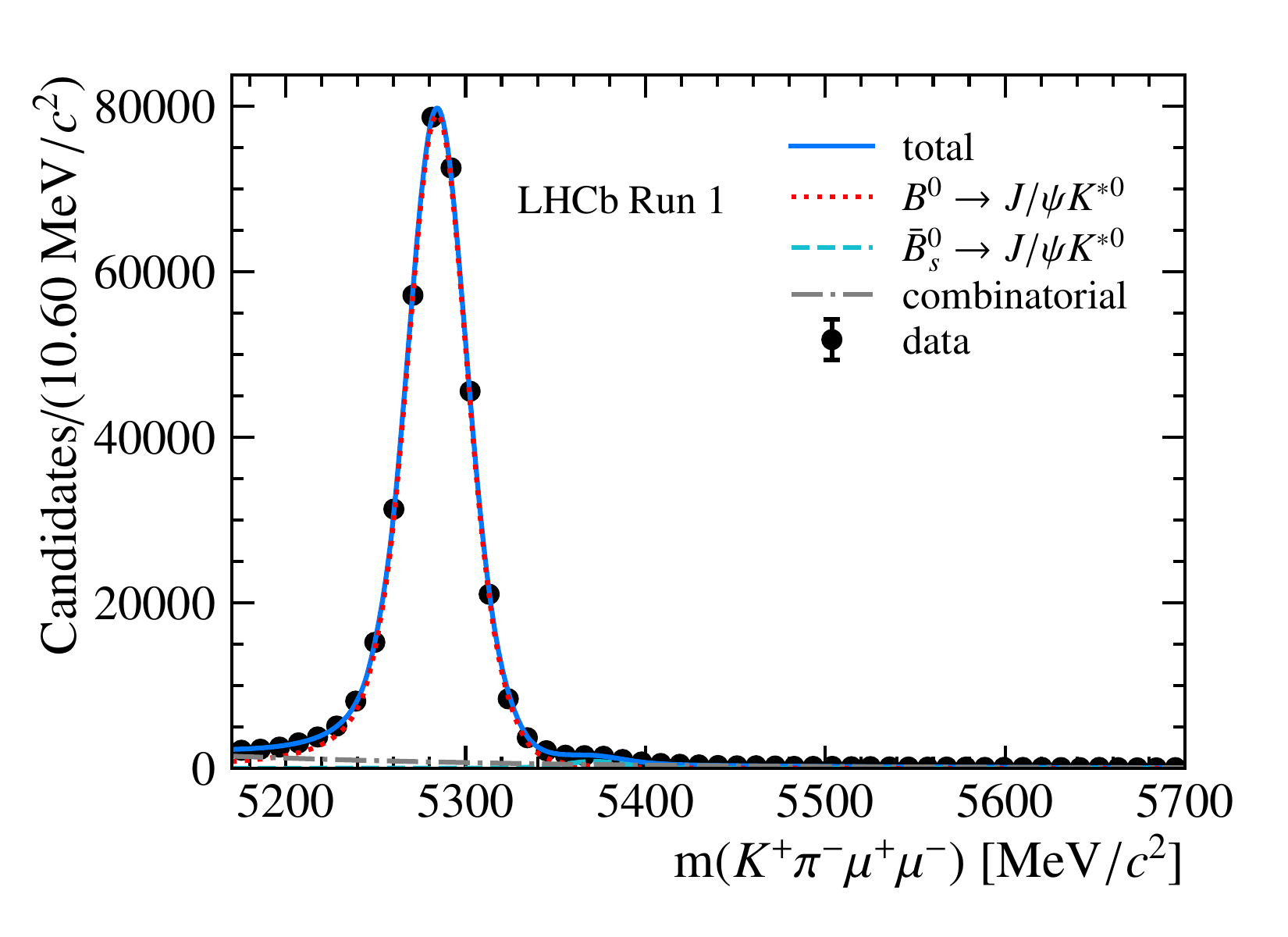}
  \includegraphics[width=0.49\textwidth]{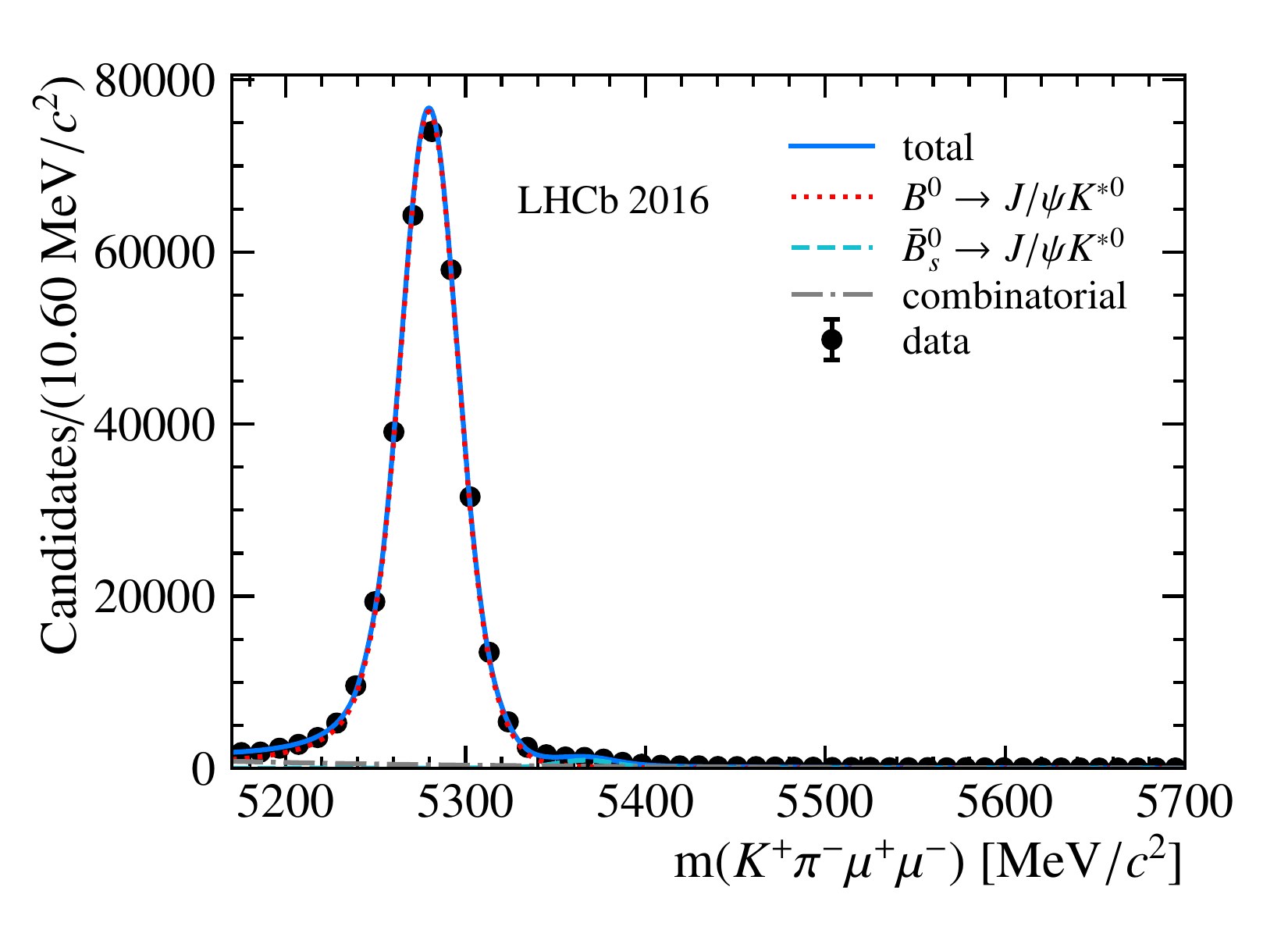}
\caption{Invariant mass distribution of $B^0 \to \jpsi K^{*0}$  decays for the (left) Run~1 and (right) 2016 datasets. 
\label{fig:Jpsi_mass}}
\end{figure}

The estimation of the signal yield can be easily incorporated in the amplitude analysis	by performing an extended unbinned maximum likelihood fit.
In addition, the observed signal yield in the $k^2$ and $q^2$ mass windows 
\begin{equation}
\label{eq:nsig}
  N_{\rm sig} = N_{\jpsi K\pi} \times \frac{{\mathcal{B}}_{\mathrm{win}}(B^0 \to K^{*0} \mu^+\mu^-) \times {\left(\frac{2}{3}\right)} \times F_{\rm P}^{\,-1}  }{\mathcal{B}(B^0 \to \jpsi K^+ \pi^-)  \times f^{B^0 \to \jpsi K \pi}_{\pm 100\mathrm{MeV}}  \times \mathcal{B}(\jpsi \to \mu^+\mu^-)}   \times R_\varepsilon 
\end{equation}
can be conveniently related to the signal branching fraction via
\begin{equation}
\label{eq:Br_Kstmm}
      {\mathcal{B}}_{\mathrm{win}}(B^0 \to K^{*0} \mu^+\mu^-)  = \frac{\tau_B}{\hbar}  \int_{q^2_{\min}}^{q^2_{\max}} \int_{k^2_{\min}}^{k^2_{\max}} \frac{{\textrm{d}}^2 \Gamma^{\rm P}}{{\textrm{d}} q^2 {\textrm{d}} k^2} {\textrm{d}} q^2 {\textrm{d}} k^2  \, ,
\end{equation}
where $\tau_B$ is the lifetime of the \Bz meson and the two-dimensional differential decay rate is obtained by integrating the full P-wave decay rate over the three angles.
The  factor $(\frac{2}{3})$ is the squared Clebsch--Gordan coefficient related to $K^{*0} \to K \pi$ that corresponds to the $K^{*0} \to K^+ \pi^-$ decay probability,
while the term $F_{\rm P}$, built from the fraction of the  P-wave over the total P- and S-wave decay rates,
is introduced to correct for the S-wave contribution observed in the signal region.
Similarly,  the  observed yield in the control channel, $N_{\jpsi K\pi}$, which 
is determined from a fit to the $K^+ \pi^- \mumu$ invariant mass distribution as shown in Fig.~\ref{fig:Jpsi_mass},
includes a combination of P- and S-wave  as well as contributions of exotic nature, such as $ B^0 \to Z_c(4430)^- K^+$  
and $ B^0 \to Z_c(4200)^- K^+$ decays,  with $Z_c^-$ decaying to $\jpsi \pi^-$.
Given that all these contributions share the same final state, they appear as single component in the fit.
The observed $N_{\jpsi K\pi}$ yield is found to be $348\,500 \pm 600$ and $328\,500 \pm 600$ for Run~1 and 2016, respectively, where uncertainties are statistical only,
while contamination from $\Bsb \to \jpsi K^{*0}$ decays are found to be of the order of 1\%.
Since an exact separation between the P-wave, S-wave and exotic components can only be achieved by a dedicated amplitude analysis 
that comprises the $\jpsi \pi^-$ invariant mass dimension~\cite{Chilikin:2013tch}, 
the signal branching fraction of Eq.~\ref{eq:Br_Kstmm} is normalised to the inclusive 
branching fraction of $B^0 \to \jpsi K^+ \pi^-$ decays.
%
%
This is taken from Ref.~\cite{Chilikin:2013tch} and is scaled 
 by a correction factor, $f^{B^0 \to \jpsi K \pi}_{\pm 100\mathrm{MeV}}$, which takes into account the fraction of
$B^0 \to \jpsi K^+ \pi^-$ decays that fall into the $m_{K\pi}$ window of this analysis.
This correction factor is estimated with a series of pseudoexperiments generated from the amplitude model of Ref.~\cite{Chilikin:2013tch}.
Unfortunately, neither the correlations nor the systematic uncertainties associated to the individual amplitudes 
are provided in Ref.~\cite{Chilikin:2013tch} and $f^{B^0 \to \jpsi K \pi}_{\pm 100\mathrm{MeV}}$ can only be 
evaluated under the conservative assumption of uncorrelated uncertainties.
Within this hypothesis, $f^{B^0 \to \jpsi K \pi}_{\pm 100\mathrm{MeV}}$ is found to be 
$ 0.644 \pm 0.010$, whose central value is fixed in the fit and the associated uncertainty is treated
as a source of systematic uncertainty.
Finally, the charmonium branching fraction, $\mathcal{B}(\jpsi \to \mu^+\mu^-)$, is taken from Ref.~\cite{PDG2022}
and the relative efficiency between the signal and control modes, $R_\varepsilon$, 
is obtained from simulated samples.
The use of simulation to model the efficiency is validated on the $B^0 \to \psi(2S) K^{*0}$ control channel,
where the ratio of branching fractions $\mathcal{B}(B^0 \to \psi(2S) K^{*0}) / \mathcal{B}(B^0 \to \jpsi K^{*0})$ 
is measured precisely and found to be consistent with the value of Ref.~\cite{PDG2022}.

It is important to note that the constraint on the branching fraction  determination sets the scale of the Wilson coefficients. Without this constraint, $|\mathcal{C}_9|$ and $|\mathcal{C}_{10}|$ would not be bounded and only ratios of Wilson coefficients could be assessed.

\subsection{Background modelling}
\label{sec:background}

The combinatorial background that passes the selection requirements of Sec.~\ref{sec:selection} must be modelled in the fit.
These candidates are described by an exponential function in $m_{K\pi\mu\mu}$, by Chebyshev polynomials of up-to second order for the decay angles and $q^2$, and by the sum of a linear function and a Breit--Wigner amplitude-squared for $k^2$. The Breit--Wigner model 
accounts for background candidates originating from a real $K^{*0}$ resonance associated with an unrelated \mup \mun pair. 
The ratio between this resonant component and the linear term  is fixed to the value observed in the upper-mass sideband.
In addition, a sculpting effect is introduced by the veto against $B^+ \to K^+ \mumu$ decays, 
strongly distorting a specific region of the $(\cos \thetak, \, q^2, \, m_{K\pi\mu\mu})$ phase space.
This effect is studied using the data:
events reconstructed in the lepton-flavour-violating channel $B^0 \to K^{*0} \mu^\pm e^\mp$ are used to study the impact 
of the veto on the variables of interest in absence of peaking $B^+ \to K^+ \mu^\pm e^\mp$ backgrounds. 
A three-dimensional efficiency map is extracted from the ratio of templates obtained with and without the veto selection applied, 
and is multiplied to the background shape as a correction factor to retrieve the original  distributions.
Besides this term, all remaining functions are assumed to factorise.
This assumption, together with the choice of the order of the polynomials, is investigated as a source of systematic uncertainty in Sec.~\ref{sec:systematics}.
Finally, the parameterisation of the combinatorial background is treated separately for the Run~1 and 2016 datasets as well as for the two considered $q^2$ regions 
$[1.1, \, 8.0]$ and $[11.0, \, 12.5]\gevgevcccc$.
All coefficients are allowed to vary in the fit.


\section{Fit to data}
\label{sec:data}

The polynomial expansion introduced in Eq.~\ref{eq:Hhat_definition} used to parameterise the non-local hadronic matrix elements
$\mathcal{H}_\lambda$ must be truncated at a certain order $z^n$.
Furthermore, the central point of the expansion $t_0$ is a free parameter of the model 
and its choice can have an impact on how fast the polynomial expansion converges.
In general, a sensible choice is a value of $t_0$ that minimises $|z|$ on the domain of the expansion.
As originally proposed by Ref.~\cite{Bobeth:2017vxj},
the choice of
\begin{equation} 
t_0 = t_+ - \sqrt{t_+ ( t_+ - M^2_{\psi(2S)})} \simeq 11.8 \gevgevcccc
\end{equation}
is the one that minimises $|z|$ in the interval $-7 \gevgevcccc \leq q^2 \leq M^2_{\psi(2S)}$;
this value is taken as the default for the fit configuration with $q^{2} > 0$ information only.
However, due to the strong constraints provided by the three $q^2$  points, 
$t_0$ is fixed to zero for the fit model with the $q^{2} < 0$ constraints
in order to best accommodate the theoretical inputs in the negative $q^2$ region.
Following this choice, the truncation order $z^n$ is determined based on a data-driven procedure:
fits are repeated with increasing truncation order for the polynomial sums, \textit{i.e.} $n=2,3,4,5$, 
and the Akaike information criterion~\cite{Akaike} is used to infer the importance of each additional set of coefficients. 
The improvement between two subsequent orders is considered to be significant if $2 \Delta \log \mathcal{L} > 2 \Delta N_{\rm pars}$, where $N_{\rm pars}$ is the number of parameters of the model and each additional order $z^{n+1}$ brings one complex coefficient for each of the three polarisations, for a total of six additional parameters.
For the fit model using only inputs at $q^{2} > 0$, it is found 
that a polynomial expansion truncated at $z^2$ is sufficient to describe the data. 
For fits with $q^{2} < 0$ constraints, a significant improvement 
is found with the inclusion of terms up to $z^4$, as shown in Table~\ref{tab:H_zn}.
The quality of the fit is assessed using an unbinned goodness-of-fit test based on point-to-point dissimilarity methods~\cite{Williams:2010vh} and the $p$-value is found to be larger than 10\%.

\begin{table}[t]
\caption{Log-likelihood differences 
between the fits to data with different truncation orders of the non-local hadronic parameterisation ${\mathcal{H}}_\lambda[z^{n}]$ for the two considered fit configurations. 
}
\begin{center}
\begin{tabular}{ c  c  c }
 \toprule
          &      \multicolumn{2}{c}{$2 \Delta \log \mathcal{L}$} \\
         &  $q^{2} < 0$ constraints	 &  $q^{2} > 0$ only  \\
\midrule
${\mathcal{H}}_\lambda[z^{3}] - {\mathcal{H}}_\lambda[z^{2}]$      &  -  &    3.6   \\
${\mathcal{H}}_\lambda[z^{4}] - {\mathcal{H}}_\lambda[z^{3}]$      &  21.2   &    -  \\
${\mathcal{H}}_\lambda[z^{5}] - {\mathcal{H}}_\lambda[z^{4}]$      &  8.6   &  -    \\
\bottomrule
\end{tabular}
\label{tab:H_zn}
\end{center}
\end{table}

\begin{figure}[t]
\centering
\includegraphics[width=0.45\textwidth]{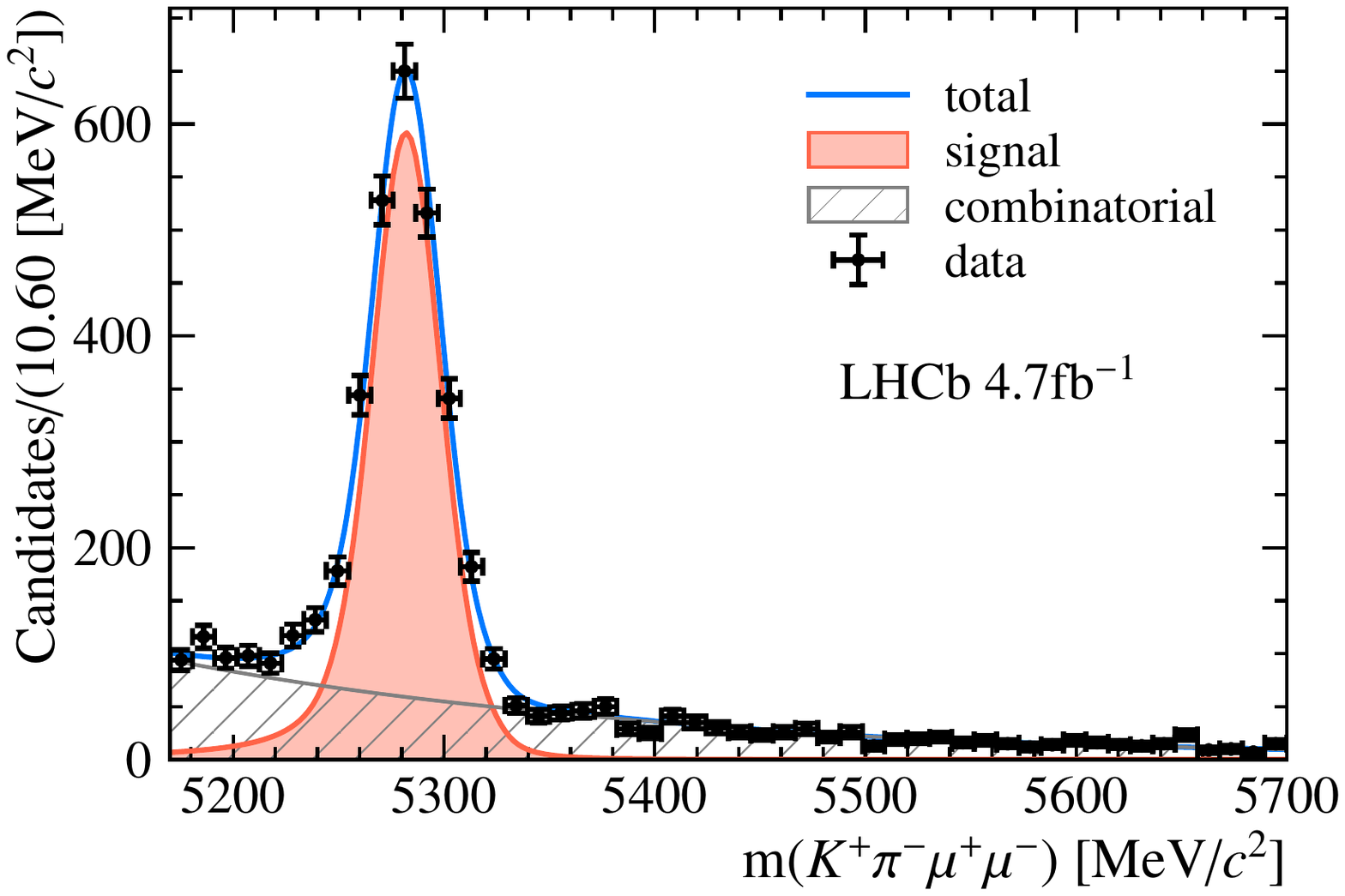}
\includegraphics[width=0.45\textwidth]{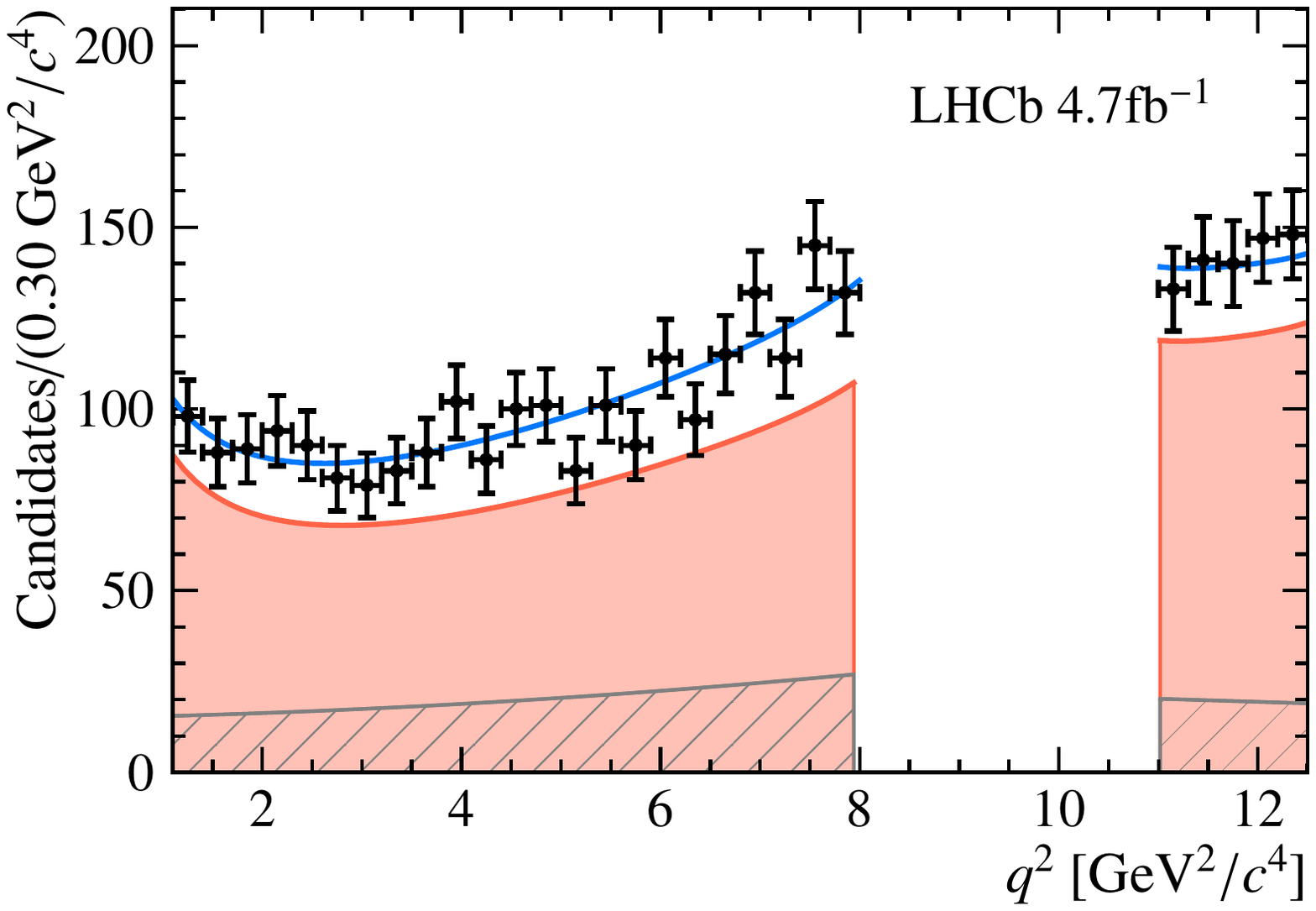} \\
\includegraphics[width=0.45\textwidth]{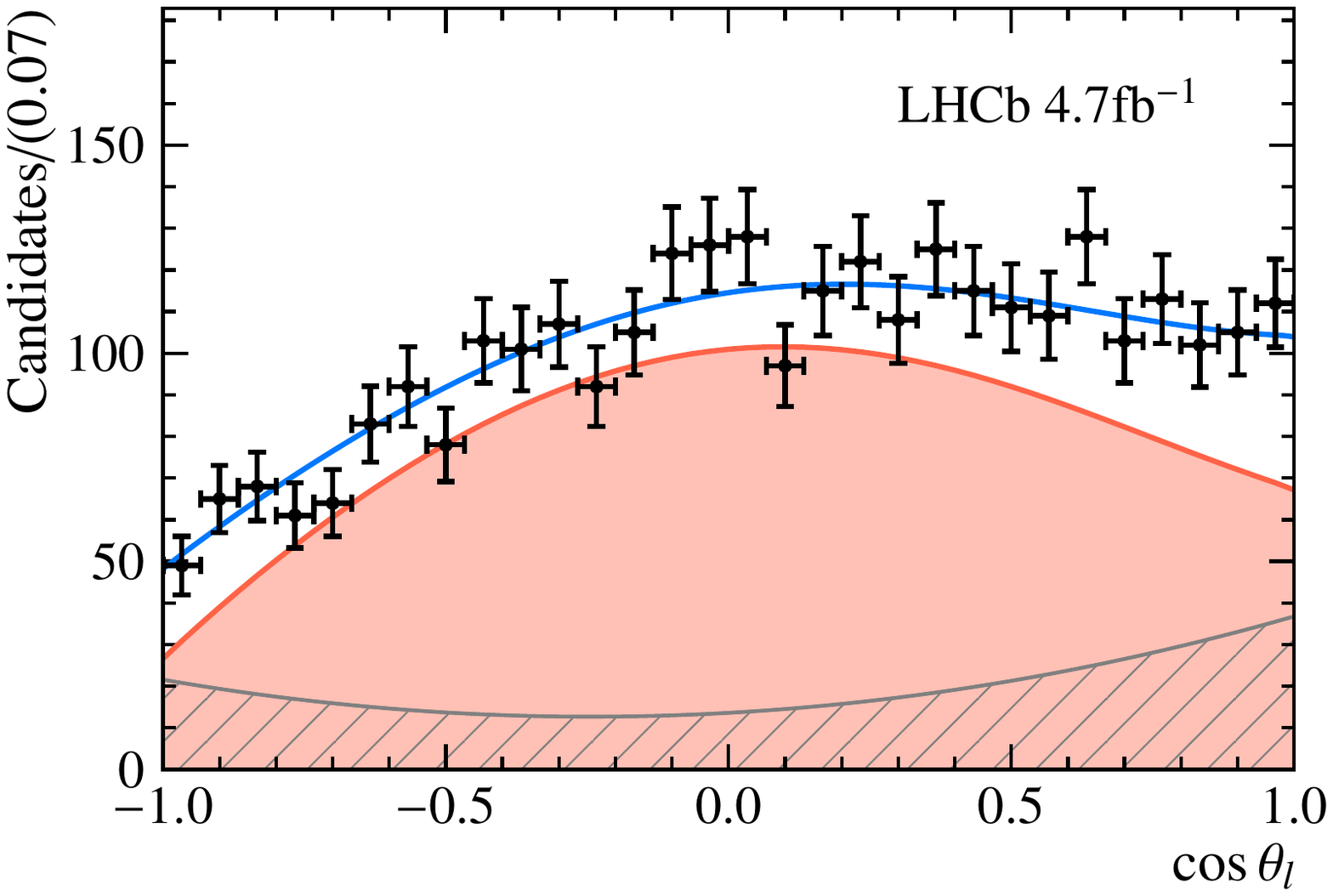} 
\includegraphics[width=0.45\textwidth]{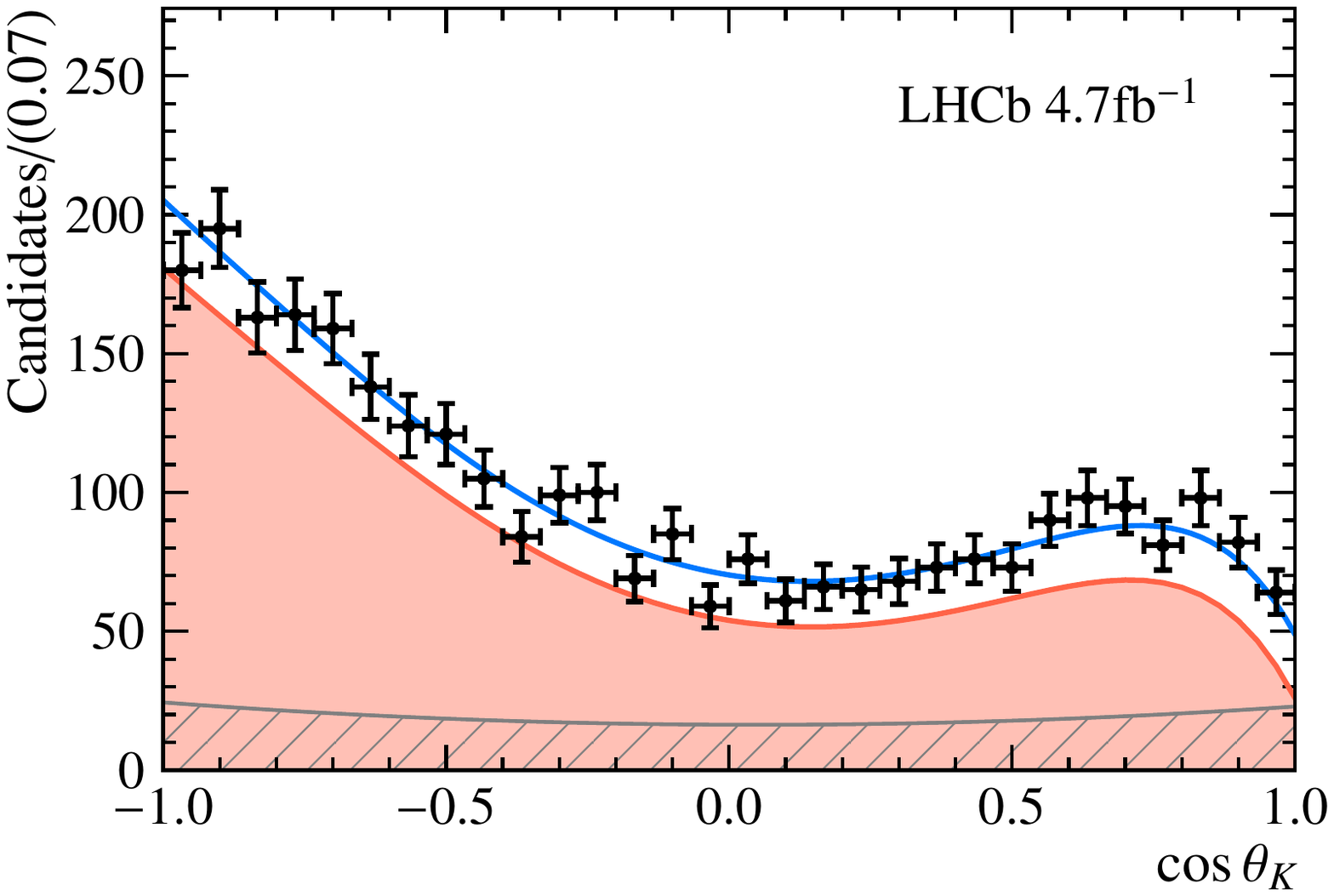} \\
\includegraphics[width=0.45\textwidth]{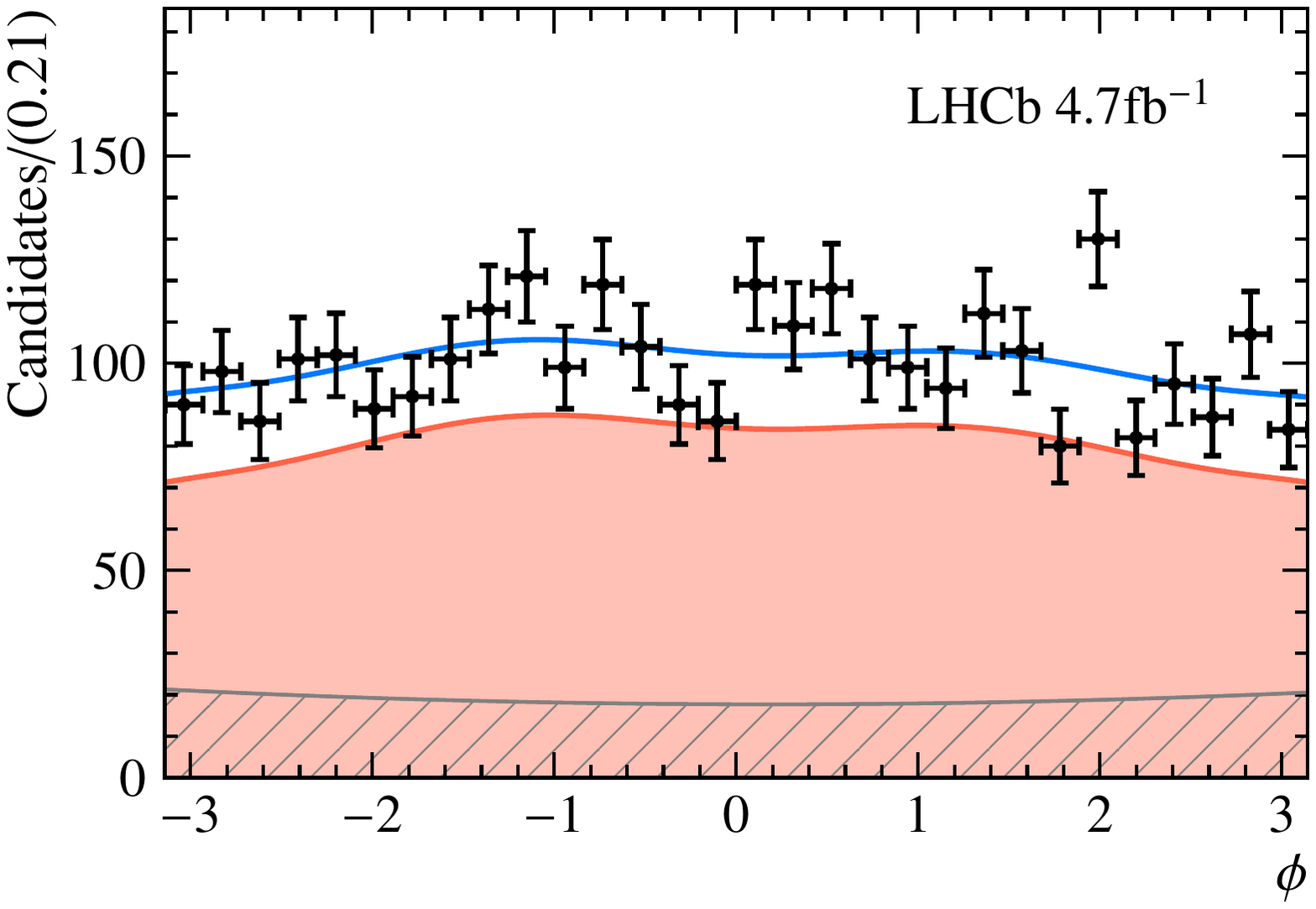}
\includegraphics[width=0.45\textwidth]{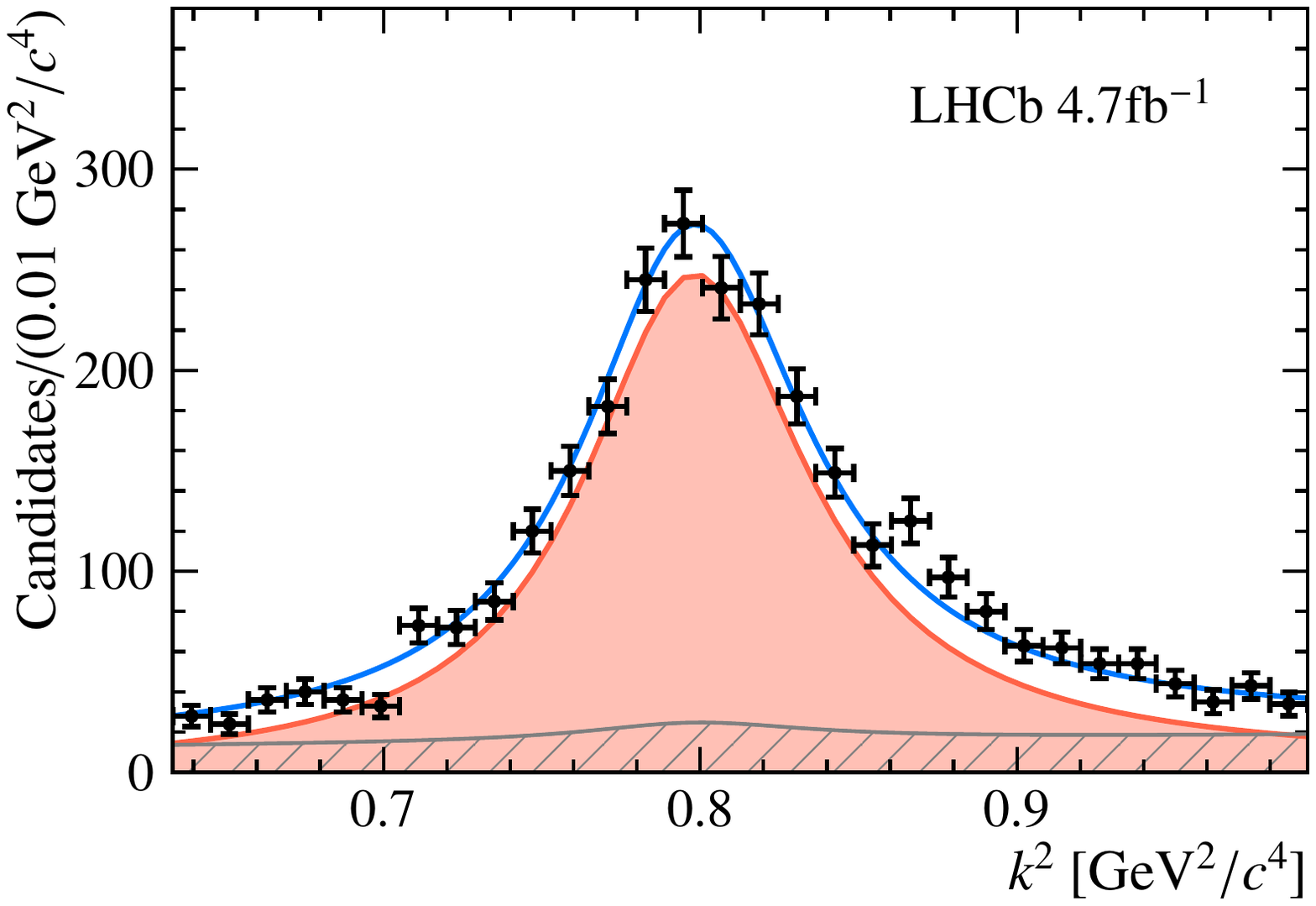} \\
\caption{Distribution of the fit variables for the combined Run~1 and 2016 datasets. 
The distributions of the three angles, $q^2$, and $k^2$ are given for candidates within $50\mevcc$ of the known $B^0$ mass.
The total fit projections together with the individual signal and background components are overlaid.
\label{fig:fit_projection}}
\end{figure}

Figure~\ref{fig:fit_projection} shows the distributions of events for the Run~1 and 2016 combined datasets.
The result of the fit without the $q^{2} < 0$ constraints are overlaid in the figure, no difference in the projections is visible when including $q^{2} < 0$ constraints.

\section{Systematic uncertainties}
\label{sec:systematics}

There are different categories of systematic uncertainties that affect the extraction of the parameters of interest, 
from the choice of the baseline amplitude model and the inclusion of external inputs in the fit,
to imperfect modelling of experimental effects.
The distinct sources of systematic uncertainties are discussed in detail below and are summarised in Table~\ref{tab:syst}. 
The size of each systematic uncertainty is estimated using pseudoexperiments generated using the observed signal and background yields. 
The parameters of interest are determined from these pseudoexperiments 
under the baseline and the systematically varied hypotheses. 
In most of the cases, the average difference between the two results is taken as an estimation of the systematic uncertainty.
Exceptions to this are the systematic uncertainties associated with the use of external inputs and the statistical uncertainty of the efficiency correction, where the standard deviation of the difference 
of the two results is used instead.

\begin{table}[t]
\caption{Summary of the systematic uncertainties on the Wilson coefficients. The individual sources are described in the text. The subtotal and total values are obtained by adding individual sources in quadrature.
}
\begin{center}
\begin{tabular}{ l   r  r  r r }
\toprule
        & $\mathcal{C}_9$\phz & $\mathcal{C}_{10}$\phz & $\mathcal{C}^\prime_9$\phz & $\mathcal{C}^\prime_{10}$\phz \\
\midrule
         Amplitude model    &     &    &   &      \\
\midrule
\hspace{5mm} S-wave form factors      &  $<0.01$ &  $<0.01$ &   $<0.01$ &       $<0.01$ \\
\hspace{5mm} S-wave non-local hadronic    & 0.02  & 0.02  & 0.14  & 0.04 \\
\hspace{5mm} S-wave $k^2$ model      &   $<0.01$ &    $<0.01$ & 0.05  & 0.03 \\ 
\midrule
\hspace{5mm} Subtotal  & 0.02  & 0.02  & 0.15  &  0.05 \\
\midrule[\heavyrulewidth]
         External inputs on $\mathcal{B}(B^0 \to \Kstarz \mumu)$  	&   &  &   &   \\
\midrule
\hspace{5mm} $\mathcal{B}(B^0 \to \jpsi K^+ \pi^-)$ 	&  0.05		 &  0.08	&   0.02	&	0.01  \\
\hspace{5mm} $f^{B^0 \to \jpsi K \pi}_{\pm 100\mathrm{MeV}}$  &  0.03 & 0.03	 &   0.01		&  $<0.01$ \\
\hspace{5mm} Others &  0.03 & 0.04	 &   0.03		&  0.01 \\
\midrule
\hspace{5mm} Subtotal & 0.07  &  0.09  &    0.04  &    0.01  \\
\midrule[\heavyrulewidth]
Background model   &         &            &                &                   \\
\midrule
\hspace{5mm} Chebyshev polynomial order &  $0.01$ &  $0.01$ &  $0.01$ &  $<0.01$ \\
\hspace{5mm} Combinatorial shape in $k^2$  & $0.02$ &  $<0.01$ & $0.02$ &  $<0.01$  \\
\hspace{5mm} Background factorisation &    $0.01$ &  $0.01$ &  $0.01$ &  $0.01$  \\
\hspace{5mm} Peaking background &   $0.01$ & $<0.01$ & $0.02$  &  $0.01$  \\
\midrule
\hspace{5mm} Subtotal  &  0.03  &  0.02  & 0.03  & 0.01  \\
\midrule[\heavyrulewidth]
         Experimental effects   &         &            &                &                   \\
\midrule
\hspace{5mm} Acceptance parameterisation  & $<0.01$ &  $<0.01$ &  $<0.01$ &  $<0.01$ \\
\hspace{5mm} Statistical uncertainty on acceptance &  $0.02$ &  $<0.01$ &  $0.02$ &  $<0.01$ \\
\midrule
\hspace{5mm} Subtotal  &   0.02 &  $<0.01$ &  0.02 &  $<0.01$ \\
\midrule[\heavyrulewidth]
Total systematic uncertainty  &  0.08 &  0.10  &  0.16  &   0.05  \\
\bottomrule
\end{tabular}
\label{tab:syst}
\end{center}
\end{table}

The main sources of systematic uncertainty on the Wilson coefficients $\mathcal{C}_9$ and $\mathcal{C}_{10}$
arise from the use of the external inputs in the determination of the signal branching fraction of Eq.~\ref{eq:nsig}.
This primarily concerns the uncertainty on the normalisation branching fraction $\mathcal{B}(B^0 \to \jpsi K^+ \pi^-) = (1.15 \pm 0.01 \pm 0.05) \times 10^{-3}$~\cite{Chilikin:2014bkk} and 
the fraction of $B^0 \to \jpsi K^+ \pi^-$ decays that fall in the $m_{K\pi}$ window of the analysis \mbox{$f^{B^0 \to \jpsi K \pi}_{\pm 100\mathrm{MeV}} = 0.644 \pm 0.010$}. 
The systematic uncertainties associated to the use of these external inputs are provided separately in view of possible future improvements on these quantities. 
Contributions from 
the uncertainty on the branching fraction of the $\jpsi\to\mumu$ decay, \mbox{$\mathcal{B}( \jpsi \to \mumu) = (5.96 \pm 0.03) \times 10^{-2}$~\cite{PDG2022}},
the uncertainty on the efficiency ratio $R_\varepsilon$, and the uncertainty on the observed yield of the control channel $N_{\jpsi K\pi}$ are also considered and reported together under ``others'' in Table~\ref{tab:syst}.
The uncertainty on $R_\varepsilon$ is due to the finite size of the simulation samples and assumptions made in the simulation model. 
The model dependence of the simulation is studied by varying the signal model in multiple ways:
hadronic parameters are extensively varied within and beyond the SM prediction, large variations of the Wilson coefficients are artificially inserted,
and an S-wave component compatible with what is observed in the fit to data is added.
The different sources are found to contribute to the relative uncertainty on $R_\varepsilon$ at the level of 1--2\%, depending on the $q^2$ region.
Finally, the measurement of the yield in the control channel is found to be systematically dominated,
with the prime sources of uncertainty associated to the choice of the signal mass model and assumptions about the residual background contribution from \decay{\Lb}{p\Km\jpsi} decays.
The different sources contribute to the relative uncertainty on $N_{\jpsi K\pi} $ at a level below 1\%.

The main source of systematic uncertainty for $\mathcal{C}_9^\prime$ comes from 
ignoring the non-local hadronic contribution in the S-wave component.
In absence of any theoretical study on non-local hadronic effects on $K\pi$-scalar amplitudes,
pseudoexperiments are generated assuming a non-local hadronic component which is identical to the one of the longitudinal P-wave amplitude.
Other sources of systematic uncertainties associated to the modelling of the S-wave amplitudes are related to the choice of the S-wave form factors and $k^2$ parameterisation.
The former is assessed by generating pseudoexperiments with the alternative model from Ref.~\cite{Doring:2013wka}, while the latter is assessed by replacing the LASS lineshape with an isobar model built from the sum of the $K^{*}_{0}(700)$ and $K^{*}_0(1430)$ resonances.

For the combinatorial background modelling, three sources of systematic uncertainty are considered.
The first is associated with the choice of second-order polynomials to model the background angular and $q^2$ distributions.
Due to the small number of background candidates, 
the BDT requirement is relaxed and the background candidates selected in the upper-mass sideband are fitted with a fourth-order polynomial in each of the angles and $q^2$. This model is used as an alternative model for the generation of pseudoexperiments.
The second is associated with the modelling of the $k^2$ distribution, where the value of the fraction of the resonant component introduced in Sec.~\ref{sec:background} is varied within its uncertainty.
The third is associated to the assumption of complete factorisation of the background distributions. This is studied in the upper-mass sideband.
A mild non-factorisation between angles $\phi$ and $\thetal$  is observed and an alternative  background model
that does not assume factorisation in these two variables is used for the generation of pseudoexperiments.
In addition, systematic uncertainties are assessed for the different sources of peaking background that are neglected in the analysis.
The distribution of residual peaking-background events is studied in data, after removing PID information from the BDT and inverting the background vetoes.
Events are then drawn from the selected background samples and injected into the pseudoexperiment data.

Finally, two sources of systematic uncertainties are associated to the determination of the acceptance function.
The first is related to the finite size of the simulated samples used to derive the acceptance coefficients and is studied by sampling the obtained coefficients within their covariance matrix.
The second is associated to the choice of the order of the Legendre polynomials used and is investigated by considering a higher-order acceptance parameterisation built from polynomials of order six, seven, eight and four for $\cos \thetal, \, \cos \thetak, \, \phi$ and $q^2$, respectively.


\section{Results}
\label{sec:results}

\subsection{Local form factors}
\label{sec:form_factors}

\begin{figure}[t]
\centering
\vspace{-5mm}
\hspace{-4mm}
\includegraphics[width=0.33\textwidth]{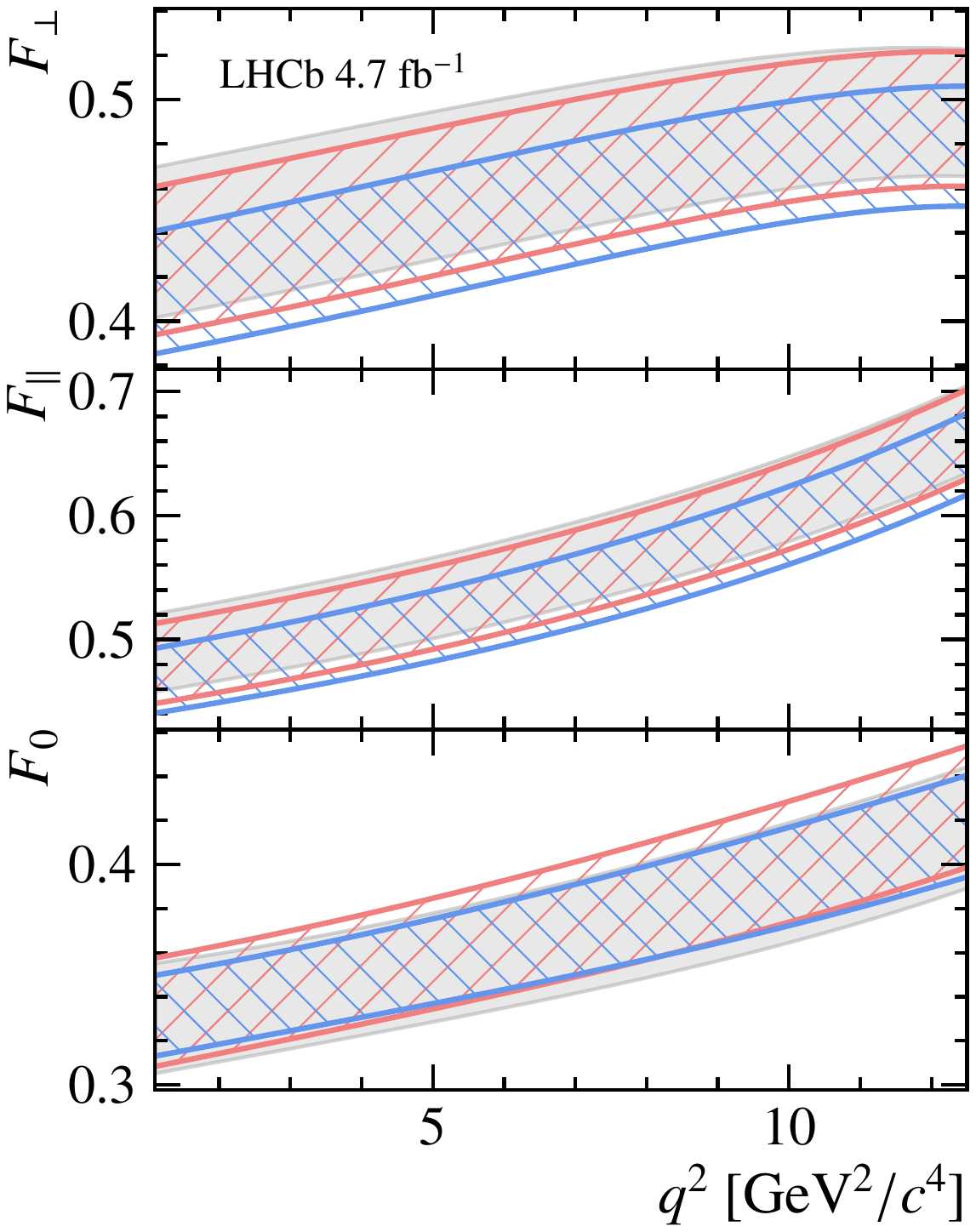}  
\includegraphics[width=0.33\textwidth]{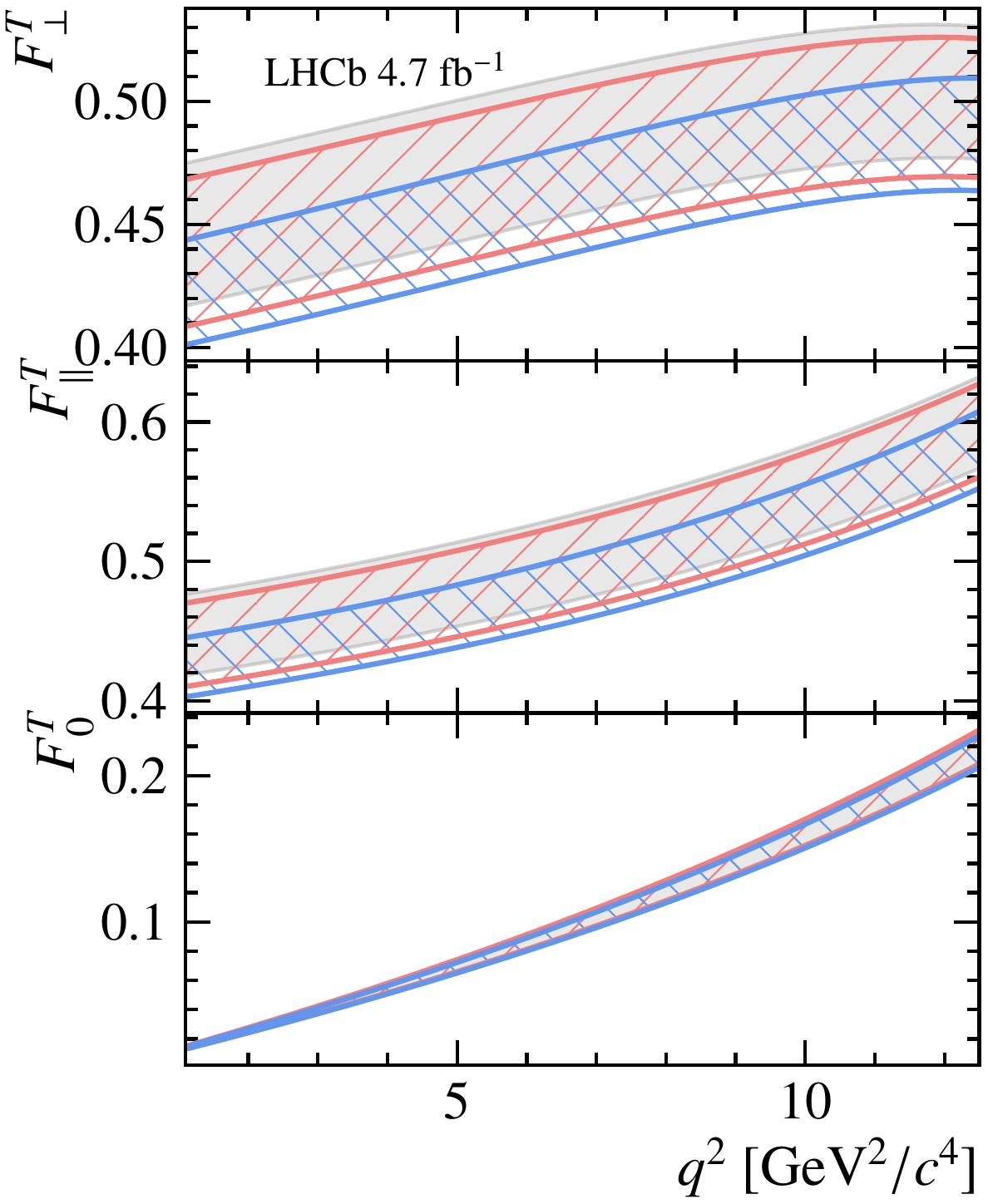}
\includegraphics[width=0.33\textwidth]{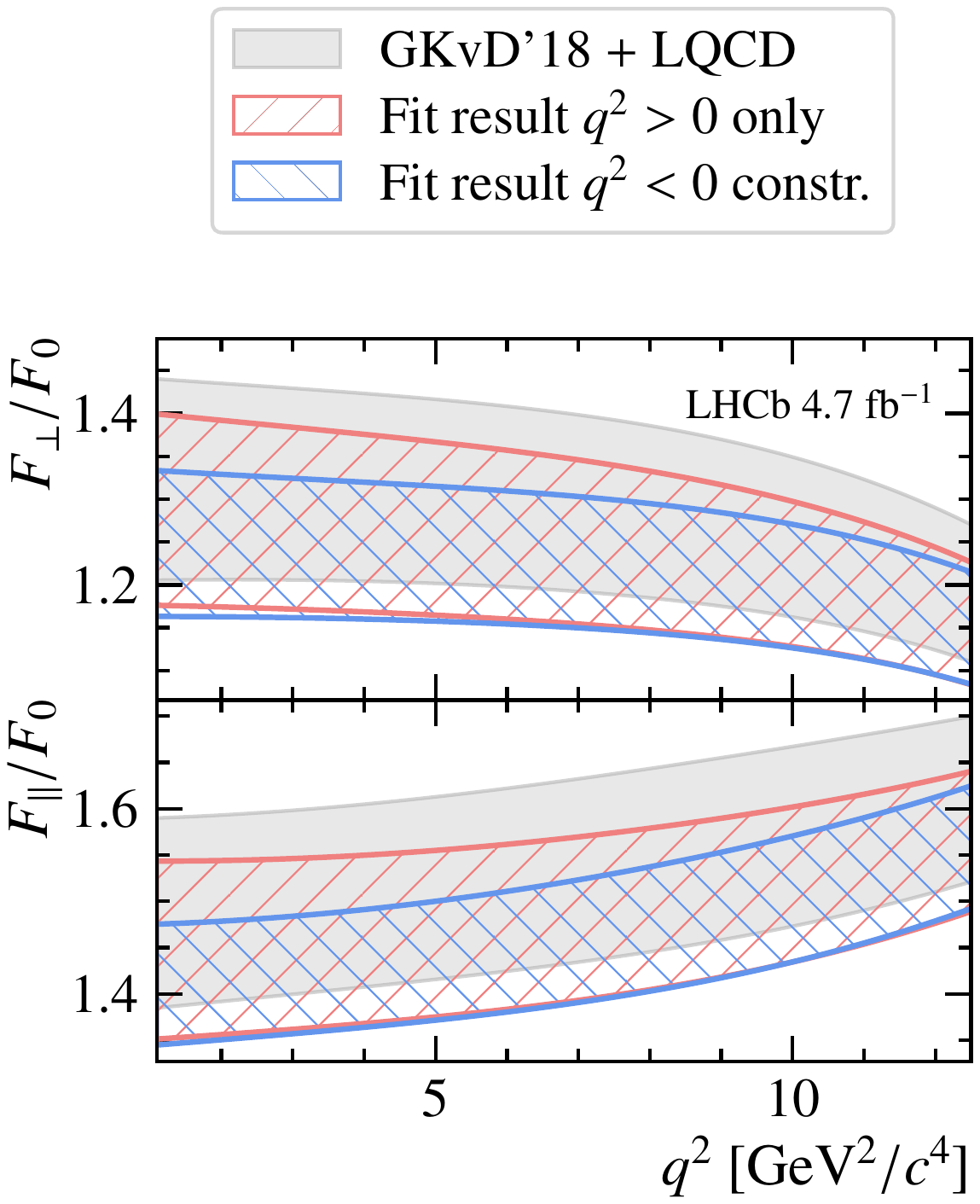}
\hspace{-4mm}
\vspace{-5mm}
\caption{Form factor results as a function of $q^2$ obtained from the amplitude fit in the two fit configurations, compared to the predictions from Refs.~\cite{Gubernari:2018wyi,Gubernari:2022hxn,Horgan:2015vla} that are used as external constraint in the fit.
 \label{fig:form_factors}}
\end{figure}

Figure~\ref{fig:form_factors} shows the form factor results obtained from the amplitude fit in the two tested configurations.
A tendency to slightly adjust the ratio $\mathcal{F}_{\perp, \parallel} / \mathcal{F}_0$  towards lower values with respect to the theoretical predictions used as external input to the fit is observed, as shown in Fig.~\ref{fig:form_factors} (right).
Both the fit configurations with and without the $q^{2} < 0$ constraints manifest this behaviour coherently.

\subsection{Non-local hadronic contributions}
\label{sec:non_local}

\begin{figure}[t]
\centering
\hspace{-8mm}
\includegraphics[width=0.36\textwidth]{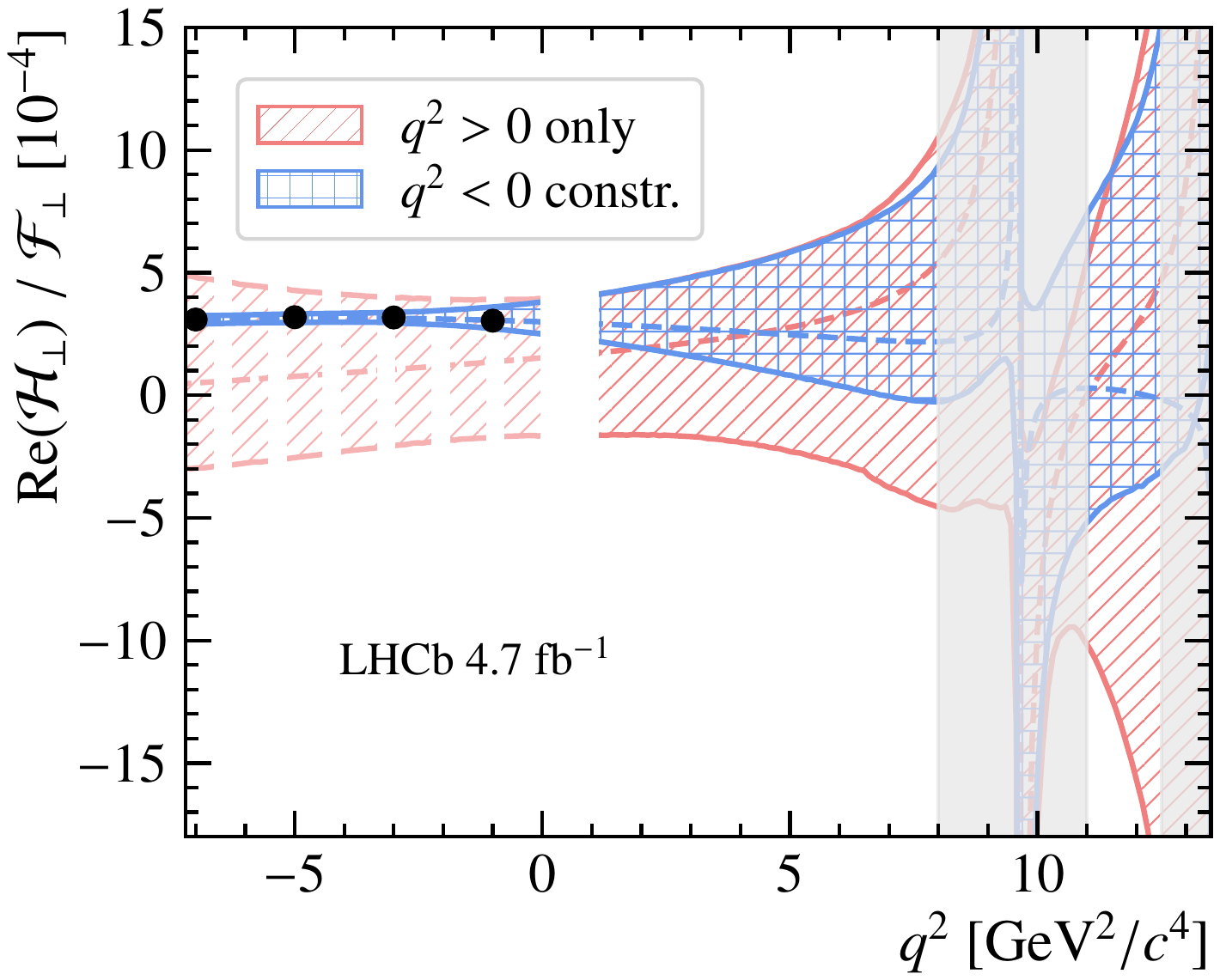} 
\hspace{-8mm}
\includegraphics[width=0.36\textwidth]{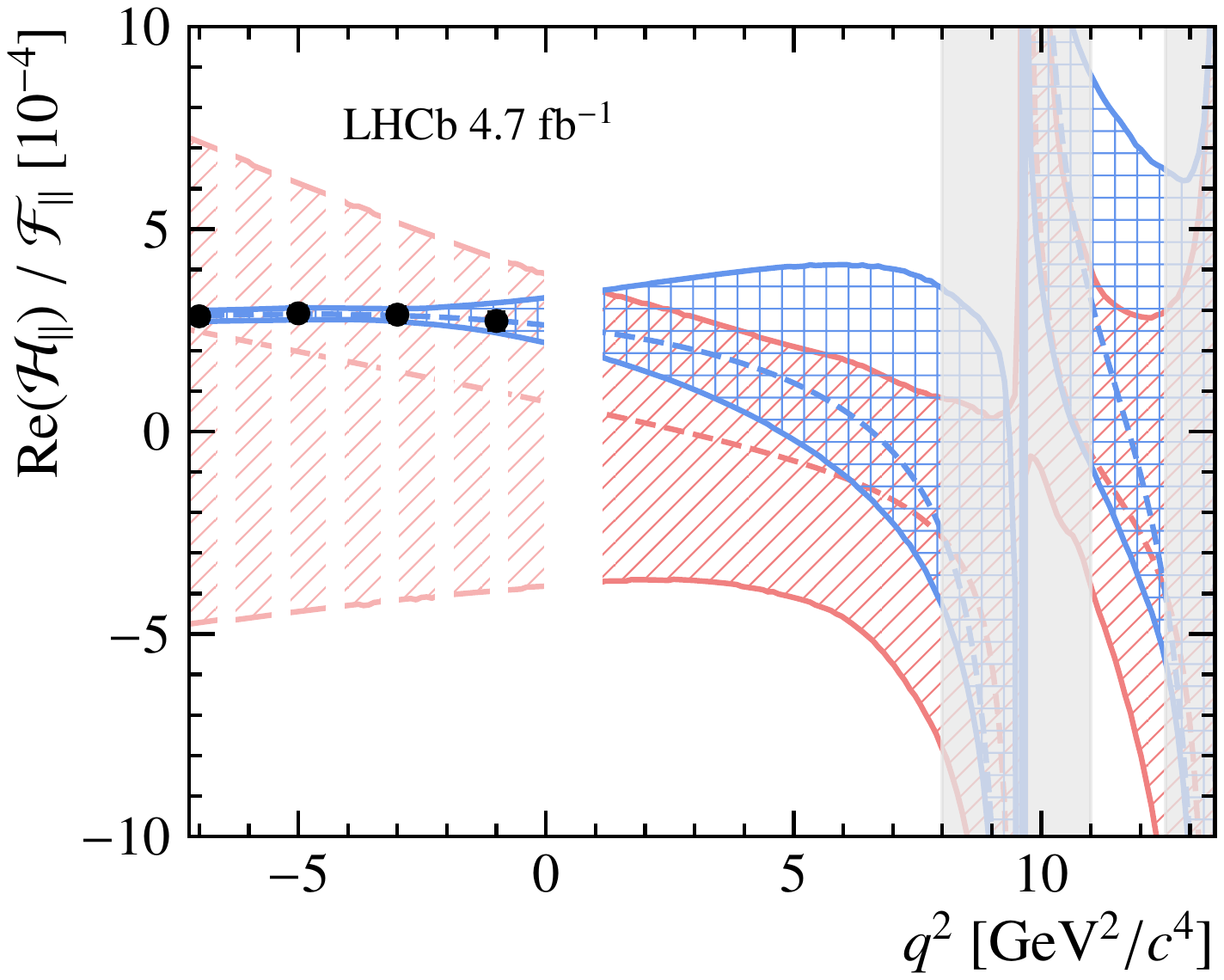} 
\hspace{-8mm}
\includegraphics[width=0.36\textwidth]{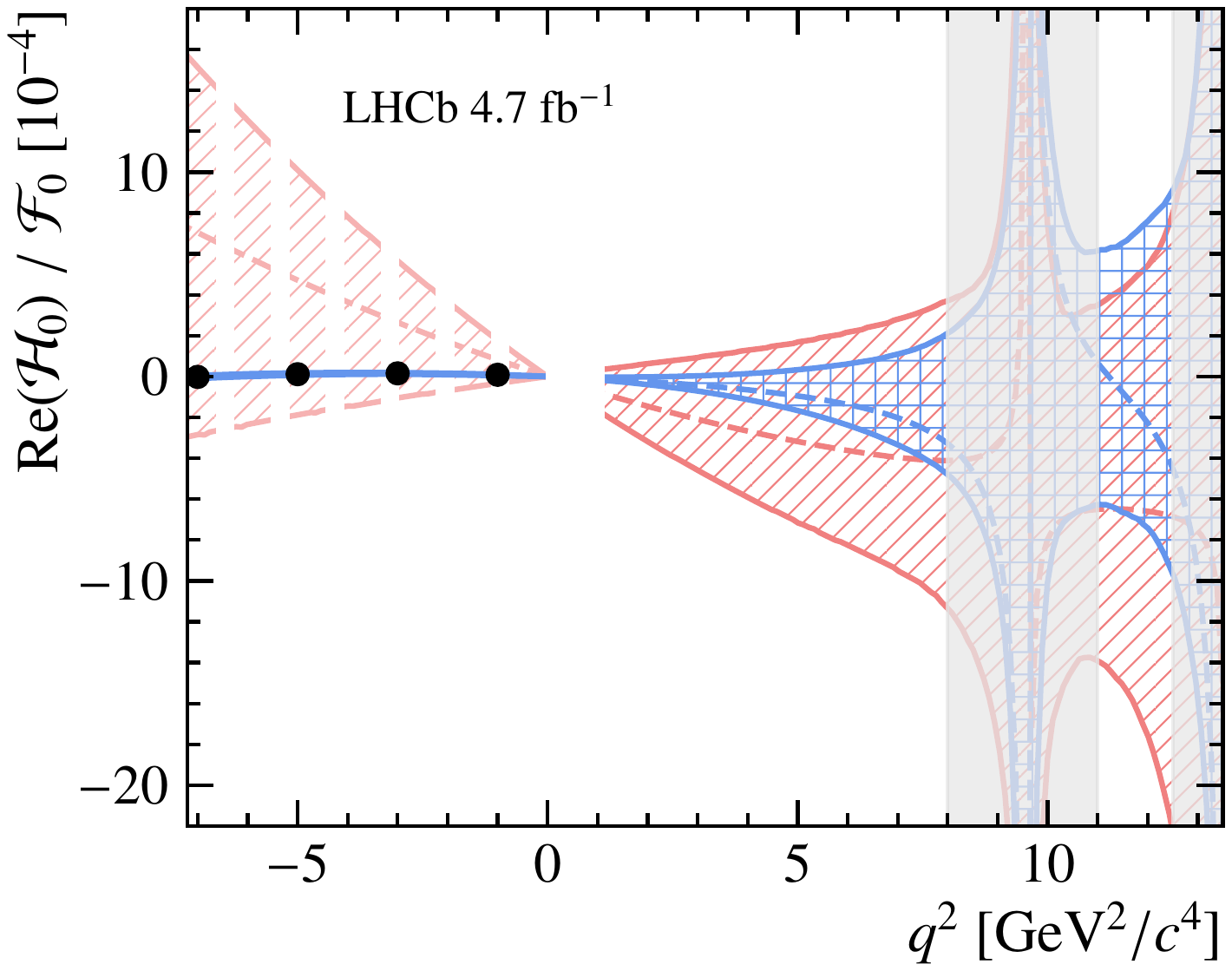}  \\
\hspace{-8mm}
\includegraphics[width=0.36\textwidth]{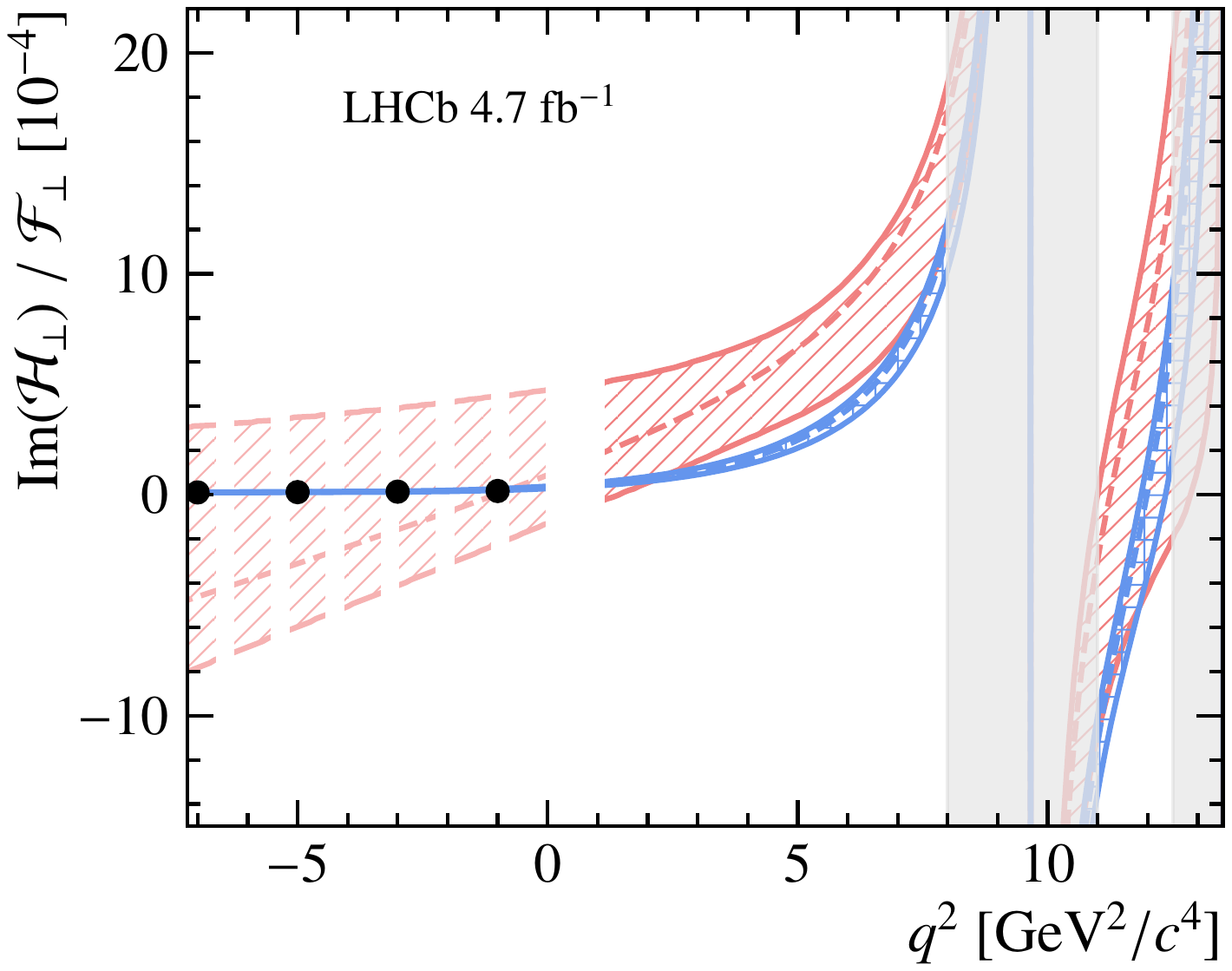} 
\hspace{-8mm}
\includegraphics[width=0.36\textwidth]{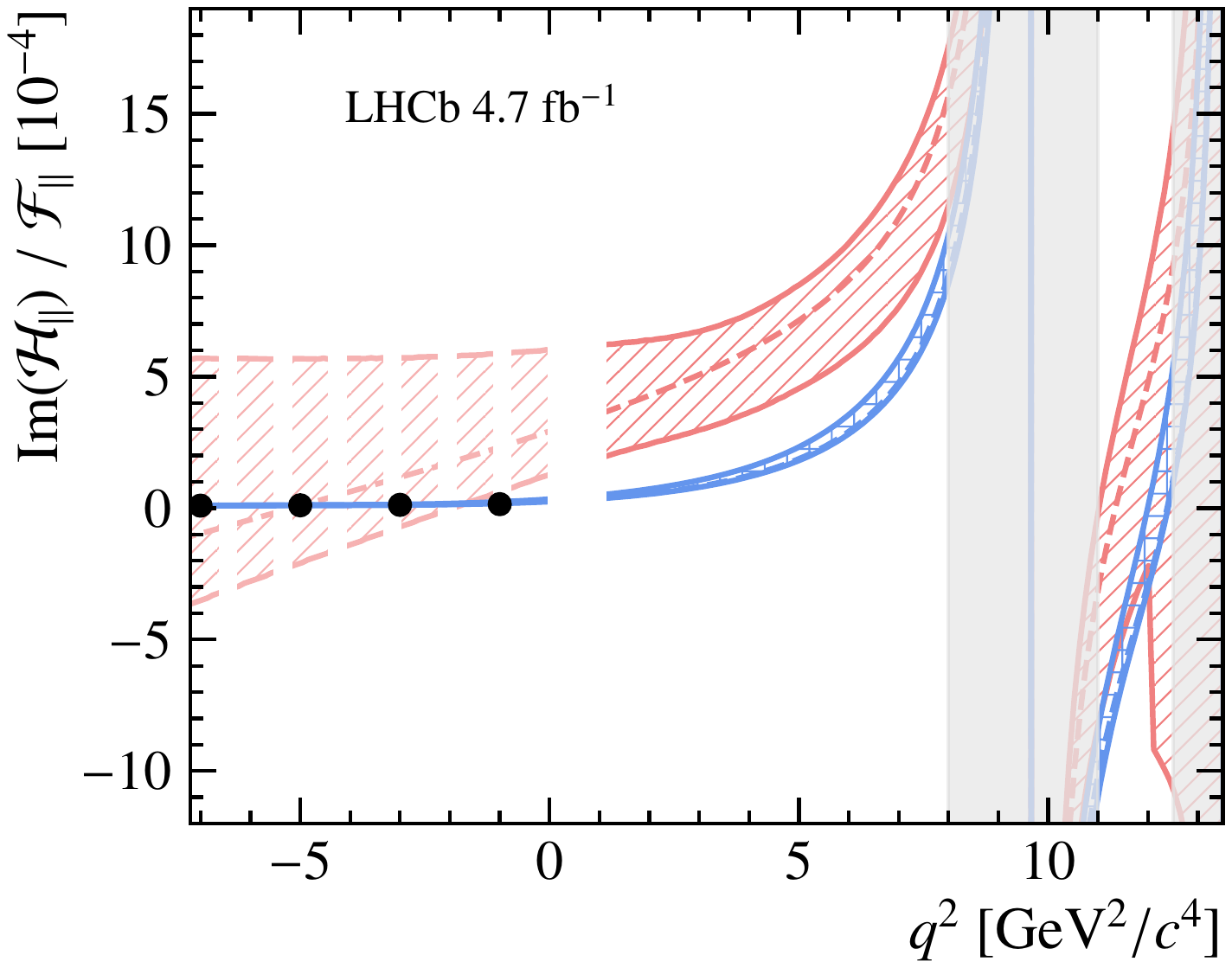} 
\hspace{-8mm}
\includegraphics[width=0.36\textwidth]{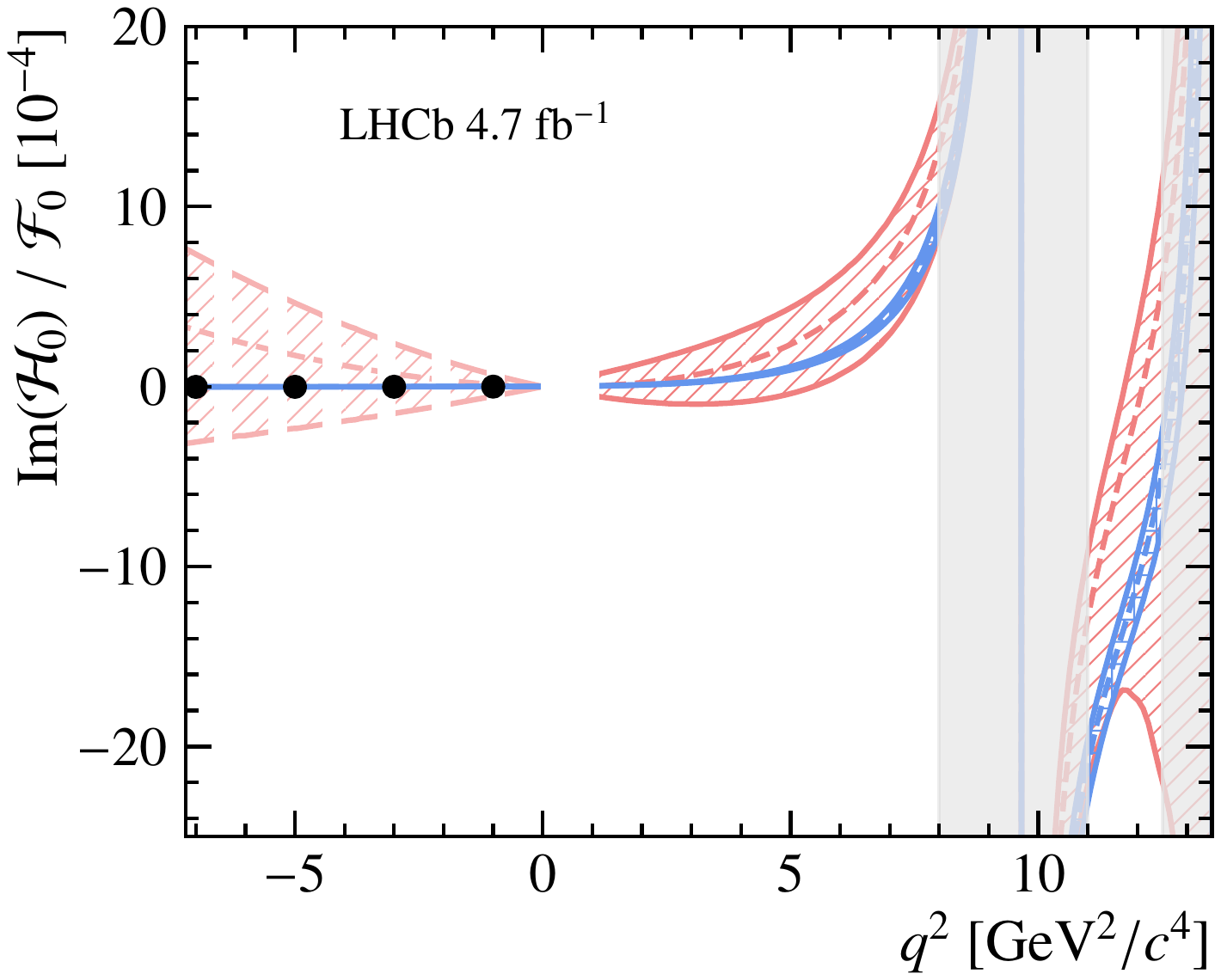} 
\caption{Real and imaginary part of the non-local contributions $\mathcal{H}_\lambda (q^2)$ normalised to the size of the local form factors $\mathcal{F}_\lambda (q^2) $ obtained 
for the two fit configurations.
The black dots correspond to the  predictions  at $q^2<0$ from~\cite{Gubernari:2022hxn,Gubernari:2020eft}.
The result obtained without the  constraints  is also extrapolated to the negative $q^2$ region  for comparison.
The shaded gray regions correspond to the vetoed $\jpsi$ and $\psi(2S)$ regions.
 \label{fig:H_q2}}
\end{figure}

Figure~\ref{fig:H_q2} shows the real and imaginary parts of the 
non-local hadronic contributions obtained for the two fit configurations
normalised to the size of the local form factors.
The two results are compatible,
however some discrepancy is visible in their imaginary parts, especially in 
$\mathrm{Im}( \mathcal{H}_\parallel)$.
The theoretical predictions at $q^2 < 0$ impose an extremely strong constraint
on the shape of these contributions,
which are in fact forced to be approximately constant 
(and have an imaginary part very close to zero)
at negative $q^2$. 
The size of $\mathrm{Im}(\mathcal{H}_\lambda(q^2))$ is then found to rise in the physical region.
At finite truncation order, the presence of the constraint at $q^2 < 0$ limits the flexibility of 
$\mathrm{Im}(\mathcal{H}_\lambda(q^2))$ in the physical region and overconstrains their contribution towards smaller values.
The behaviour of these functions in the transition between the unphysical and physical regions of $q^2$ is further investigated in Appendix~\ref{app:ImHq2} and the imaginary part of $\mathcal{H}_\lambda(q^2)$ is found to rise more rapidly than the theoretical predictions.
It is interesting to note that, while phase differences between the amplitudes are predicted to be tiny at low $q^2$, significant differences are measured between the amplitudes for  $B^0 \to \jpsi K^{*0}$ decays.

One of the advantages of the parameterisation proposed in 
Refs.~\cite{Gubernari:2020eft,Gubernari:2022hxn}
is the introduction of a dispersive bound 
to provide control over the systematic truncation errors on the $z$-expansion.
This states that, under a particular choice of polynomial functions, the sum of the coefficients squared over all $b\to s \ell \ell$ processes must be less than unity.
However, the dispersive bound is 
found to be irrelevant for this analysis since it is very far from being fulfilled, as 
the sum of the coefficients squared, after the appropriate basis transformation, is found to be  
$\mathcal{O}(10^{-3})$, for the fit result without the constraints at negative $q^{2}$.

Finally, a good compatibility between the input values and corresponding fit results
is observed on all the $B^0 \to \psi_n K^{*0}$ observables. 
Moreover, in addition to the differences of phases 
provided by $B^0 \to \psi_n K^{*0}$ external measurements,
this analysis introduces another phase difference that can be determined from the model, namely the difference between the phase of $\mathcal{A}^{\psi_n}_0$ and the local amplitudes. 
The phase difference of the $\jpsi$ longitudinal amplitude (at the $\jpsi$ mass pole) with respect to the rare mode is found to be 
$-1.55^{+0.22}_{-0.18}$ for the fit result with the $q^{2} < 0$ constraints
and $-1.61^{+0.22}_{-0.20}$ for the one without these constraints,\footnote{The fit result with $q^{2} > 0$ only information shows a second solution at about
$\phi_0^{\jpsi} \mapsto \phi_0^{\jpsi} +\pi$, which is however excluded at more than $3\,\sigma$.}
showing a good agreement between the two fit configurations.
This result is also compatible with one of the two solutions obtained in the measurement of the phase difference between $B^+ \to K^+ \mumu$ and $B^+ \to \jpsi K^+$ decays~\cite{LHCb-PAPER-2016-045}, which are ruled by the same rare-electroweak and tree-level underlying transitions, respectively, but with a different spectator quark.
The phase difference of $\mathcal{A}^{\psi(2S)}_0 $ with respect to the rare mode
shows an almost complete degeneracy and cannot be determined precisely from the fit.

\subsection{Wilson coefficients}
\label{sec:Wilson}

\begin{figure}[t]
\centering
\hspace{-5mm}
\includegraphics[width=0.49\textwidth]{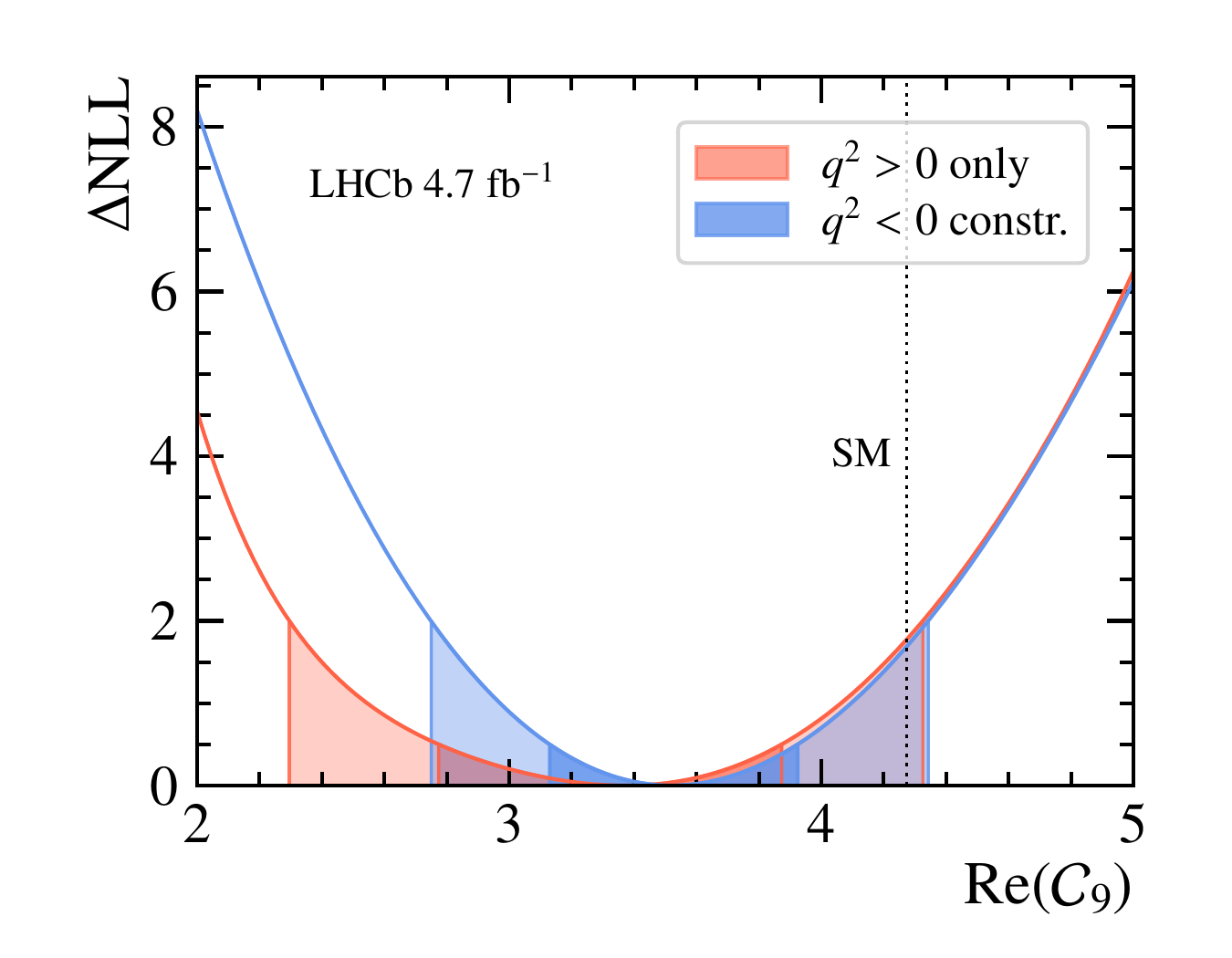}
\includegraphics[width=0.49\textwidth]{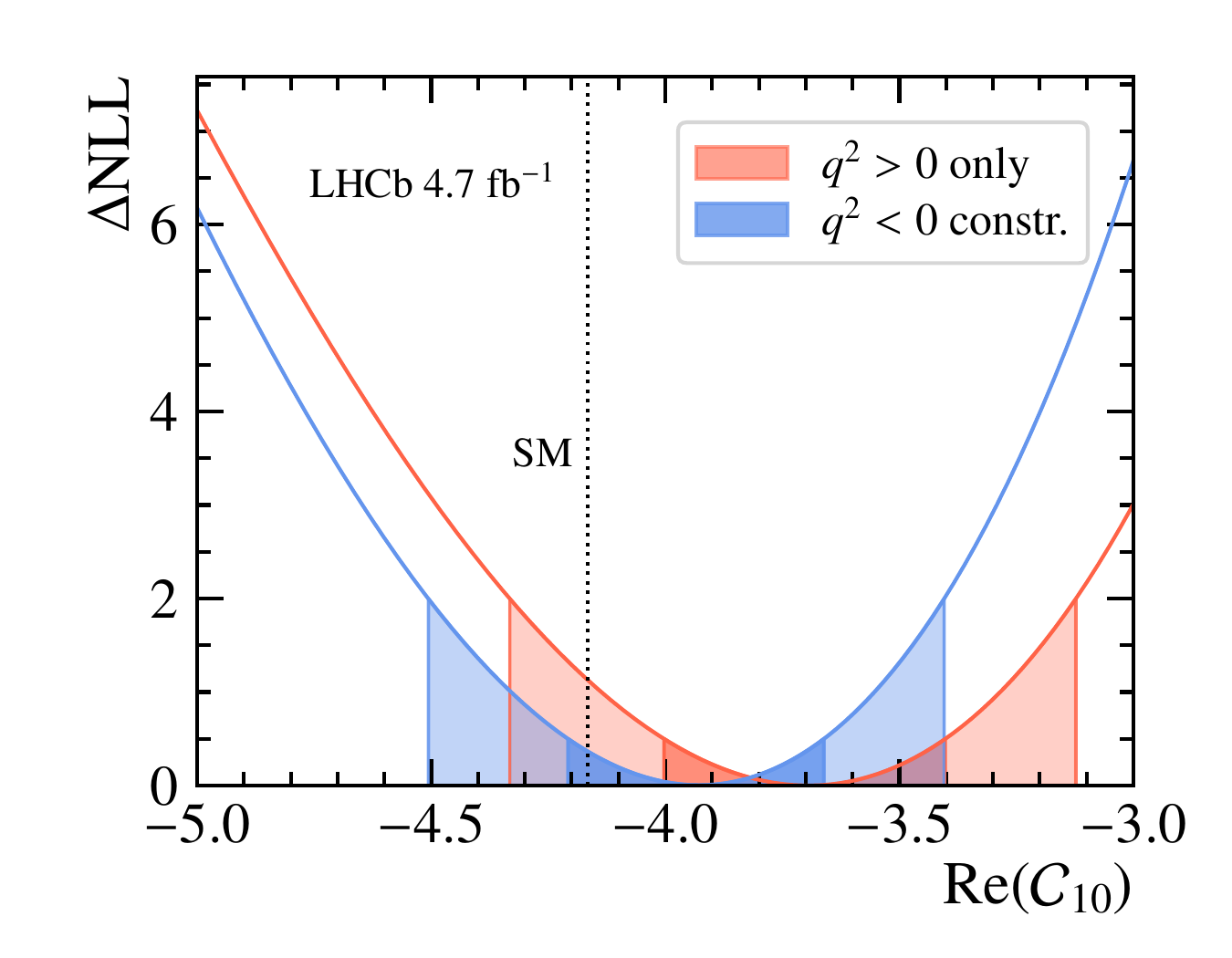} \\
\hspace{-5mm}
\includegraphics[width=0.49\textwidth]{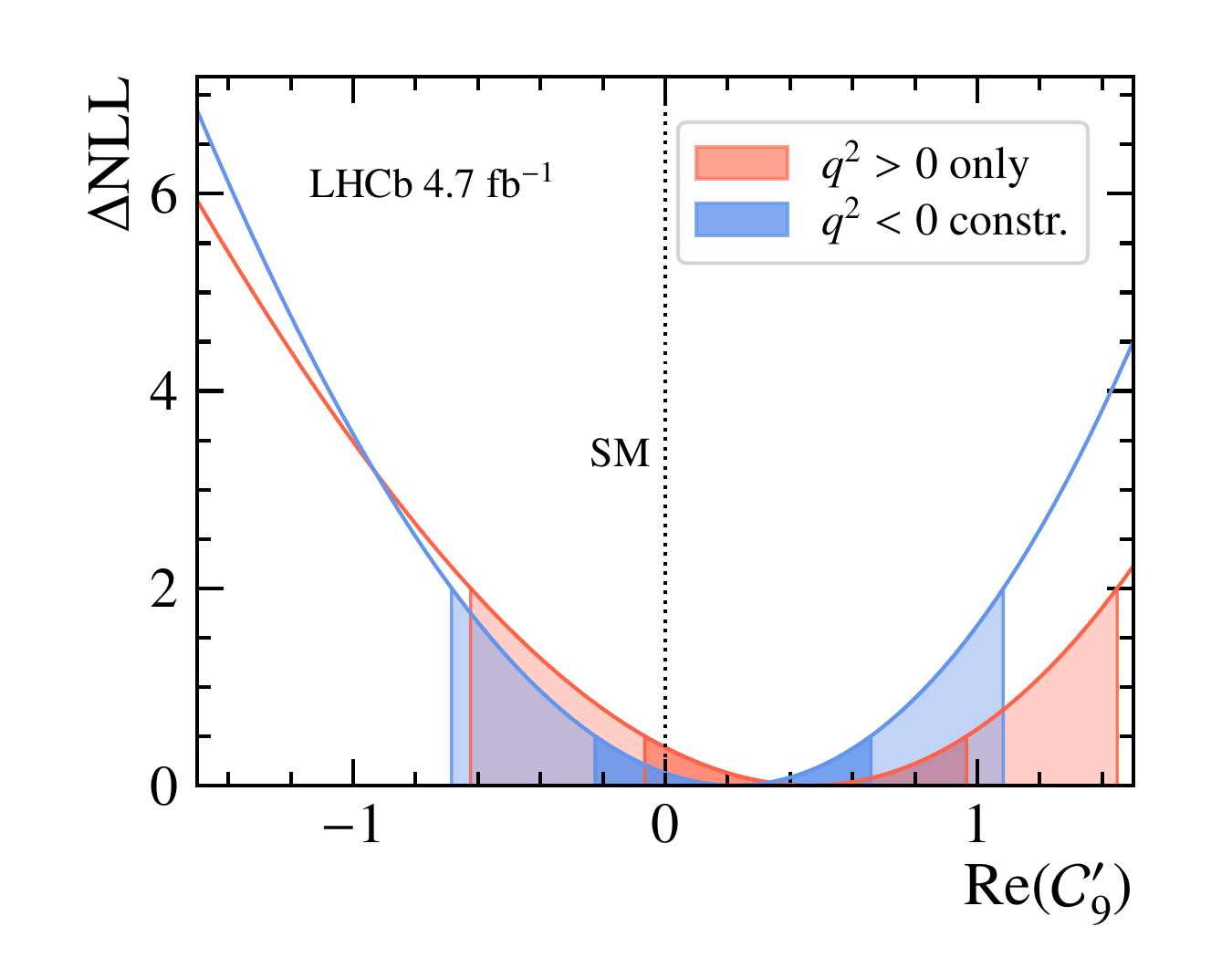}
\includegraphics[width=0.49\textwidth]{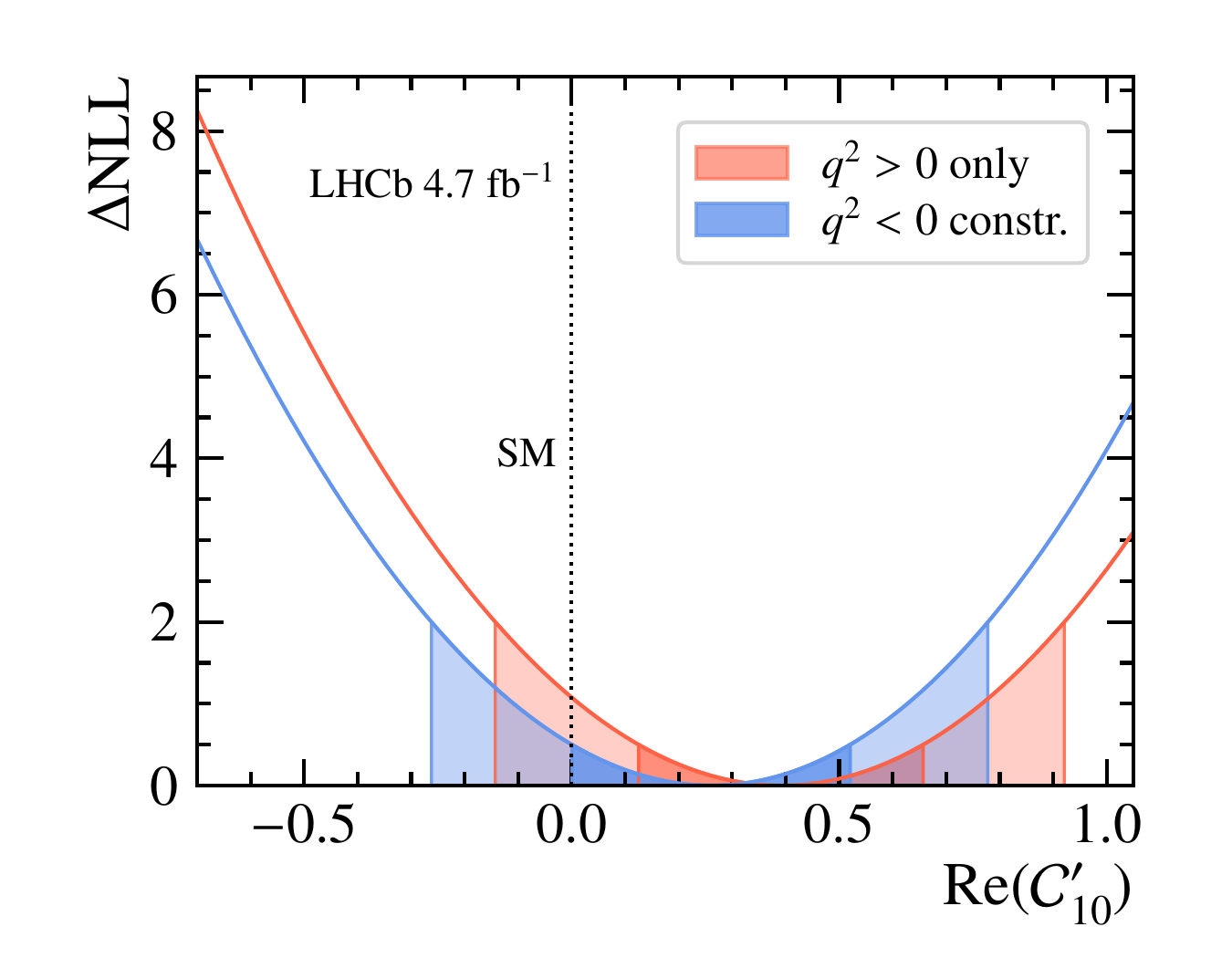}
\caption{One-dimensional profile likelihood scan of the Wilson coefficients.  
Shaded regions correspond to the 68\%  and 95\%  confidence intervals.}
\label{fig:NLL_WCs_1D}
\end{figure}

Table~\ref{tab:result_WCs} reports the values of the Wilson coefficients for the two fit configurations, 
together with their confidence intervals (C.I.) and compatibility with the Standard Model.
For each of the four Wilson coefficients,  confidence intervals are built from the one-dimensional profile likelihood scans shown in Fig.~\ref{fig:NLL_WCs_1D}.  
The 68\% (95\%) C.I. range is identified with the interval where the negative log-likelihood difference, $\Delta \textrm{NLL}$, is smaller than 0.5 (2).
The difference between the best fit values and the corresponding SM predictions obtained are
\begin{alignat}{4}
& \Delta \mathcal{C}_9 && = - && 0.93^{+0.53}_{-0.57}  \;\;  (- && 0.68^{+0.33}_{-0.46}\, )  \, , \nonumber \\
& \Delta \mathcal{C}_{10} && = && 0.48^{+0.29}_{-0.31}     \;\;  ( && 0.24^{+0.27}_{-0.28}\, ) \, , \nonumber\\
& \Delta \mathcal{C}_{9}^{\prime} && = && 0.48^{+0.49}_{-0.55}    \;\; ( && 0.26^{+0.40}_{-0.48} \, ) \, , \nonumber\\
& \Delta \mathcal{C}_{10}^{\prime} && = && 0.38^{+0.28}_{-0.25}   \;\; (   && 0.27^{+0.25}_{-0.27} \, ) \, , \nonumber
\end{alignat}
for the fit configuration without (with) constraints at negative $q^{2}$, where the SM prediction  at the $b$-quark energy scale is taken to be 
$\mathcal{C}^{\rm{SM}}_9 = 4.27, \, \mathcal{C}^{\rm{SM}}_{10} = -4.17$ and $\mathcal{C}^{\prime\, \rm{SM}}_{9, 10} = 0$~\cite{Bobeth:1999mk,Gorbahn:2004my}.
The coefficient that shows the largest difference with respect to the SM is $\mathcal{C}_9$, whose compatibility with the SM is found to be at the level of 1.9 and 1.8 standard deviations, for fit models using only $q^{2} > 0$  information and with the $q^{2} < 0$ constraints, respectively.

\begin{table}[t]
\renewcommand{\arraystretch}{1.2}
\renewcommand\cellalign{cc}
\caption{Best fit value, confidence intervals and deviation from the SM predictions~\cite{Bobeth:1999mk,Gorbahn:2004my} for the four Wilson coefficients and the two fit configurations.
For each Wilson coefficient, the likelihood has been profiled over the  other coefficients. The SM predictions  at the $b$-quark energy scale~\cite{Bobeth:1999mk,Gorbahn:2004my} are also reported for reference. }
\begin{center}
\begin{tabular}{%
 l 
S[table-format=3.2]  
S[table-format=3.2, table-text-alignment=center]    
@{\hspace{0ex}}                                  	
c                                                  	
@{\hspace{0ex}}                                     
S[table-format=3.2, table-text-alignment=center]    
S[table-format=3.2, table-text-alignment=center]    
@{\hspace{0ex}}                                  	
c                                                  	
@{\hspace{0ex}}                                     
S[table-format=3.2, table-text-alignment=center]    
S[table-format=3.2]   
c   
}
\toprule
			&  \multicolumn{9}{c}{$q^{2} > 0$ only}  	\\ 
			&		{\makecell{best fit \\ value}}  	&		\multicolumn{3}{c}{68\% C.I.}		&		\multicolumn{3}{c}{95\% C.I.}	&   {SM value}    &	{\makecell{deviation \\ from SM}}      \\  
\midrule
$\mathcal{C}_9$				& 		3.34	& 		[2.77 & , & 3.87] 	& [2.30 & , & 4.33] 	&   4.27   & 1.9 $\sigma$	   \\ 
$\mathcal{C}_{10}$			&		-3.69	&		[-4.00 & , & -3.40] & [-4.33 & , & -3.12] 	&   -4.17  & 1.5 $\sigma$	 \\
$\mathcal{C}_9^\prime$		&		0.48	&		[-0.07 & , & 0.97] 	& [-0.62 & , & 1.45]	&   0   & 0.9 $\sigma$	 \\ 
$\mathcal{C}_{10}^\prime$	&		0.38	&		[0.13 & , & 0.66] 	& [-0.14 & , & 0.92] 	&   0   & 1.5 $\sigma$	  \\ 
\midrule
			&  \multicolumn{9}{c}{$q^{2} < 0$ constraints}  	\\ %
\midrule
$\mathcal{C}_9$				 &	 3.59	&	[3.13 & , & 3.92] 		& [2.75 & , & 4.34] 	&	4.27      & 1.8 $\sigma$	  \\
$\mathcal{C}_{10}$			 &	-3.93	&	[ -4.21 & , & -3.66]    & [-4.51 & , & -3.40] 	&	-4.17     & 0.9 $\sigma$	  \\
$\mathcal{C}_9^\prime$		 &	0.26	&	[-0.22 & , & 0.66] 	    & [-0.68 & , & 1.08] 	&	0     & 0.5 $\sigma$	      \\
$\mathcal{C}_{10}^\prime$	 &	0.27	&	[0.00 & , & 0.52]		& [-0.26 & , & 0.78]	&	0     & 1.0 $\sigma$		  \\
\bottomrule
\end{tabular}
\label{tab:result_WCs}
\end{center}
\end{table}

\begin{figure}[t]
\centering
\hspace{-5mm}
\includegraphics[width=0.49\textwidth]{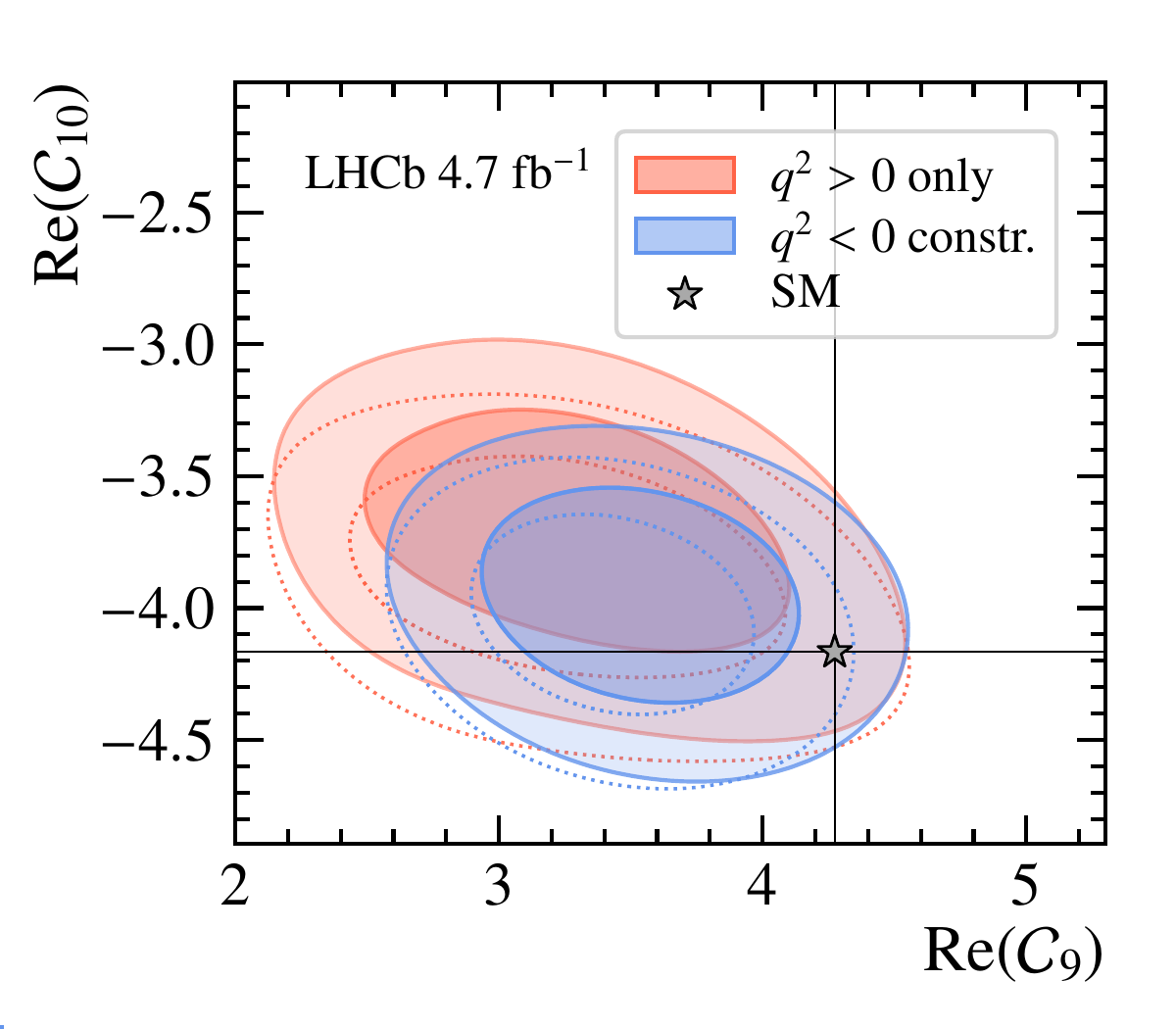} 
\includegraphics[width=0.49\textwidth]{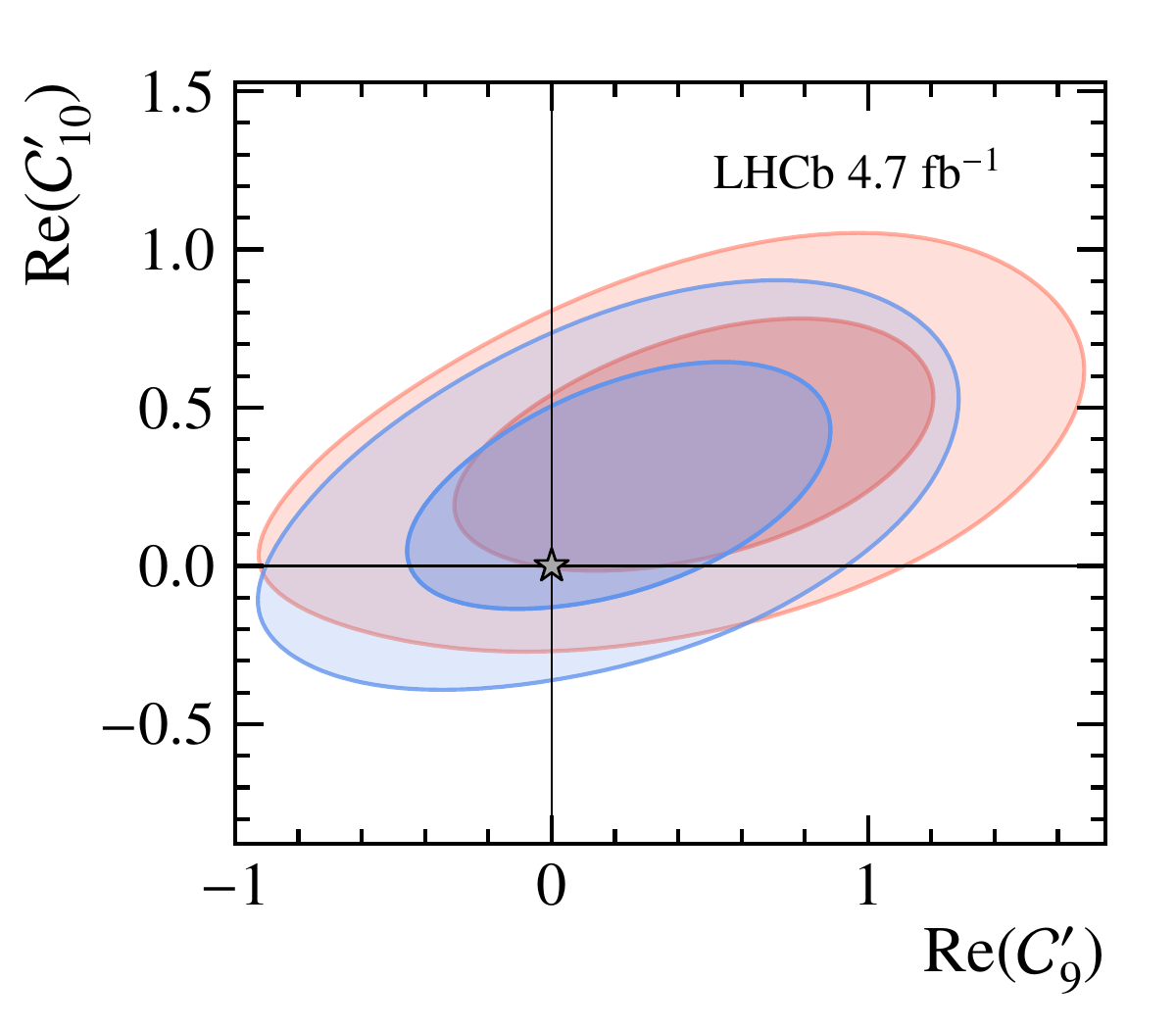} \\
\hspace{-5mm}
\includegraphics[width=0.49\textwidth]{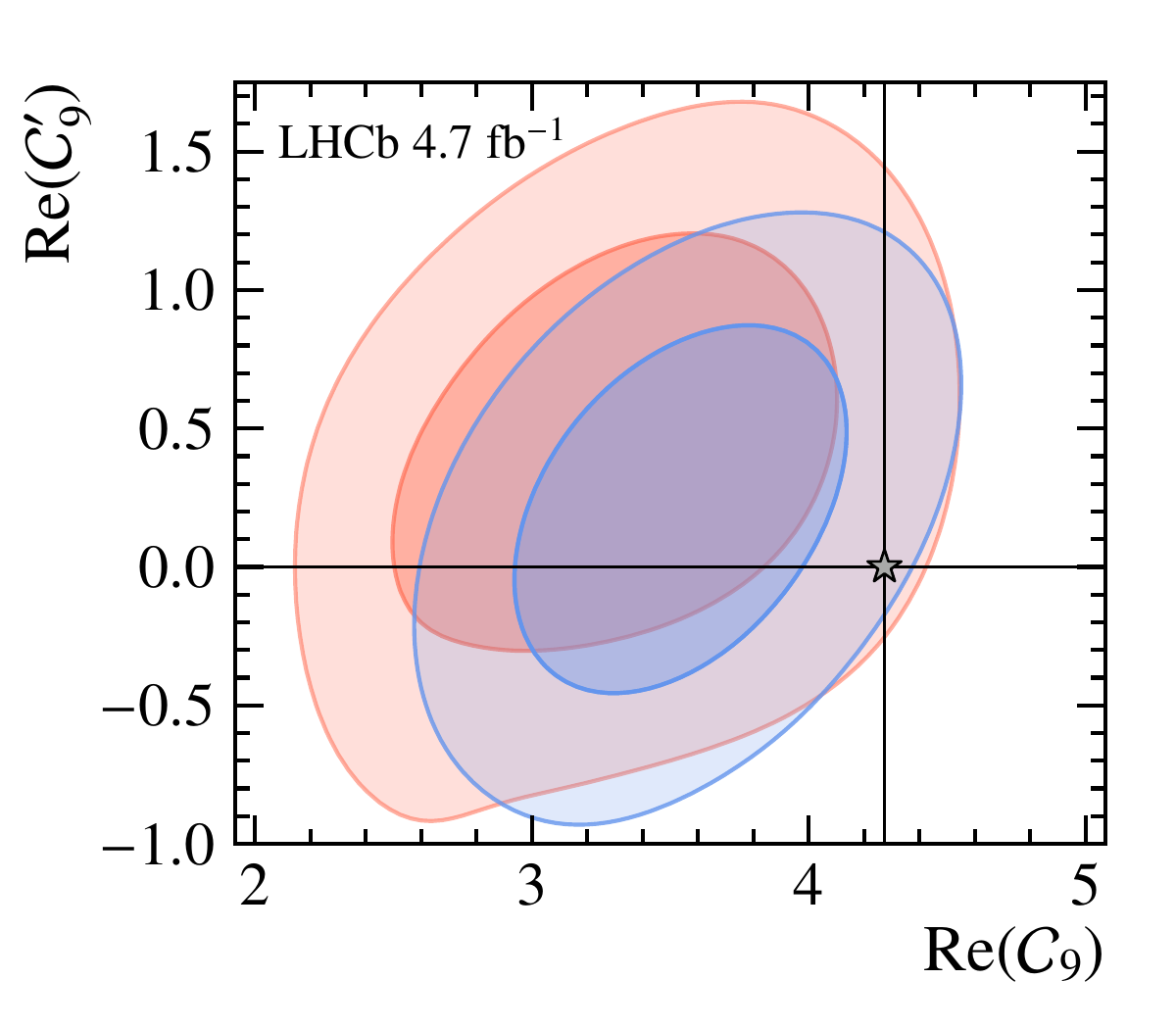}
\includegraphics[width=0.49\textwidth]{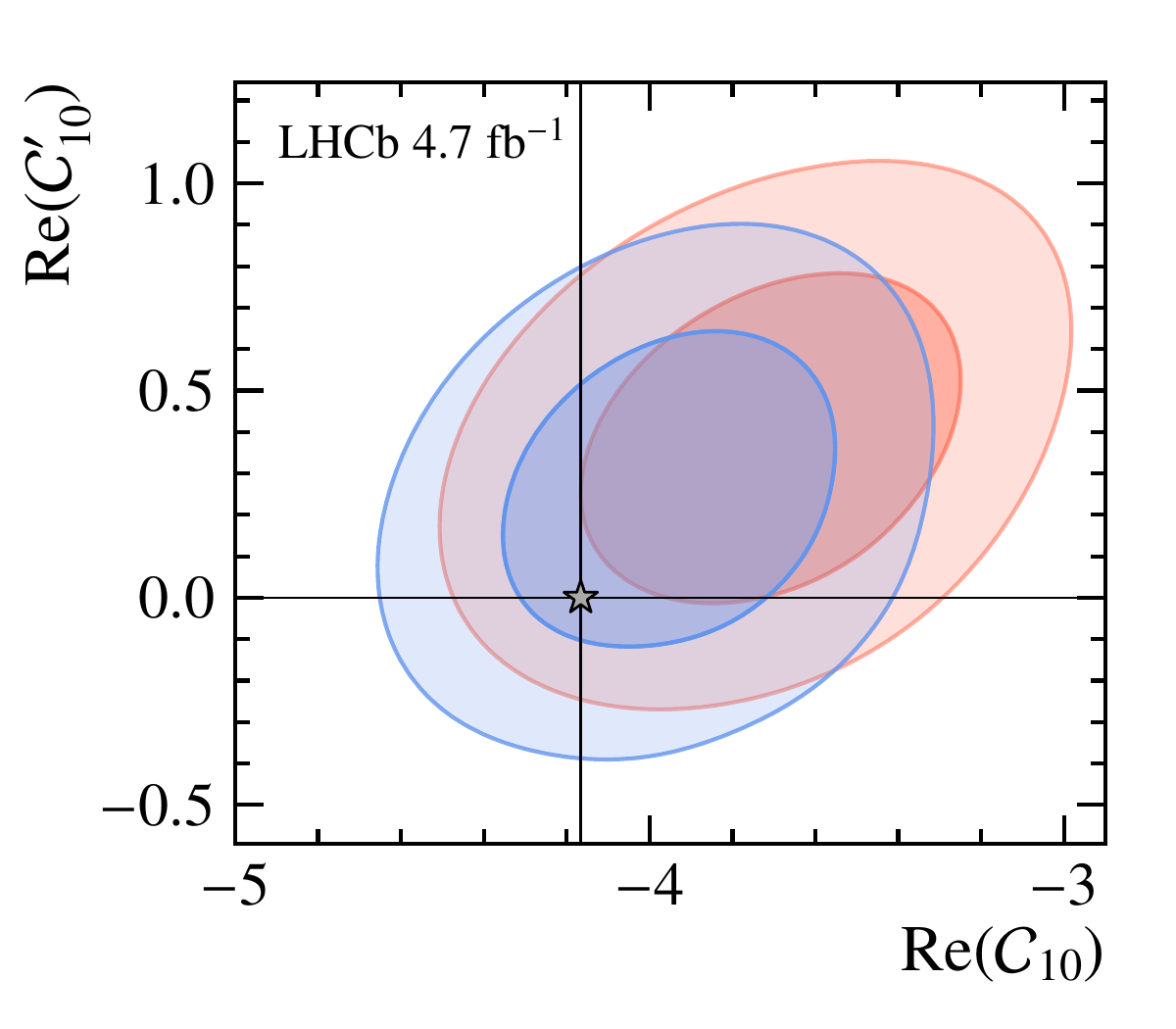}
\caption{Two-dimensional profile likelihood scan of the Wilson coefficients.  
Shaded areas correspond to the 68\% C.I. and 95\% C.I.  contour regions.
Dotted contours in the top left plot assume right-handed Wilson coefficients fixed to their SM values, \textit{i.e.} $\mathcal{C}_9^\prime = \mathcal{C}_{10}^\prime = 0$.}
\label{fig:NLL_WCs_2D}
\end{figure}

Two-dimensional profile likelihood contours for the Wilson coefficients are shown in Fig.~\ref{fig:NLL_WCs_2D}, 
where the 68\% (95\%) C.I. range is identified with the
region where the $\Delta \textrm{NLL}$ is smaller than 1.15 (3.09).
A shift of approximately $0.2$ is observed in the central values of all the Wilson coefficients between the two fit configurations,
with the fit result with the $q^{2} < 0$ constraints being closer to the SM.
While from a theoretical perspective one could expect that non-local hadronic contributions would only affect $\mathcal{C}_9$,
the experimental determination of the Wilson coefficients is affected by the strong correlations of the system:  
a modification of the non-local hadronic contributions is found to influence the result on the form factors (as shown in Fig.~\ref{fig:form_factors}), 
which in turn have an impact on the Wilson coefficients.
This behaviour has been studied with pseudoexperiments, where the same generated dataset is fitted with and without the constraints at negative $q^{2}$ replicating the procedure adopted on data, 
and the variation measured in data is found to be compatible with what is observed in the pseudoexperiments.

Finally, the global compatibility with respect to the SM is evaluated by inspecting the likelihood difference in the four-dimensional space given by the four considered Wilson coefficients.
Taking into account the systematic uncertainties, the observed difference in twice the log-likelihood between the best fit and SM point is found to be 2.99 (3.25). Considering the four degrees of freedom of the system, this corresponds to 1.3 (1.4) standard deviations with respect to the SM, for the fit without (with) the negative $q^{2}$ constraints.

\subsection{Comparison to binned observables}
\label{sec:ang_obs}

\begin{figure}[t]
\centering
\hspace{6mm}
\includegraphics[width=0.81\textwidth]{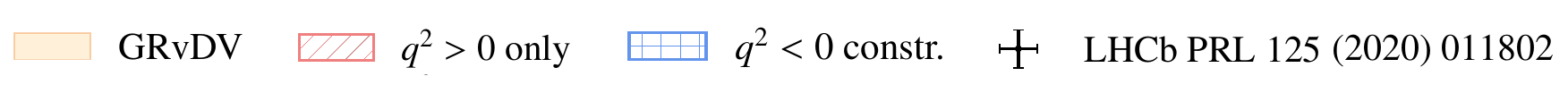} \\
\includegraphics[width=0.47\textwidth]{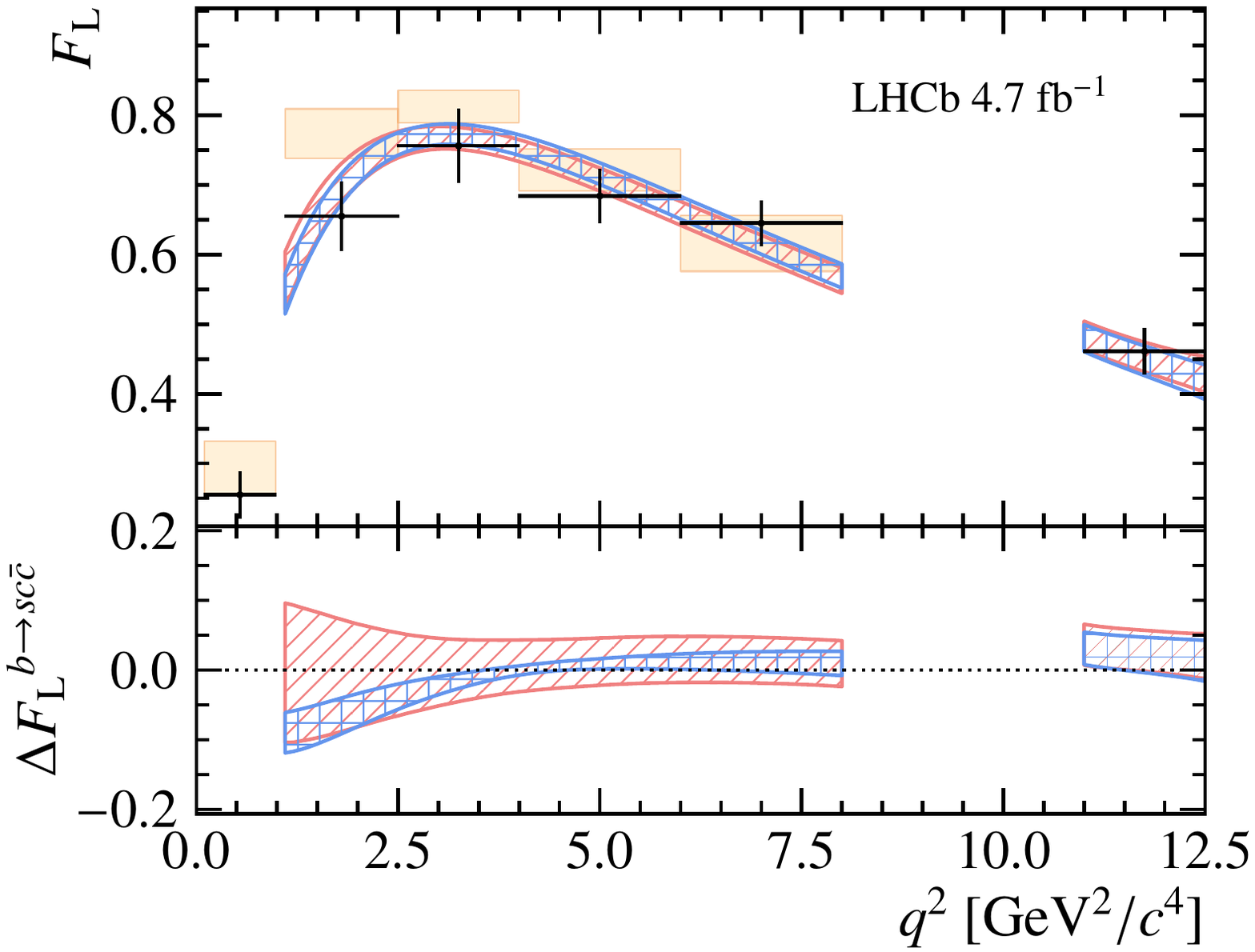}
\includegraphics[width=0.47\textwidth]{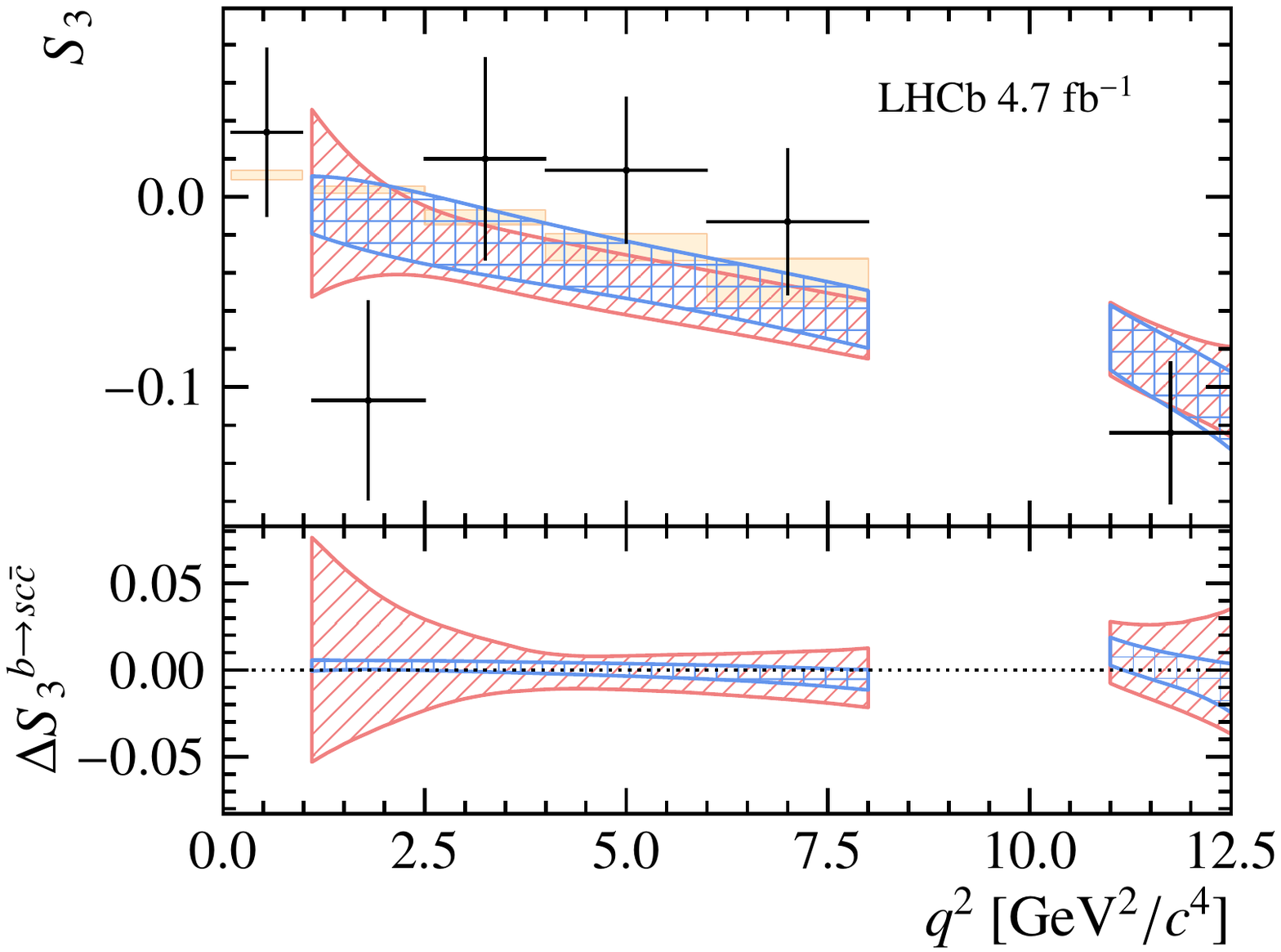}  \\
\vspace{1mm}
\includegraphics[width=0.47\textwidth]{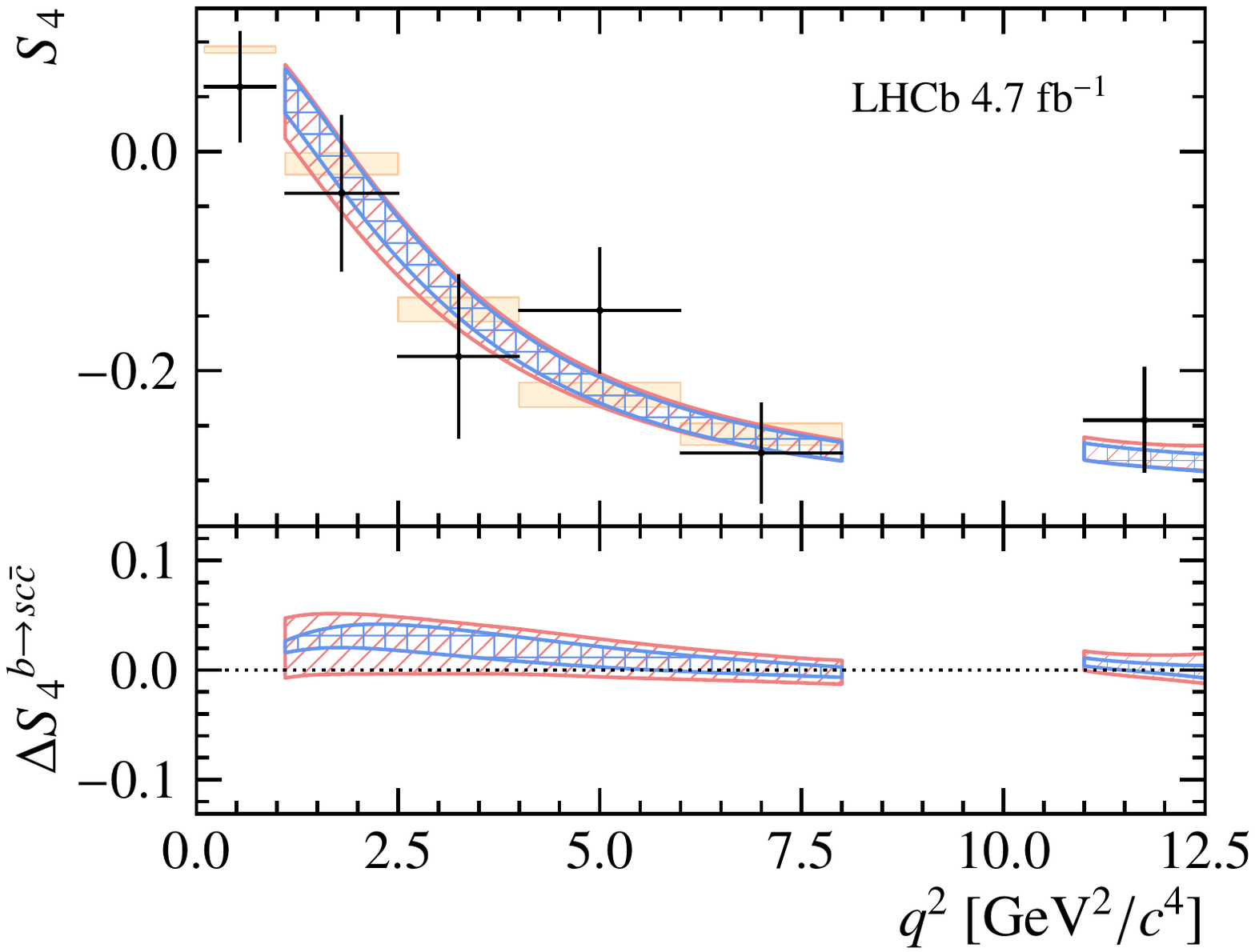}
\includegraphics[width=0.47\textwidth]{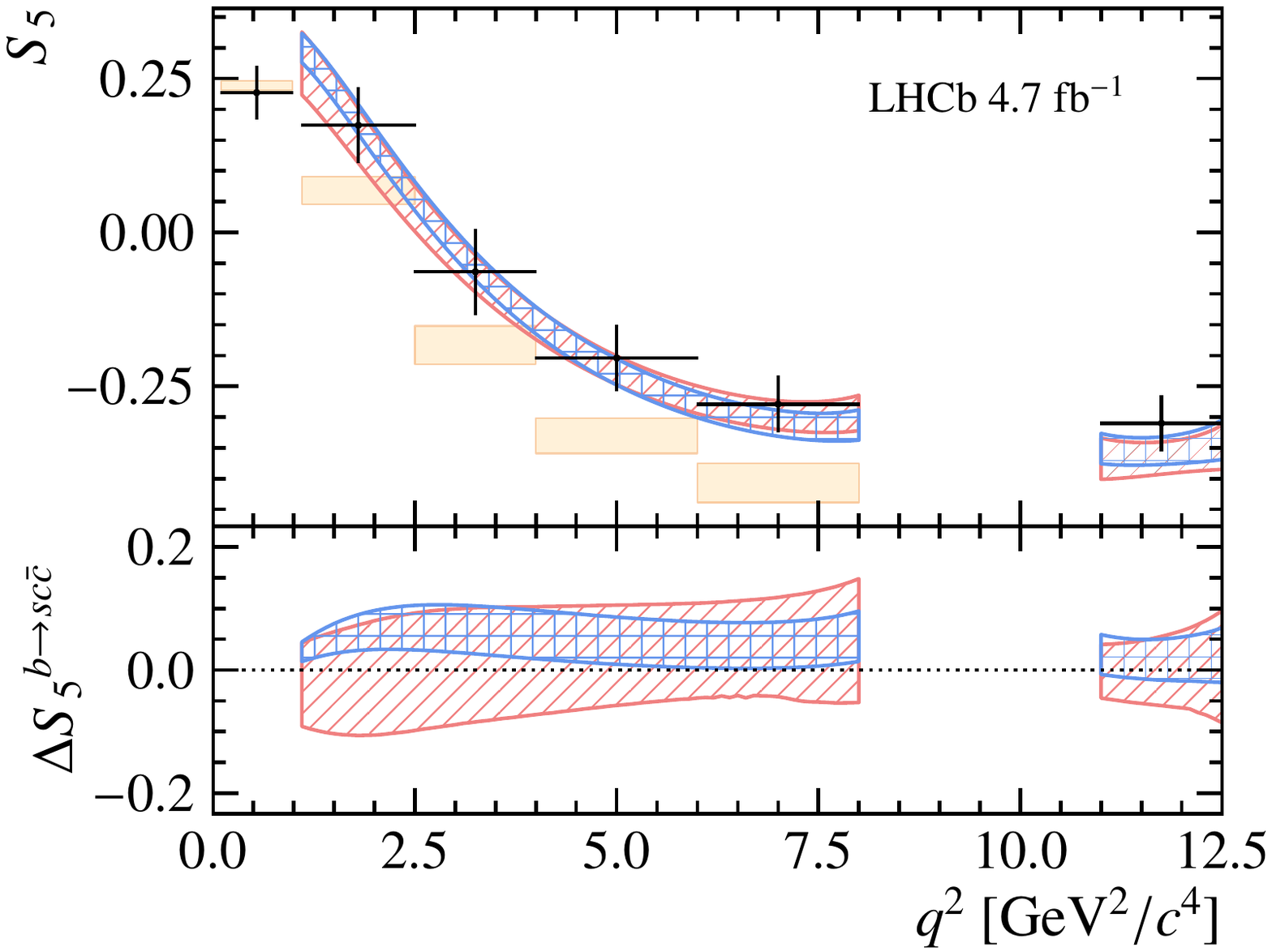}  \\
\vspace{1mm}
\includegraphics[width=0.47\textwidth]{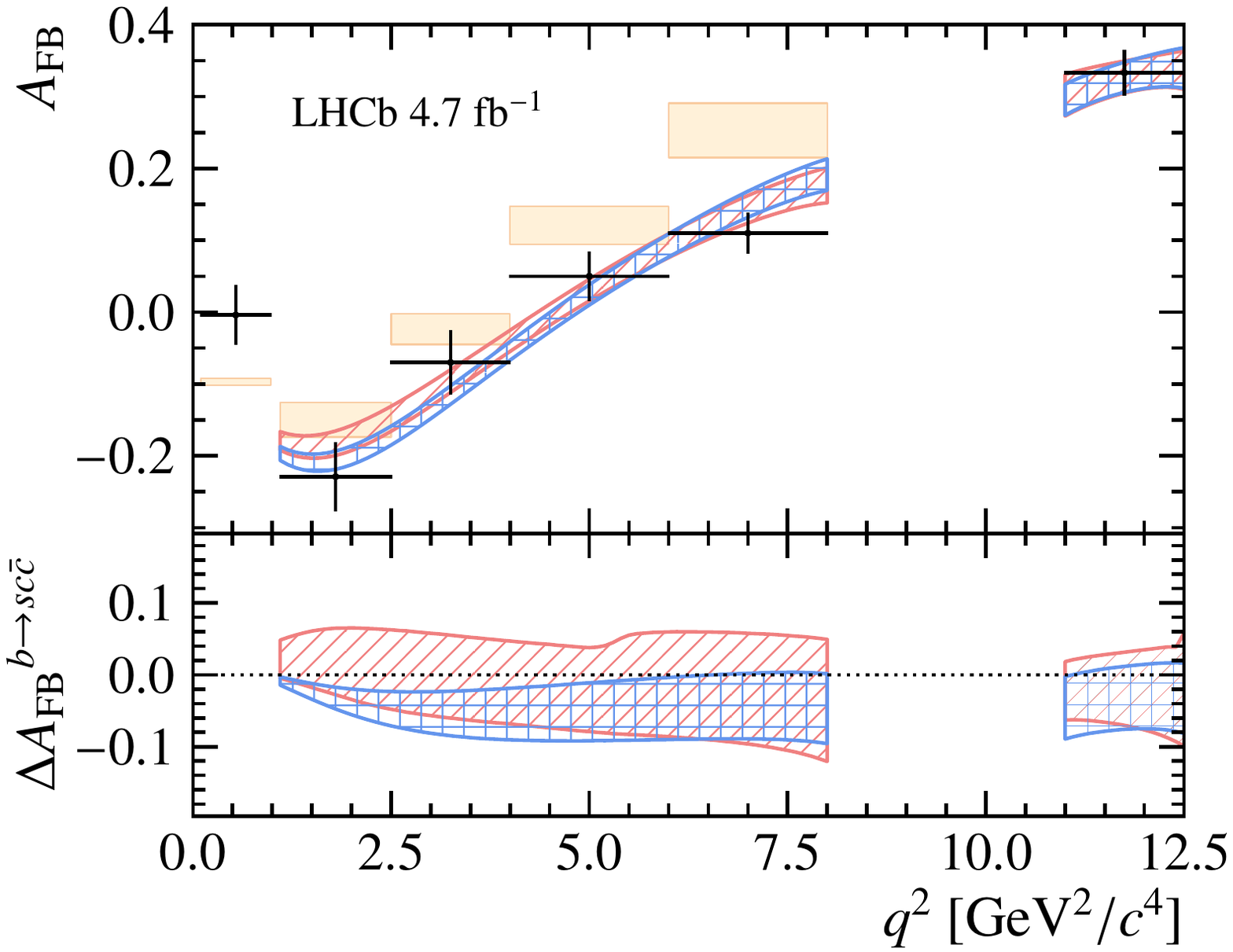}
\includegraphics[width=0.47\textwidth]{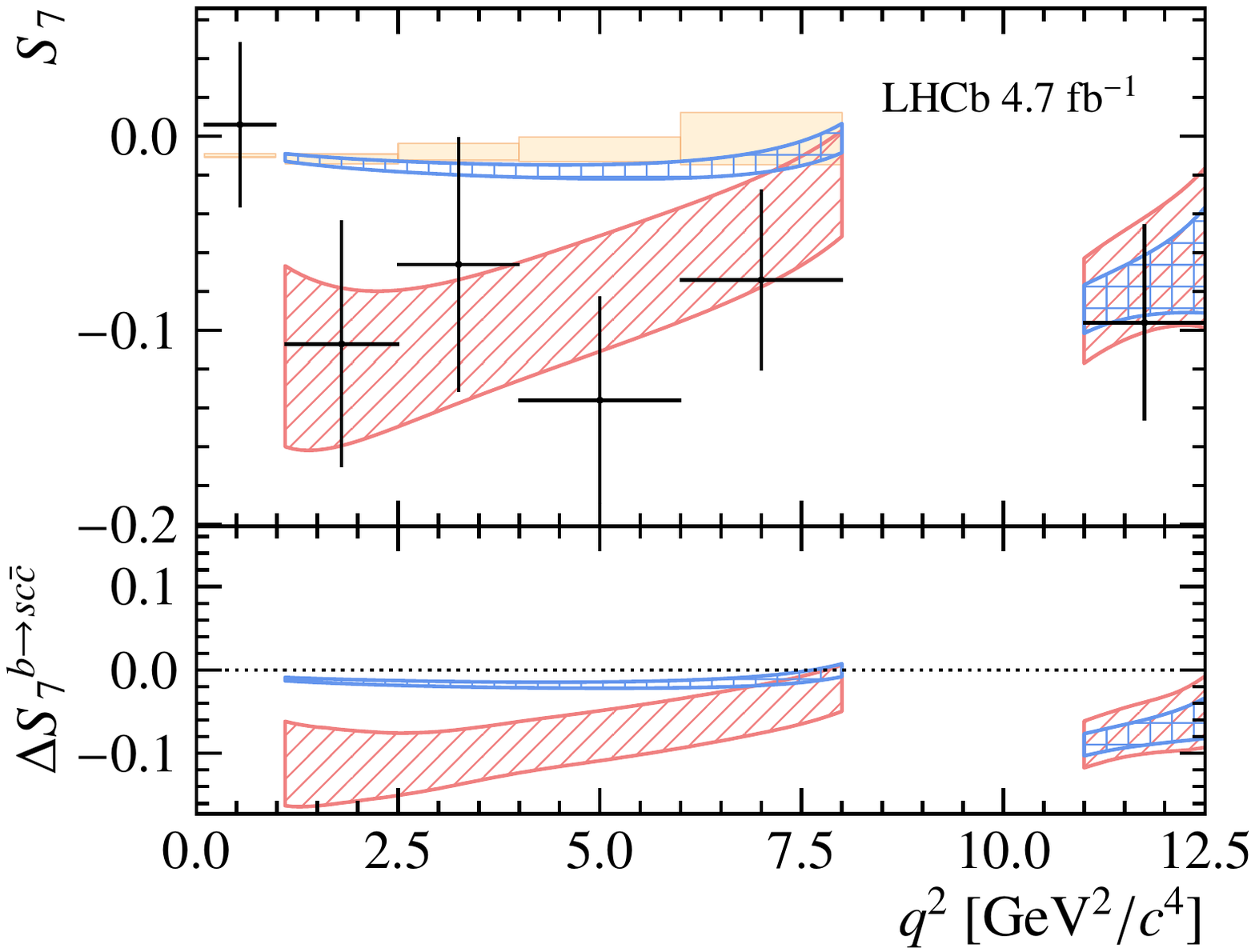} \\
\vspace{1mm}
\includegraphics[width=0.47\textwidth]{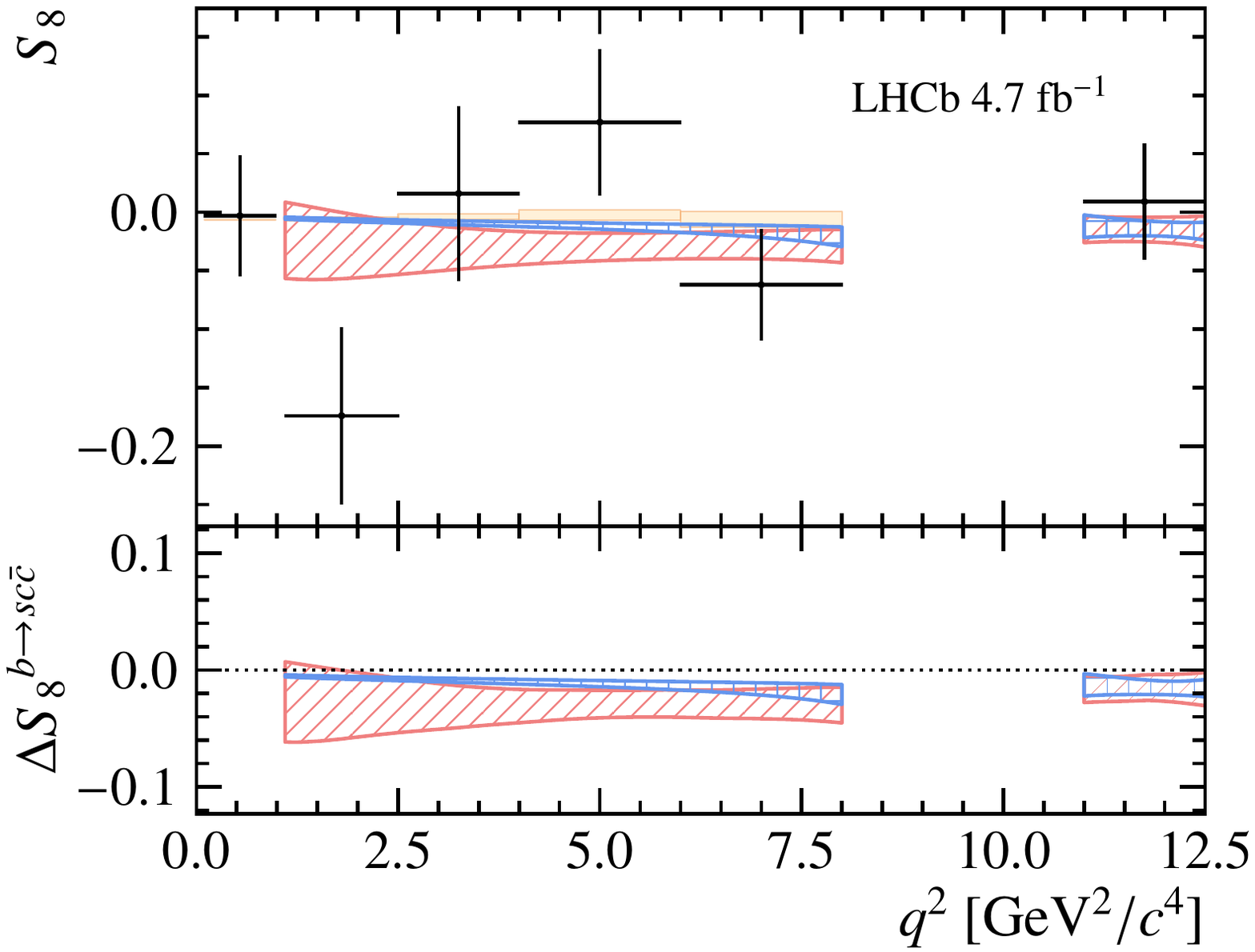}
\includegraphics[width=0.47\textwidth]{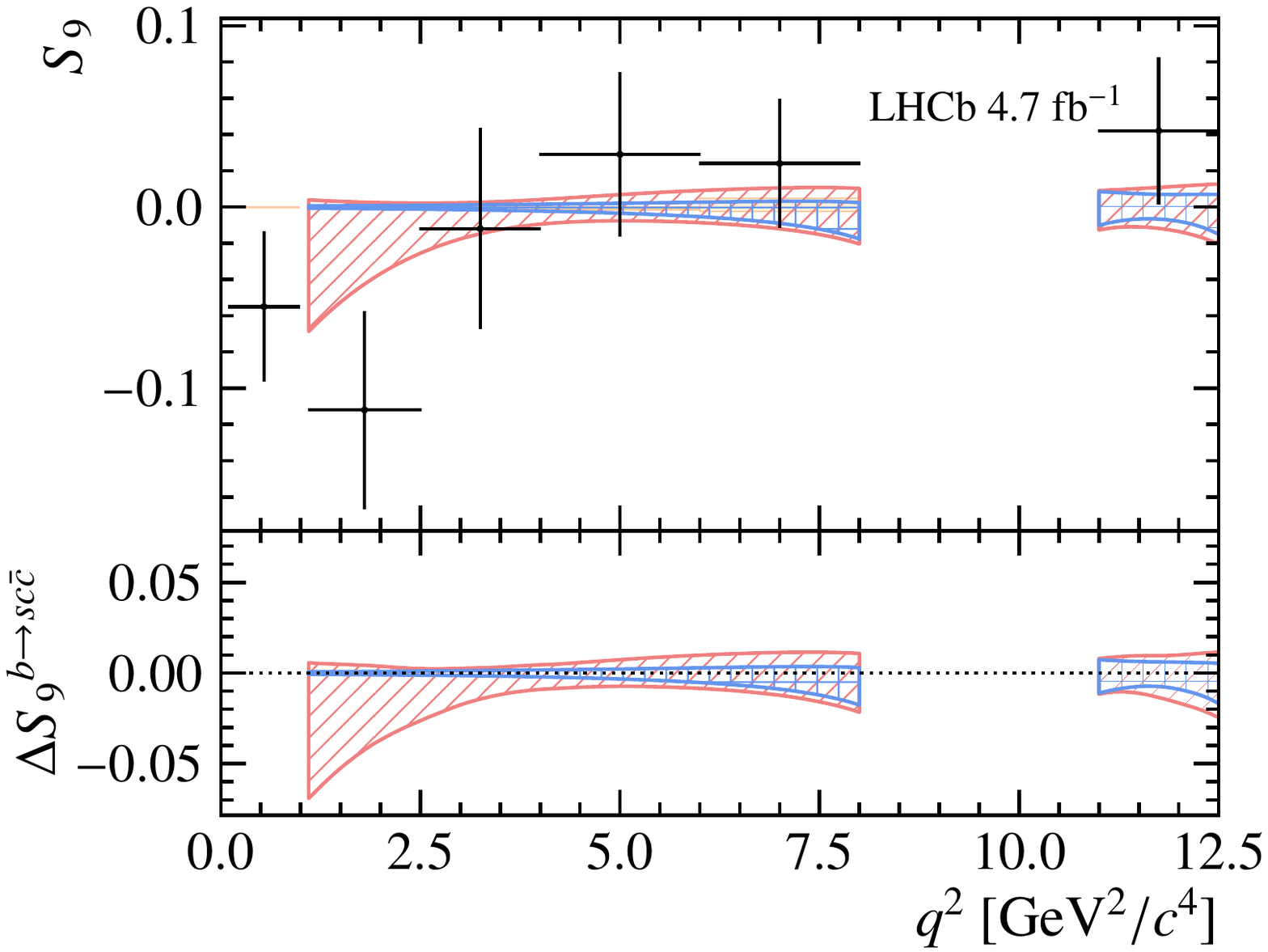}
\caption{Angular observables ($S$-basis) obtained \textit{a posteriori} from the fit results of the two fit configurations; 
the subfigures isolate the contribution from non-local effects to the given angular observables.
The LHCb result from Ref.~\cite{LHCb-PAPER-2020-002} is overlaid for comparison,  
together with the SM  prediction from GRvDV~\cite{Gubernari:2022hxn}.  }
\label{fig:ang_obs_S}
\end{figure}

\begin{figure}[t]
\centering
\includegraphics[width=0.47\textwidth]{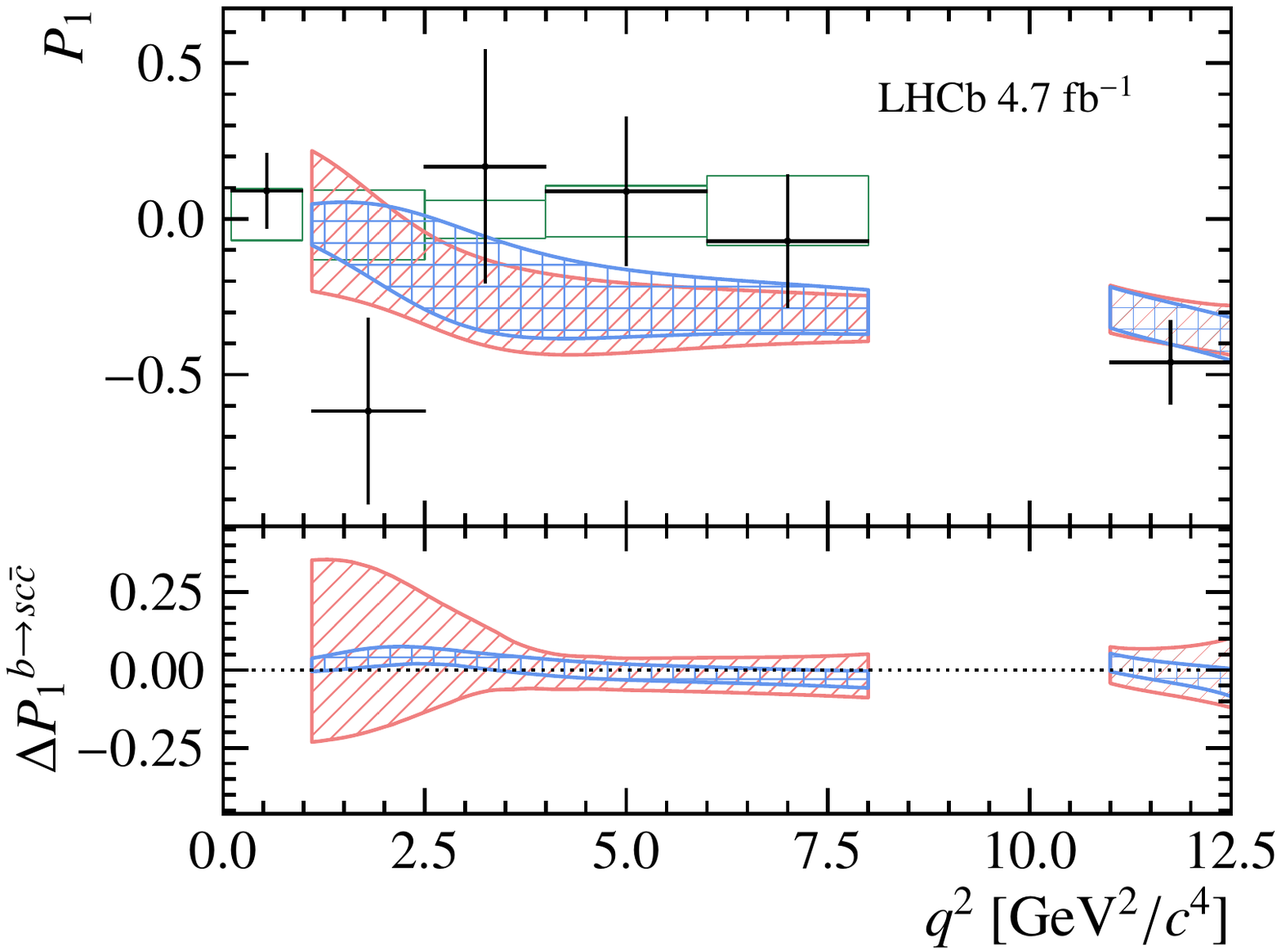}
\includegraphics[width=0.47\textwidth]{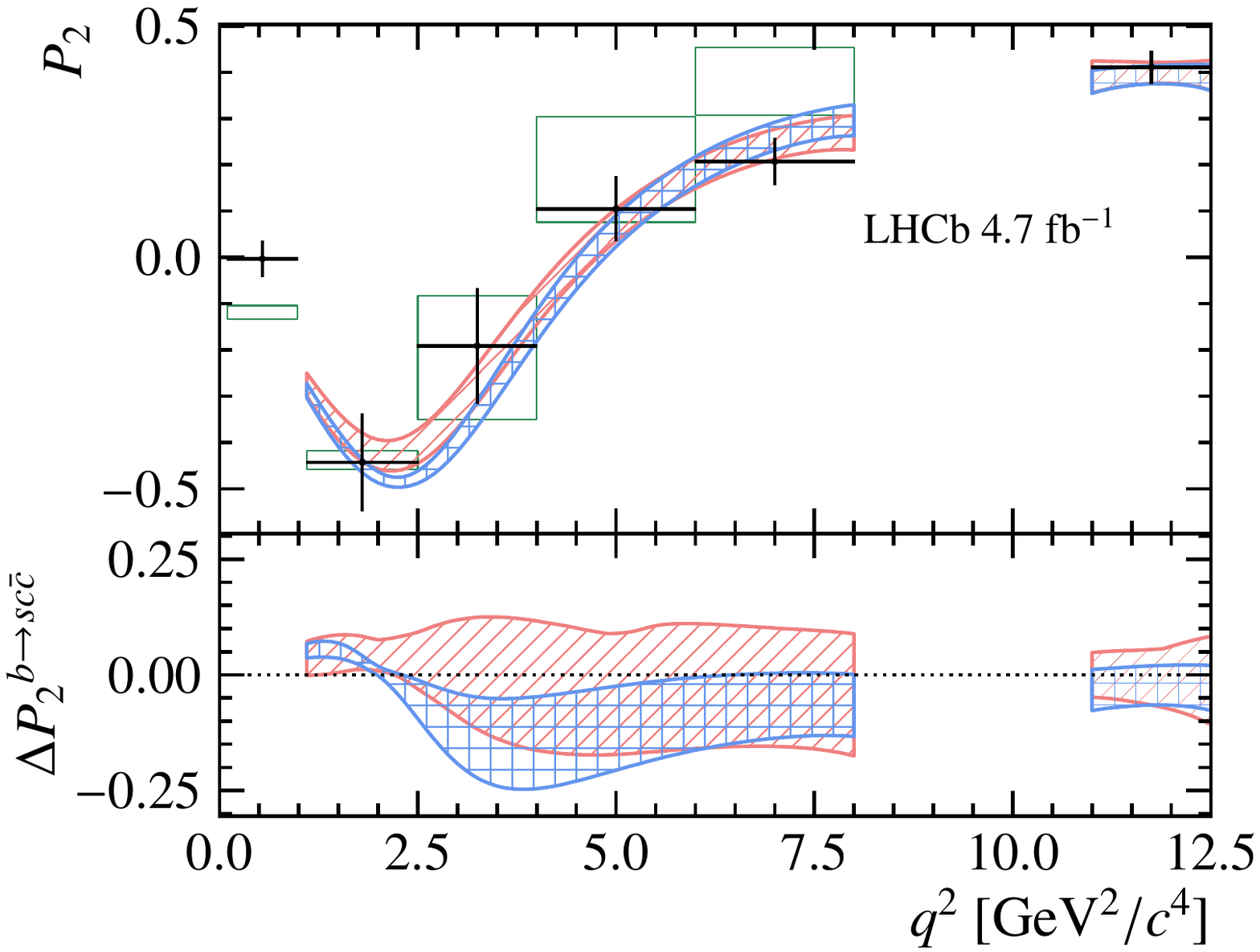}  \\
\vspace{2mm}
\includegraphics[width=0.47\textwidth]{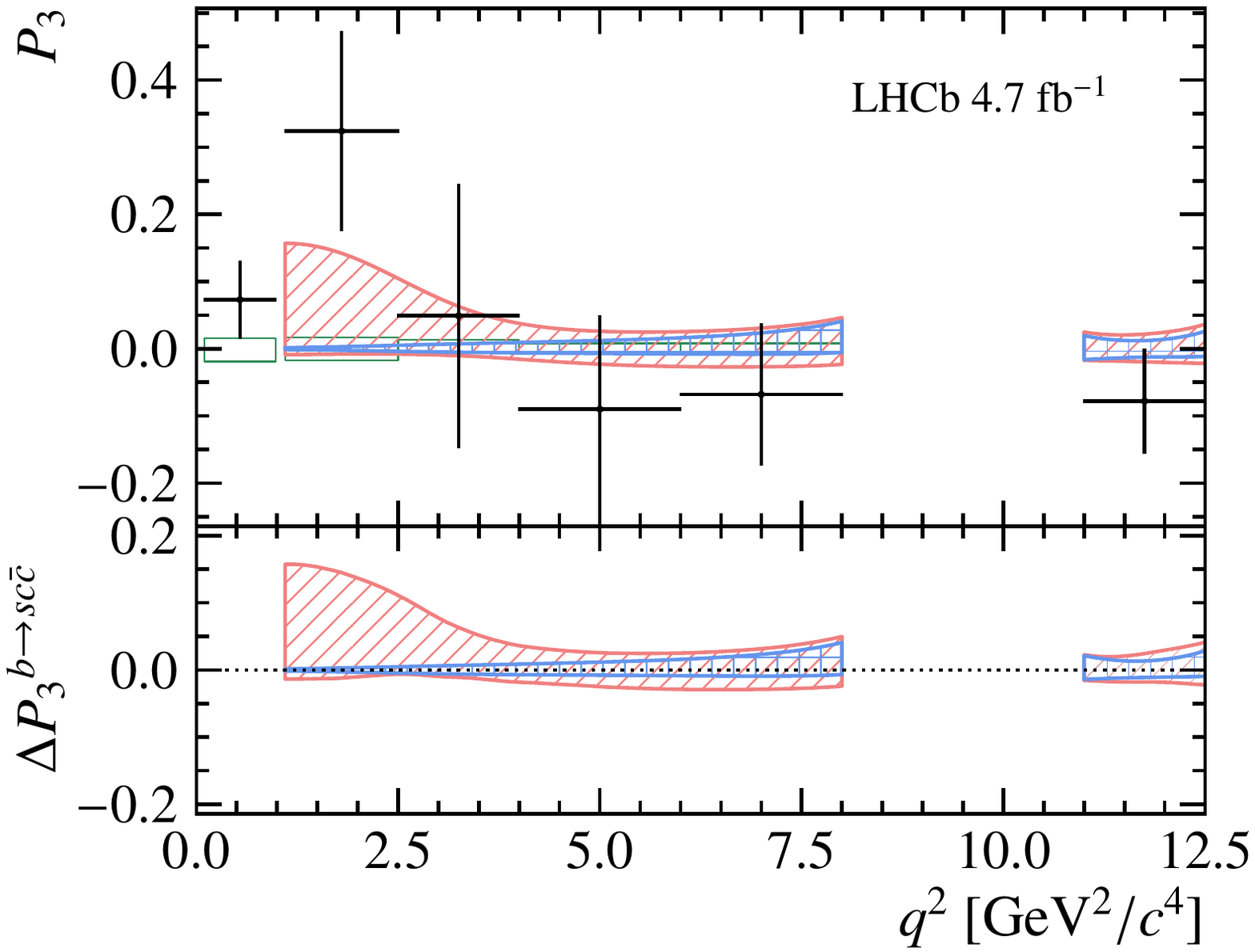} 
\includegraphics[width=0.47\textwidth]{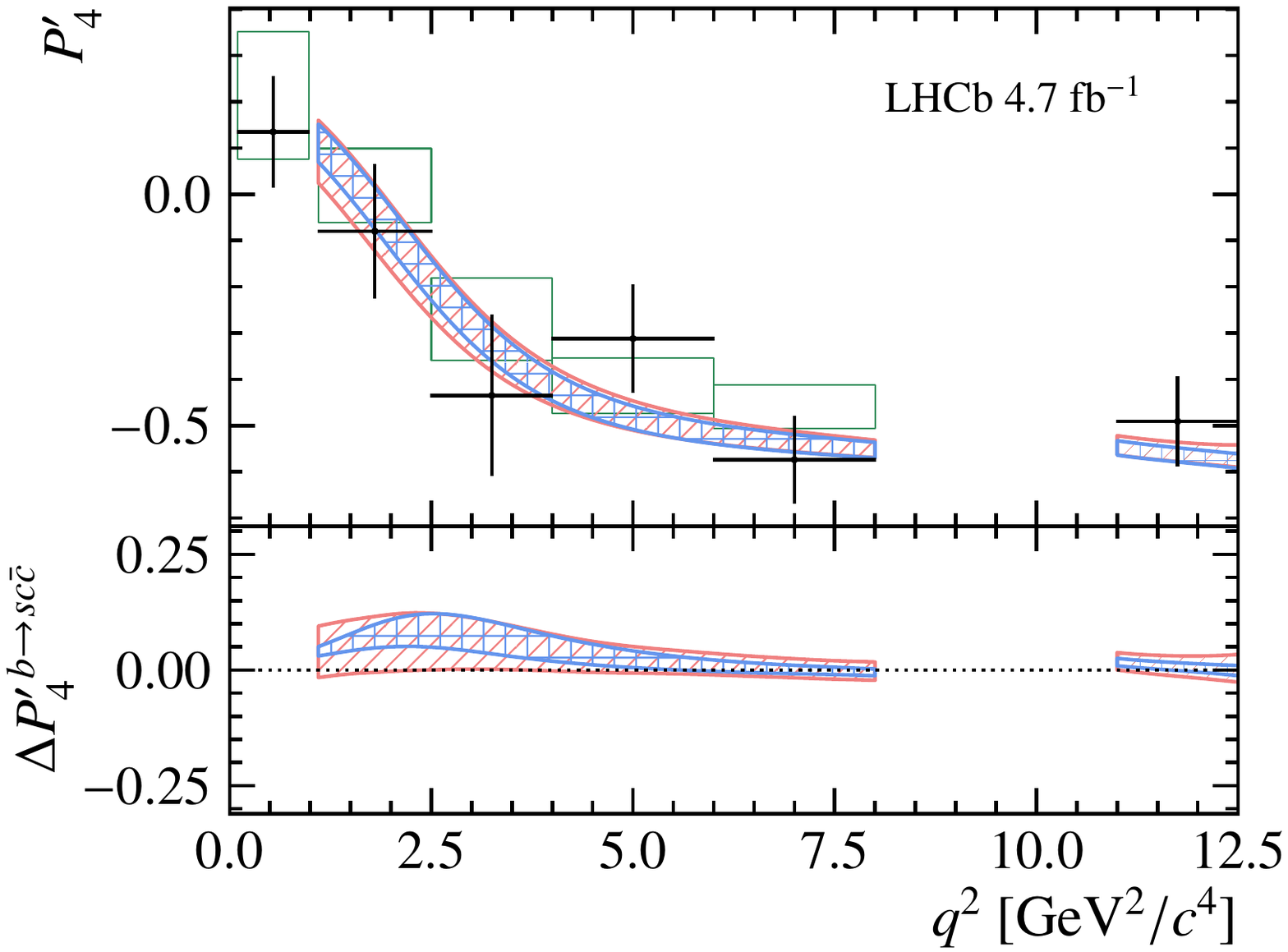}  \\ 
\vspace{2mm}
\includegraphics[width=0.47\textwidth]{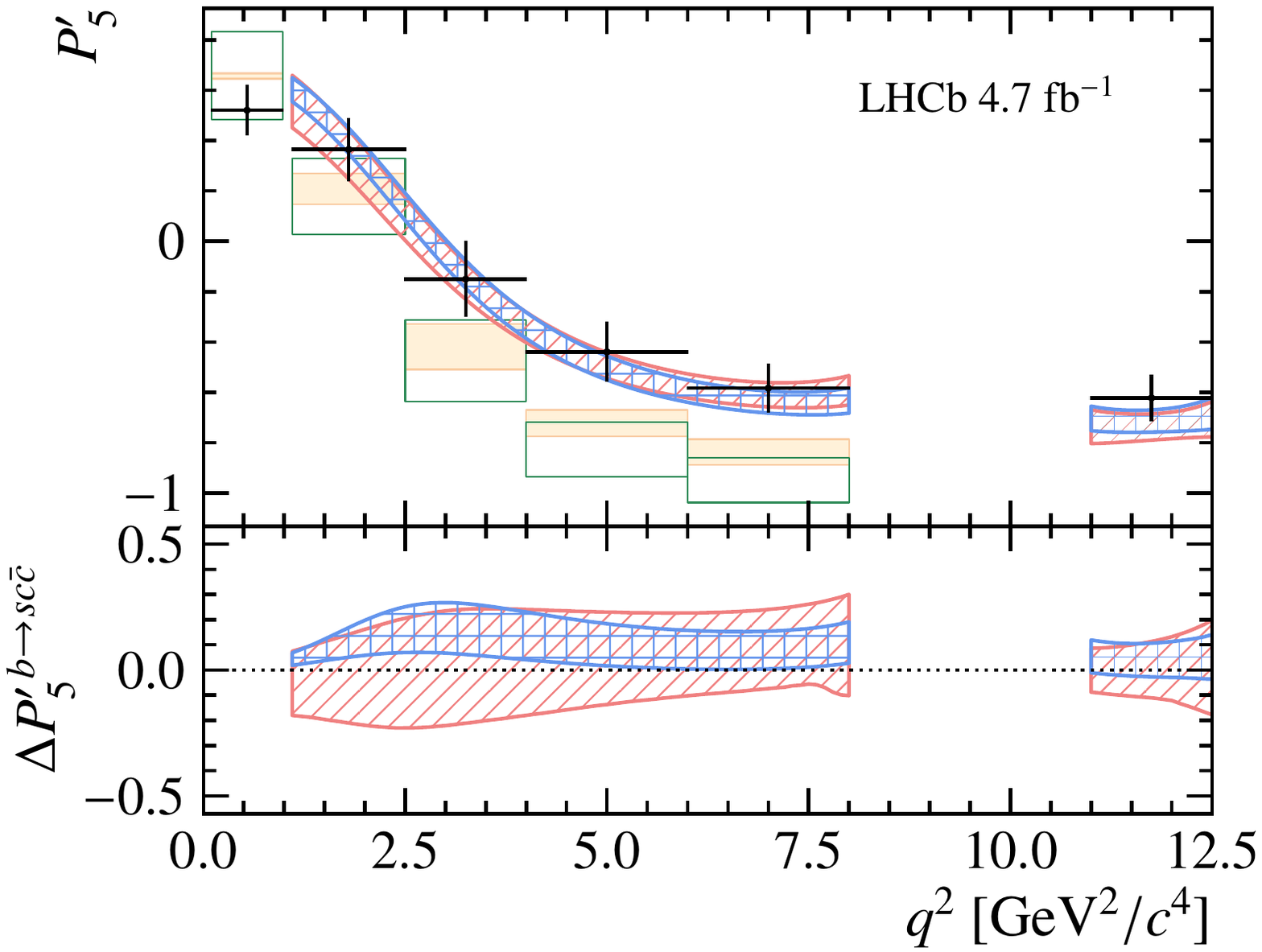}
\includegraphics[width=0.47\textwidth]{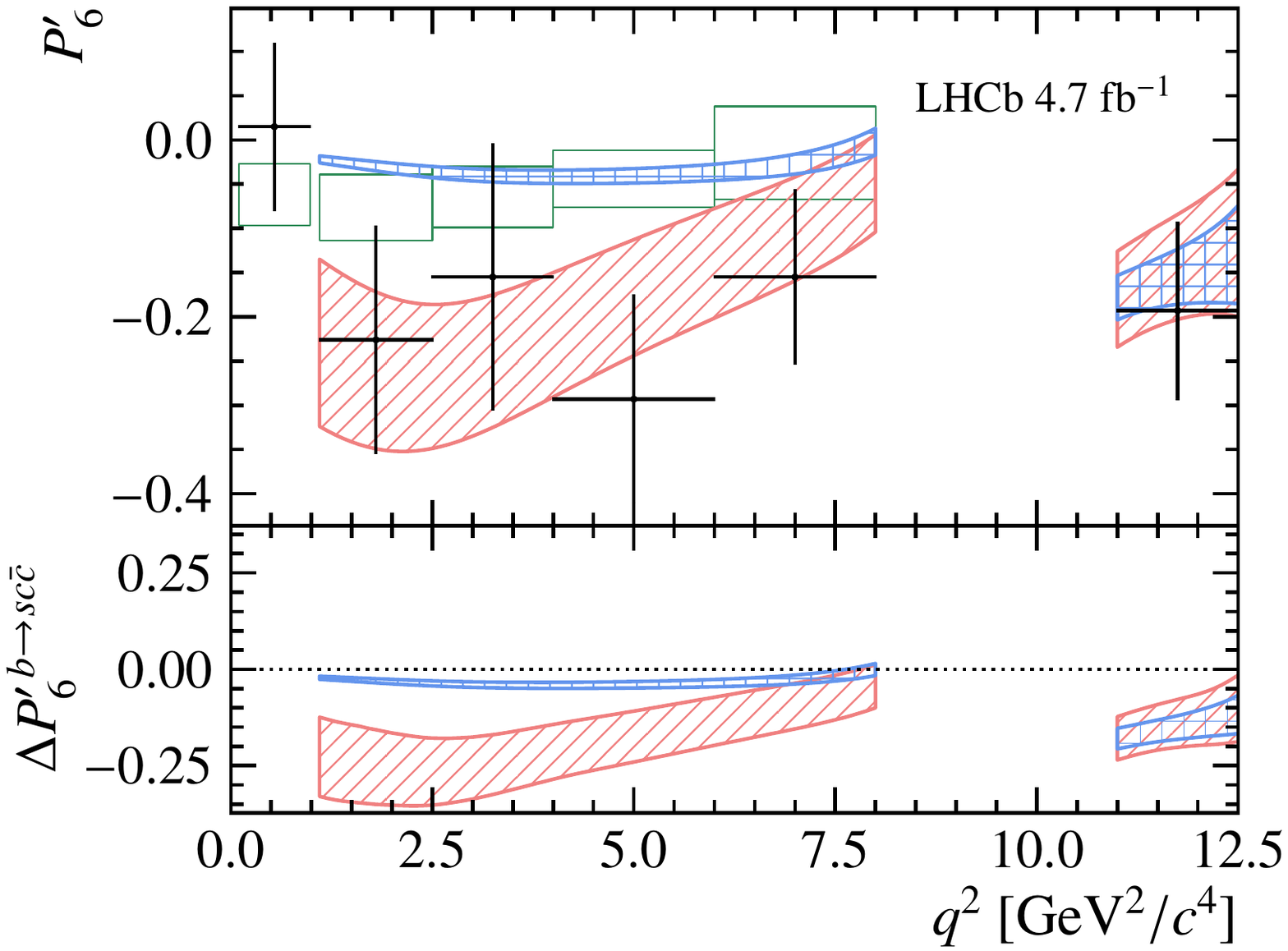} \\
\vspace{2mm}
\includegraphics[width=0.47\textwidth]{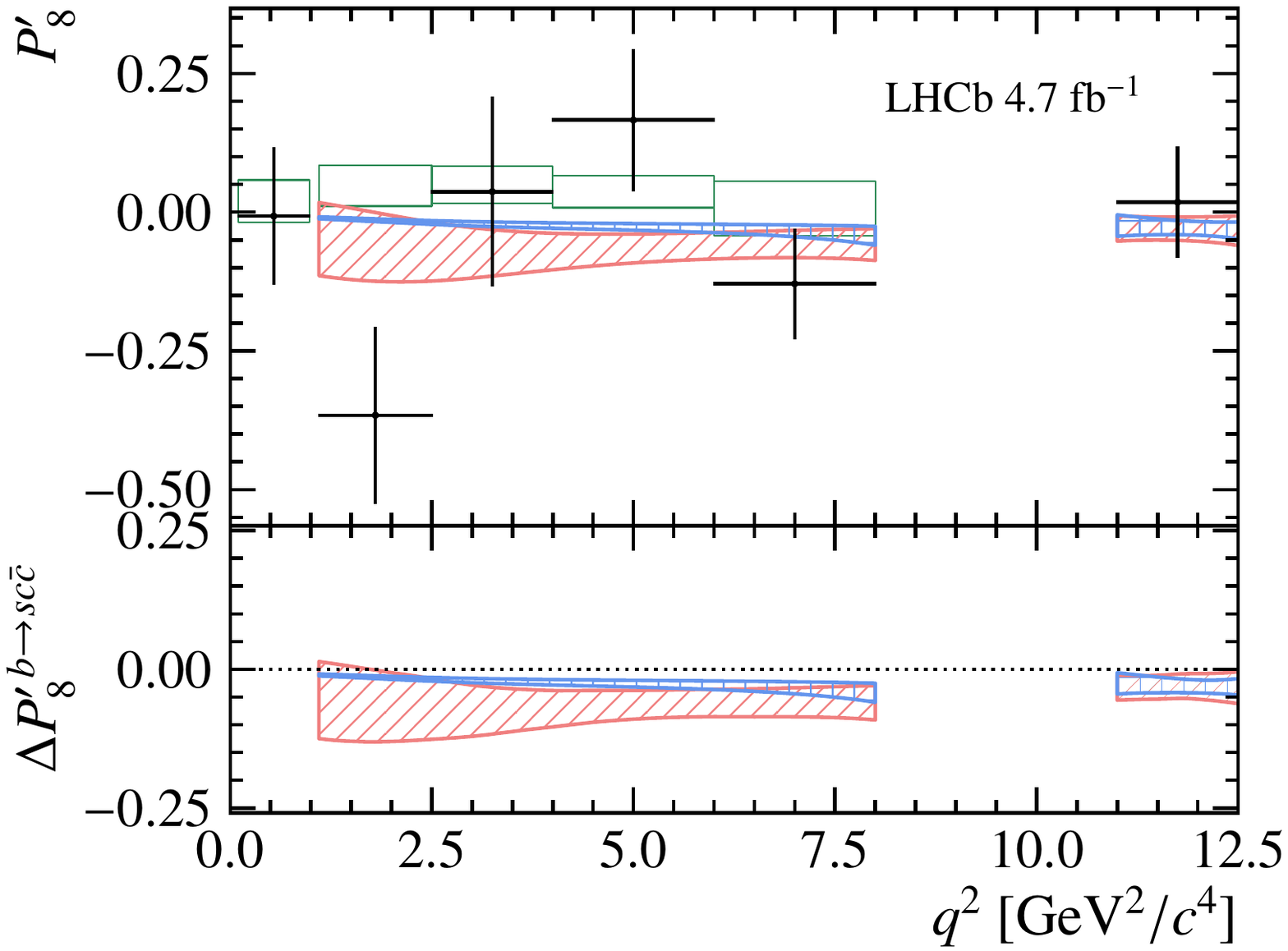}
\includegraphics[width=0.47\textwidth]{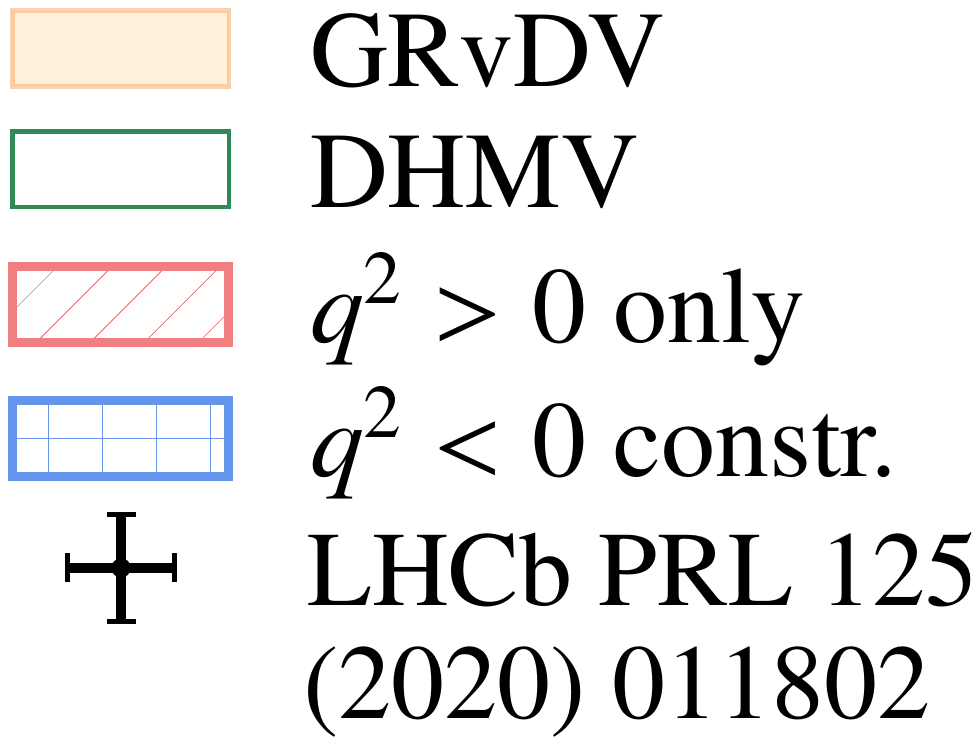}
\caption{Angular observables ($P$-basis) obtained \textit{a posteriori} from the fit results of the two fit configurations; 
the subfigures isolate the contribution from non-local effects to the given angular observables.
The LHCb result from Ref.~\cite{LHCb-PAPER-2020-002} is overlaid for comparison, together  with the SM  predictions from DHMV~\cite{Khodjamirian:2010vf,Descotes-Genon:2014uoa} and (for $P_5^\prime$) GRvDV~\cite{Gubernari:2022hxn}.}
\label{fig:ang_obs_P}
\end{figure}

Conventional angular observables accessed by binned angular 
analyses~\cite{LHCb-PAPER-2013-019,LHCb-PAPER-2015-051,LHCb-PAPER-2020-002}
can be determined from the fit results by dividing the angular coefficients, $I_i(q^2, k^2)$, by the differential decay rate, 
$\textrm{d}^2 \Gamma^{\mathrm{P}} / \textrm{d} q^2 \textrm{d} k^2$,  
both integrated over $k^2$. 
The determination of these angular observables offers an important perspective for 
the validation and interpretation of the results.
Figures~\ref{fig:ang_obs_S} and~\ref{fig:ang_obs_P}
show the $q^2$-dependent angular observables derived from the amplitude fit results.
The contributions from non-local effects to the 
so-called \textit{CP}-averaged $S_{i}$~\cite{Altmannshofer:2008dz} and corresponding optimised $P_{i}$~\cite{Descotes-Genon:2012isb} series of observables, 
$\Delta S(P)_i^{bsc\bar{c}}$, 
is also illustrated in the plots.
In general, the post-fit determination of the angular observables agrees very well 
with the dedicated measurement of Ref.~\cite{LHCb-PAPER-2020-002}
and the overall impact of non-local hadronic contributions on the angular observables 
is found to be compatible between the two tested fit configurations.
The only exception is observed in the $S_7$ (and the related $P^\prime_6$) observable, 
which is related to the imaginary part of the product of the longitudinal and parallel amplitudes,
where the fit result that includes the theory points at $q^2<0$
does not have enough freedom to fully accommodate the shape observed in the physical region.
This is a reflection of the different behaviour of the imaginary part of $\mathcal{H}_\lambda(q^2)$ 
between the two fit configurations observed in Sec.~\ref{sec:non_local}.
In addition, a closer look at the $P^\prime_5$ observable indicates that 
non-local hadronic contributions are responsible for a positive shift in 
$P^\prime_5$ of the order of $0.1 \pm 0.1$ 
in the region between $4$ and $8\gevgevcccc$.
This is found to be true for both the fit configurations with and without the $q^{2} < 0$ constraints, 
with the latter characterised by a  naturally larger uncertainty.

\begin{figure}[t]
\centering
\hspace{-5mm}
\includegraphics[width=0.49\textwidth]{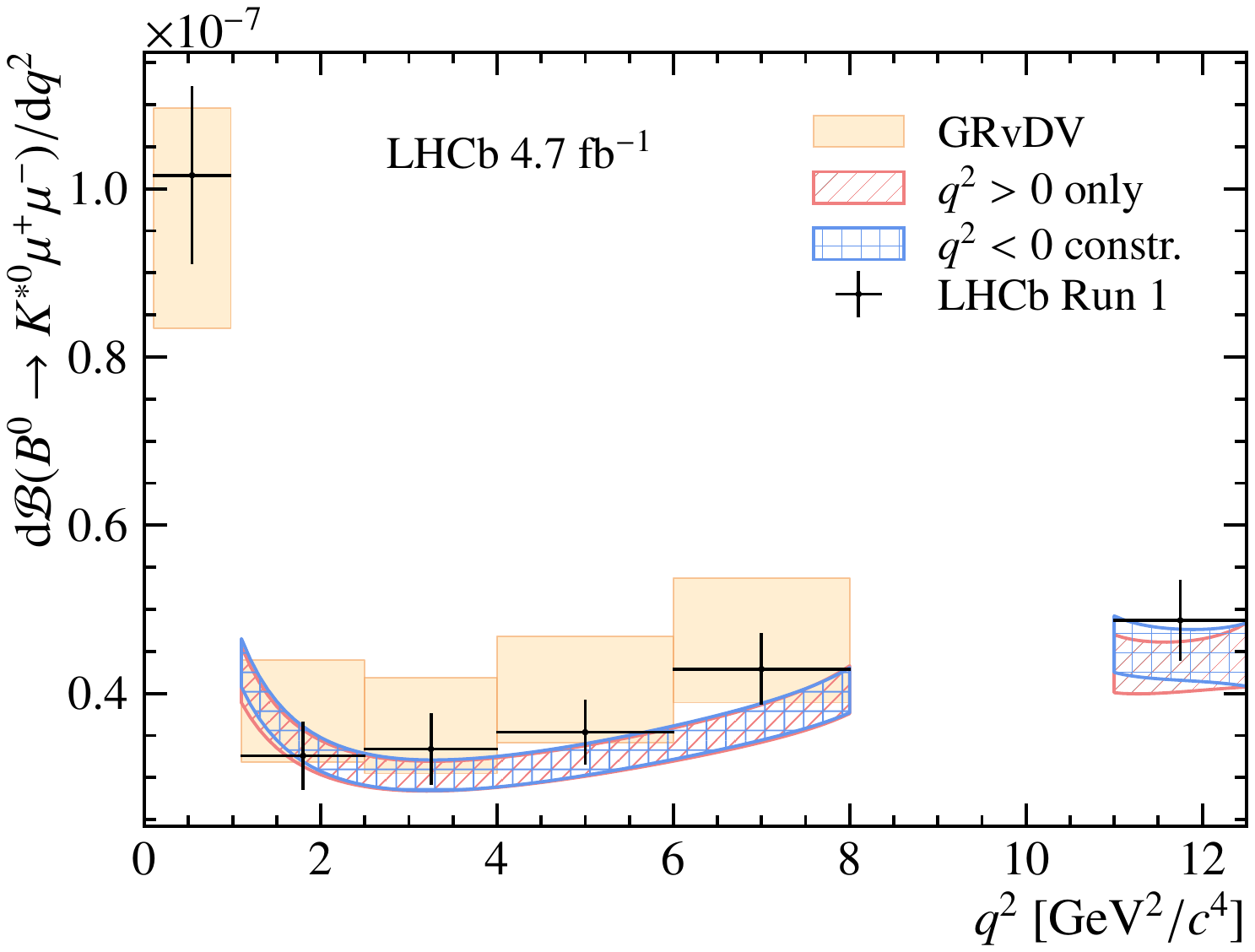}
\hspace{-5mm}
\includegraphics[width=0.49\textwidth]{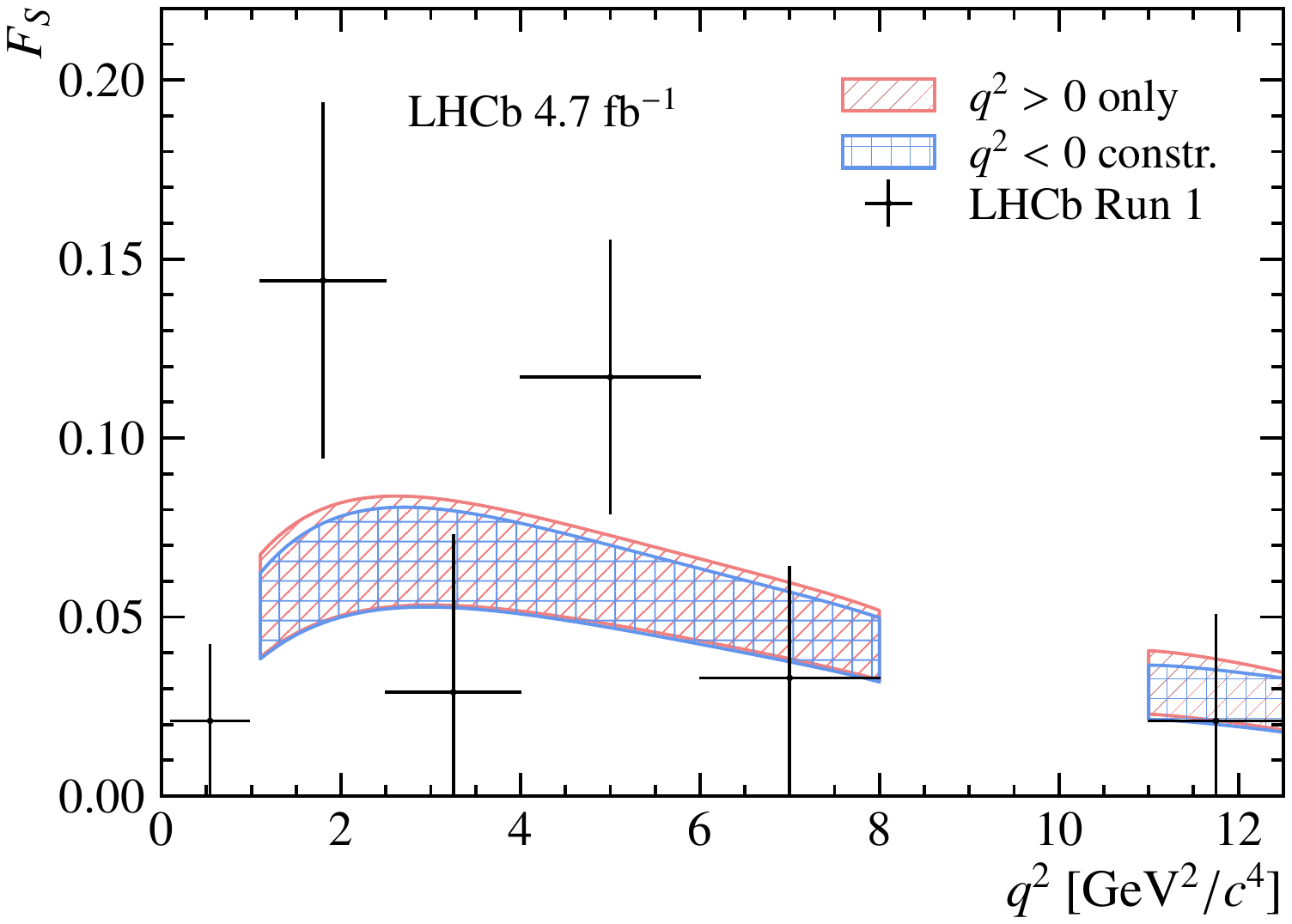} 
\caption{Branching fraction and S-wave fraction obtained \textit{a posteriori} from the fit results of the two fit configurations.
The published Run~1 measurement from LHCb~\cite{LHCb-PAPER-2016-012} has been overlaid for comparison.
The SM branching fraction prediction from GRvDV~\cite{Gubernari:2022hxn} is also reported.}
\label{fig:Br_result}
\end{figure}

Similarly, the signal branching fraction can be derived from the amplitude fit parameters through Eq.~\ref{eq:Br_Kstmm}.
Figure~\ref{fig:Br_result} shows the unbinned determination of the $B^0 \to K^{*0} \mumu$ differential branching fraction obtained from the amplitude fit compared to the values reported in Ref.~\cite{LHCb-PAPER-2016-012}. 
Despite the larger sample analysed in this measurement, 
a visible systematic shift toward lower values is observed in the analysis presented here.
This difference is due to the improved understanding of the
$B^0 \to \jpsi K^+ \pi^-$ normalisation channel from Ref.~\cite{Chilikin:2013tch},
which provides the external inputs required for the branching fraction determination as in Eq.~\ref{eq:nsig}.
It is observed that when the same external inputs of Ref.~\cite{LHCb-PAPER-2016-012} are used, the agreement improves dramatically.
Finally, Fig.~\ref{fig:Br_result} also shows the obtained fraction of S-wave in the signal region, 
which is found to be in line with previous estimations.

\section{Discussion}
\label{sec:discussion}

This analysis provides an innovative investigation of $B^0 \to K^{*0} \mumu$ decays
that enables the determination of the signal amplitude parameters directly from data.
Non-local hadronic contributions are parameterised through a $z$-expansion, where the truncation of the series unavoidably introduces some model dependence.
Two complementary fit configurations are developed with the intention to provide the highest
level of information that can be extracted from the data.
The first one provides a direct determination of non-local hadronic contributions which relies entirely on experimental data;
the second one includes the best theory knowledge on those matrix elements available to date~\cite{Gubernari:2022hxn}.

Both models provide a result which is in line with global fits to $b \to s \mumu$ 
mediated decays available in the literature~\cite{Ciuchini:2022wbq,Greljo:2022jac,Alguero:2023jeh,Gubernari:2022hxn,Capdevila:2023yhq,Hurth:2023jwr},
favouring a shift in $\mathcal{C}_9$ of about $-0.9$ or $-0.7$, depending on the theory assumptions.
The angular analysis of Ref.~\cite{LHCb-PAPER-2020-002}, with which this analysis shares the analysed dataset,
evaluated the $\mathcal{C}_9$ tension with respect to the SM at $3.3\, \sigma$ with the \flavio software package~\cite{Straub:2018kue}.
This evaluation was, however, based on binned inputs and dimensional estimates of the hadronic uncertainties.
In addition, the sole variation of $\mathcal{C}_9$ was considered, \textit{i.e.} all other Wilson coefficients were fixed to their SM values.
On the other hand, Ref.~\cite{Gubernari:2022hxn}, which employs the same parameterisation used in this analysis, considers a simultaneous variation of the Wilson coefficients $\mathcal{C}_9$ and $\mathcal{C}_{10}$ which results in a similar compatibility with the SM as the one obtained in this analysis.
The main difference between this analysis and the work in Ref.~\cite{Gubernari:2022hxn} is the use of the data on an event-by-event basis.
This unbinned approach allows one to explore experimental data in its full complexity and therefore achieves the highest possible precision on the physics parameters of interest.
In addition, one of the advantages of the analysis presented in this paper is a more flexible investigation of the non-local contributions, which can be studied in relation to what is observed in the data. 
Some tension is observed in the imaginary part of those contributions where the theory points seem to overconstrain the behaviour of $\mathrm{Im}(\mathcal{H}_\lambda)$ to the physical region.
The effect is mostly visible in the $S_7$ angular observable, where non-zero values can only originate from strong phase differences in the non-local hadronic amplitudes.
The fit model with only $q^{2} > 0$ information follows naturally the measurement provided by the binned angular analysis, while the fit with theory cannot depart significantly from zero.

Concerning the $P_5^\prime$ observable, 
non-local hadronic contributions  are found to reduce the tension with respect to the Standard Model.
However,
despite the high degree of freedom allowed by the inclusion of non-local effects in the model, the fit still prefers to insert a shift in $\mathcal{C}_9$ rather than in the hadronic parameters, assuming the theoretical inputs on local form factors are reliable.
The use of predictions at $q^2 < 0$ does not significantly change the level of agreement between the Wilson coefficients and their SM predictions.

Finally, local form factor predictions are currently the limiting factor for the understanding of the tension observed in the branching fraction measurements of many $b \to s \mu\mu$ decay channels.
Any further indication on the contribution of these elements to the decay rate is therefore extremely valuable.
This analysis favours a small shift in the ratio of the form factors with respect to the theory predictions.


\section{Conclusions}
\label{sec:conclusions}

This paper presents the first unbinned amplitude analysis of $B^0 \to \Kstarz \mumu$ decays.
This analysis provides a rich and complementary 
set of information compared to more standard analyses of the same decay that considered only binned observables~\cite{LHCb-PAPER-2016-012,LHCb-PAPER-2020-002}. 
By exploiting the $q^2$ dependence of the decay amplitudes, an experimental determination of the impact of both short- and long-distance contributions to the decay is obtained for the first time.
This provides the most accurate parameterisation of the impact of long-distance effects on the $B^0 \to \Kstarz \mumu$ decay process to date.
To favour possible reinterpretation of the analysis, the signal amplitude parameters are made publicly available as supplemental material~\cite{Suppl}. 
The fit results indicate a shift in the $\mathcal{C}_9$ coefficient between $-0.7$ and $-0.9$, 
depending on the theory assumptions employed.
This shift is consistent with results of global analyses of binned observables.
The contribution from non-factorisable corrections alone cannot fully explain the SM discrepancies seen in these decays but does reduce the tension observed in $\mathcal{C}_9$ to the level of $1.8$ and $1.9$ standard deviations.
The global compatibility with the Standard Model is also evaluated by simultaneously considering the Wilson coefficients $\mathcal{C}_9, \, \mathcal{C}_{10}, \, \mathcal{C}_9^\prime$ and $\mathcal{C}_{10}^\prime$  and is found to be between $1.3$ and $1.4$ standard deviations, where the two sets of numbers reflect the choice of theoretical assumptions employed in the analysis.

\clearpage

\section*{Acknowledgements}
%
%
\noindent We are very grateful to Nico Gubernari, M\'{e}ril Reboud, Danny van Dyk, 
and Javier Virto for the many helpful discussions.
We express our gratitude to our colleagues in the CERN
accelerator departments for the excellent performance of the LHC. We
thank the technical and administrative staff at the LHCb
institutes.
We acknowledge support from CERN and from the national agencies:
CAPES, CNPq, FAPERJ and FINEP (Brazil); 
MOST and NSFC (China); 
CNRS/IN2P3 (France); 
BMBF, DFG and MPG (Germany); 
INFN (Italy); 
NWO (Netherlands); 
MNiSW and NCN (Poland); 
MCID/IFA (Romania); 
MICINN (Spain); 
SNSF and SER (Switzerland); 
NASU (Ukraine); 
STFC (United Kingdom); 
DOE NP and NSF (USA).
We acknowledge the computing resources that are provided by CERN, IN2P3
(France), KIT and DESY (Germany), INFN (Italy), SURF (Netherlands),
PIC (Spain), GridPP (United Kingdom), 
CSCS (Switzerland), IFIN-HH (Romania), CBPF (Brazil),
and Polish WLCG (Poland).
We are indebted to the communities behind the multiple open-source
software packages on which we depend.
Individual groups or members have received support from
ARC and ARDC (Australia);
Key Research Program of Frontier Sciences of CAS, CAS PIFI, CAS CCEPP, 
Fundamental Research Funds for the Central Universities, 
and Sci. \& Tech. Program of Guangzhou (China);
Minciencias (Colombia);
EPLANET, Marie Sk\l{}odowska-Curie Actions, ERC and NextGenerationEU (European Union);
A*MIDEX, ANR, IPhU and Labex P2IO, and R\'{e}gion Auvergne-Rh\^{o}ne-Alpes (France);
AvH Foundation (Germany);
ICSC (Italy); 
GVA, XuntaGal, GENCAT, Inditex, InTalent and Prog.~Atracci\'on Talento, CM (Spain);
SRC (Sweden);
the Leverhulme Trust, the Royal Society
 and UKRI (United Kingdom).


\section*{Appendices}

\appendix

\section{Formalism}
\label{app:formalism}

\subsection{Angular coefficients}
\label{app:angular_coefficients}

In the SM operator basis,
the P-wave angular coefficients entering in Eq.~\ref{eq:d4Gamma_P} are
\begin{eqnarray}
\label{eq:ang_coeff}
I_{1s}  & =  &  \frac{2+\beta_l^2}{4} \Big[ |\mathcal{A}_\perp^L |^2 + |\mathcal{A}_\parallel^L |^2  +  (L \to R) \Big]    
  + \frac{4 m_l^2}{q^2} \re \Big( \mathcal{A}_\perp^L \mathcal{A}_\perp^{R*} +\mathcal{A}_\parallel^L {\mathcal{A}_\parallel^R}^* \Big)  ,   \nonumber  \\
I_{1c}  & =  & \Big[ |\mathcal{A}_0^L |^2 + |\mathcal{A}_0^R |^2 \Big]  
                      + \frac{4 m_l^2}{q^2}  \Big[ |\mathcal{A}_t |^2 + 2 \re (\mathcal{A}_0^L {\mathcal{A}_0^R}^* ) \Big]  ,   \nonumber  \\ 
I_{2s}  & =  &  \frac{\beta_l^2}{4} \Big[ |\mathcal{A}_\perp^L |^2 + |\mathcal{A}_\parallel^L |^2  +  (L \to R) \Big]  ,  \nonumber \\ 
I_{2c}  & =  &  -\beta_l^2 \Big[ |\mathcal{A}_0^L |^2 + |\mathcal{A}_0^R |^2  \Big]   ,  \nonumber\\ 
I_3  & =  &  \frac{\beta_l^2}{2} \Big[ |\mathcal{A}_\perp^L |^2 - |\mathcal{A}_\parallel^L |^2  +  (L \to R)  \Big]   ,  \\ 
I_4  & =  & -  \frac{\beta_l^2}{\sqrt{2}} \re \Big[ \mathcal{A}_0^L   {\mathcal{A}_\parallel^L}^*  +  (L \to R)  \Big]   ,  \nonumber\\ 
I_5  & =  &  \sqrt{2} \beta_l \re \Big[ \mathcal{A}_0^L   {\mathcal{A}_\perp^L}^*  -  (L \to R)  \Big]   , \nonumber \\ 
I_{6s}  & =  & -   2\beta_l \re \Big[ \mathcal{A}_\parallel^L   {\mathcal{A}_\perp^L}^*  -  (L \to R)  \Big]  ,  \nonumber \\ 
I_{7}  & =  & -   \sqrt{2}\beta_l \im \Big[ \mathcal{A}_0^L   {\mathcal{A}_\parallel^L}^*  -  (L \to R)  \Big]   , \nonumber \\ 
I_{8}  & =  &  \frac{\beta_l^2}{\sqrt{2}} \im \Big[ \mathcal{A}_0^L   {\mathcal{A}_\perp^L}^*  +  (L \to R)  \Big]  ,   \nonumber\\ 
I_{9}  & =  & -   \beta_l^2 \im \Big[ \mathcal{A}_\perp^L   {\mathcal{A}_\parallel^L}^*  +  (L \to R)  \Big]   ,   \nonumber
\end{eqnarray}
where the negative sign  in front of $I_{4,6s,7,9}$ is due to the adoption of the LHCb
angular convention that differs from the theory convention of Refs.~\cite{Bobeth:2012vn,Bobeth:2017vxj} by the relations given in Ref.~\cite{Gratrex:2015hna}
and $\beta_l = \sqrt{1-4m_l^2/q^2}$, with $m_l$ the mass of the lepton.
Similarly, the introduction of the S-wave contribution gives origin to the following additional 
set of angular coefficients
\begin{eqnarray}
\label{eq:Js_Swave}
I_{1c}^S  & =  & \frac{1}{3}  \Big\{   \Big[ |\mathcal{A}_{S0}^L |^2 + |\mathcal{A}_{S0}^R |^2 \Big]  
                      + \frac{4 m_l^2}{q^2}  \Big[ |\mathcal{A}_{St} |^2 + 2 \re (\mathcal{A}_{S0}^L {\mathcal{A}_{S0}^R}^* ) \Big] \Big\}  ,   \nonumber  \\ 
I_{2c}^S  & =  &  -\frac{1}{3}  \beta_l^2 \Big[ |\mathcal{A}_{S0}^L |^2 + |\mathcal{A}_{S0}^R |^2  \Big]   ,  \nonumber\\ 
\tilde{I}_{1c}  & =  & \frac{2}{\sqrt{3}}  \re  \Big[ \mathcal{A}_{S0}^L {\mathcal{A}_0^L}^*  
	                 +  \mathcal{A}_{S0}^R {\mathcal{A}_0^R}^*  + \frac{4 m_l^2}{q^2}  \Big( \mathcal{A}_{S0}^L  {\mathcal{A}_0^R}^*
	                 +  \mathcal{A}_0^L  {\mathcal{A}_{S0}^R}^*  +  \mathcal{A}_{St}  {\mathcal{A}_t}^*  \Big) \Big]  ,   \nonumber  \\ 
\tilde{I}_{2c}  & =  &  -  \frac{2}{\sqrt{3}} \beta_l^2   \re  \Big[ \mathcal{A}_{S0}^L {\mathcal{A}_0^L}^*  +  \mathcal{A}_{S0}^R {\mathcal{A}_0^R}^*  \Big]   ,  \\ 
\tilde{I}_4  & =  & - \sqrt{\frac{2}{3}} \beta_l^2 \re \Big[ \mathcal{A}_{S0}^L   {\mathcal{A}_\parallel^L}^*  +  (L \to R)  \Big]   ,  \nonumber 
\end{eqnarray}
\begin{eqnarray}
\tilde{I}_5  & =  &  \sqrt{\frac{8}{3}} \beta_l^2 \re \Big[ \mathcal{A}_{S0}^L   {\mathcal{A}_\perp^L}^*  -  (L \to R)  \Big]   , \nonumber \\ 
\tilde{I}_{7}  & =  & -  \sqrt{\frac{8}{3}} \beta_l^2  \im \Big[ \mathcal{A}_{S0}^L   {\mathcal{A}_\parallel^L}^*  -  (L \to R)  \Big]   , \nonumber \\ 
\tilde{I}_{8}  & =  &   \sqrt{\frac{2}{3}} \beta_l^2 \im \Big[ \mathcal{A}_{S0}^L   {\mathcal{A}_\perp^L}^*  +  (L \to R)  \Big]  ,   \nonumber
\end{eqnarray}
where the first two are pure S-wave contributions while those denoted $\tilde{I}_i$ arise from interference terms.
As above, the negative sign in front of $ \tilde{I}_{4,7}$ results from the transformation 
from the theory to the LHCb angular convention.

\subsection{Form factor parameterisation}
\label{app:form_factors}

The form factor basis employed in this analysis follows closely the one proposed in Ref.~\cite{Gubernari:2020eft} 
which can be translated to the one commonly used in the literature 
(see \textit{e.g.} Refs.~\cite{Bobeth:2010wg,Ball:2004rg}) via
\begin{eqnarray}\label{eq:form_factors}
\mathcal{F}_\perp    & \mapsto &  \frac{ \sqrt{2 \lambda(M_B^2, q^2,k^2)} }{ M_B (M_B + M_{K^{*0}}) }  V    \,,     \nonumber   \\
\mathcal{F}_\parallel  & \mapsto  & \frac{ \sqrt{2} (M_B + M_{K^{*0}}) }{ M_B }  A_1          \, ,      \nonumber   \\
\mathcal{F}_0  & \mapsto  & \frac{ (M_B^2 - q^2 - M_{K^{*0}}^2) (M_B + M_{K^{*0}})^2 A_1  - \lambda(M_B^2, q^2,k^2) A_2 }{ 2 M_{K^{*0}} M_B^2 (M_B + M_{K^{*0}} )  } \,,  \nonumber \\ 
\mathcal{F}^T_\perp  & \mapsto &  \frac{ \sqrt{2 \lambda(M_B^2, q^2,k^2)} }{ M_B^2 }  T_1   \,,     \\
\mathcal{F}^T_\parallel  & \mapsto &  \frac{ \sqrt{2} (M_B^2 - M_{K^{*0}}^2) }{ M_B^2 }  T_2  \, ,      \nonumber    \\
\mathcal{F}^T_0   & \mapsto &  \frac{ q^2  ( M_B^2 + 3 M_{K^{*0}}^2 - q^2) }{2 M_B^3 M_{K^{*0}} }  T_2   -  \frac{  q^2  \lambda(M_B^2, q^2,k^2) }{2 M_B^3 M_{K^{*0}} (M_B^2 - M_{K^{*0}}^2 )  }  T_3 \, , \nonumber   \\
\mathcal{F}_t  & \mapsto &  \frac{\sqrt{\lambda(M_B^2, q^2,k^2)}}{M_B \sqrt{q^2}}  A_0 \, .     \nonumber   
\end{eqnarray}
The definition of the scalar form factors $f_+, f_T$ and $f_0$ follows Ref.~\cite{Becirevic:2012dp} with the exception of the transformation $f_T \mapsto \frac{q^2}{M_B ( M_B + k)} f_T$.

\subsection{P- and S-wave \texorpdfstring{$k^2$}{k2} lineshapes}
\label{app:mKpi_lineshapes}

The $k^2$ invariant mass distribution of the signal candidates is modelled separately for the P- and S-wave contributions.
For the P-wave component, a relativistic Breit--Wigner function is used, given by
\begin{equation}\label{eq:BW}
  f_{\rm BW} (k^2) =   \frac{\lambda^{1/4}_{K\pi}}{k} \,  B^\prime_{L=1}(p,p_{892},d) \, \bigg( \frac{p}{k} \bigg) \, \frac{1}{k^2 - m^2_{892} - i m_{892} \Gamma_{892} (k)}  \,,
\end{equation}
where $\lambda_{K\pi} = \lambda(k^2, M_K^2, M_\pi^2)$, $m_{892}$ is the pole mass of the $K^{*0}$ resonance, $p \,(p_{892})$ is the momentum of the $K^+$ in the rest frame of the resonance evaluated at a given $k^2 \, (m_{892}^2$), the running width $\Gamma_{892}(k)$ is given by
\begin{equation}
  \Gamma_{892}(k) = \Gamma_{892} B_{L=1}^{\prime 2}(p,p_{892},d) \left( \frac{p}{p_{892}} \right)^3 \left( \frac{m_{892}}{k} \right)\,,
\end{equation}
and $B^\prime_L$ is the Blatt--Weisskopf barrier function~\cite{Blatt}, which depends on the momentum of the decay products and on the  size of the decaying particle, known as the meson radius parameter, which is fixed to $d=1.6 \gev^{-1} c$~\cite{Chilikin:2013tch}.
The systematic uncertainty associated with the choice of this value is negligible.
In Eq.~\ref{eq:BW}, the first term is a pure kinematic phase space factor while $(B_L^\prime p^L)$ is the orbital angular momentum barrier factor that accounts for spin-dependent effects in the conservation of the angular momentum for the $K^{*0} \to K^+ \pi^-$ decay.
The angular momentum between the $K^{*0}$ meson and the dimuon system is considered to be zero.

The S-wave component of the signal is modelled using the LASS parametrisation~\cite{Aston:1987ir}, given by
\begin{equation}\label{eq:LASS}
  f_{\rm LASS} (k^2) =   \frac{\lambda^{1/4}_{K\pi}}{k}  \Bigg[  \frac{ k }{ p \cot \delta_B - i p} 
   + e^{2i\delta_B} \frac{ m_0 \Gamma_0  \frac{m_0 }{ p_0 }  }{ (m_0^2 - k^2 )  - i m_0 \Gamma_0 \frac{p}{k } \frac{ m_0 }{p_0} }  \Bigg]  \,  ,
\end{equation}
where $m_0$ and $\Gamma_0$ are the pole mass and width of the $K^*_0(1430)^0$ resonance and 
\begin{equation}
\cot \delta_B = \frac{1}{ak} + \frac{rk}{2} \, ,
\end{equation}
where $a$ and $r$ are empirical parameters whose values 
are fixed to  $a = 1.95 \gev^{-1}  c$ 
and $r = 1.78 \gev^{-1}  c$ as from the LASS experiment~\cite{Aston:1987ir}.
In order to assess the systematic effect of this choice, these parameters are also fixed to values used in Ref.~\cite{LHCb-PAPER-2016-012}, $a = 3.83 \gev^{-1}  c$  and $r = 2.86 \gev^{-1}  c$ and the resulting systematic uncertainty is found to be negligible.
The second term of Eq.~\ref{eq:LASS} is equivalent to a Breit--Wigner function for the $K^*_0(1430)$ resonance.
Phase space and orbital angular momentum barrier factors associated to $B^0 \to K^+\pi^- \mumu$ decays employed in Refs.~\cite{LHCb-PAPER-2016-012,LHCb-PAPER-2015-051} have been omitted in Eqs.~\ref{eq:BW} and \ref{eq:LASS} since these terms are already included in the form factors and amplitude normalisation of Eqs.~\ref{eq:Ampl}, \ref{eq:ampl_norm} and \ref{eq:Ampl_Swave}.

Finally, the P- and S-wave $k^2$-dependent lineshapes to be included in the decay amplitudes are defined as
\begin{equation}
\label{eq:f_P_S_waves}
\begin{aligned}
  f_{P}(k^2)  &  = \hat{f}_{\rm BW}(k^2)  \, , \\ 
  f_{S}(k^2)  &  = |g_S| \, e^{i\delta_S} \, \hat{f}_{\rm LASS}(k^2)     \, ,   
\end{aligned}
\end{equation}
where $\hat{f}_{\rm BW}$ ($\hat{f}_{\rm LASS}$) is the the Breit--Wigner (LASS) function of Eq.~\ref{eq:BW}(\ref{eq:LASS}) normalised to unity and the coefficients $g_S$ and $\delta_S$ determine respectively the relative magnitude and phase between the two P- and S-wave contributions.

\section{Non-local  contributions at $\boldmath{q^2 = -1 \gevgevcccc}$}
\label{app:ImHq2}

\begin{figure}[t]
\centering
\includegraphics[width=0.49\textwidth]{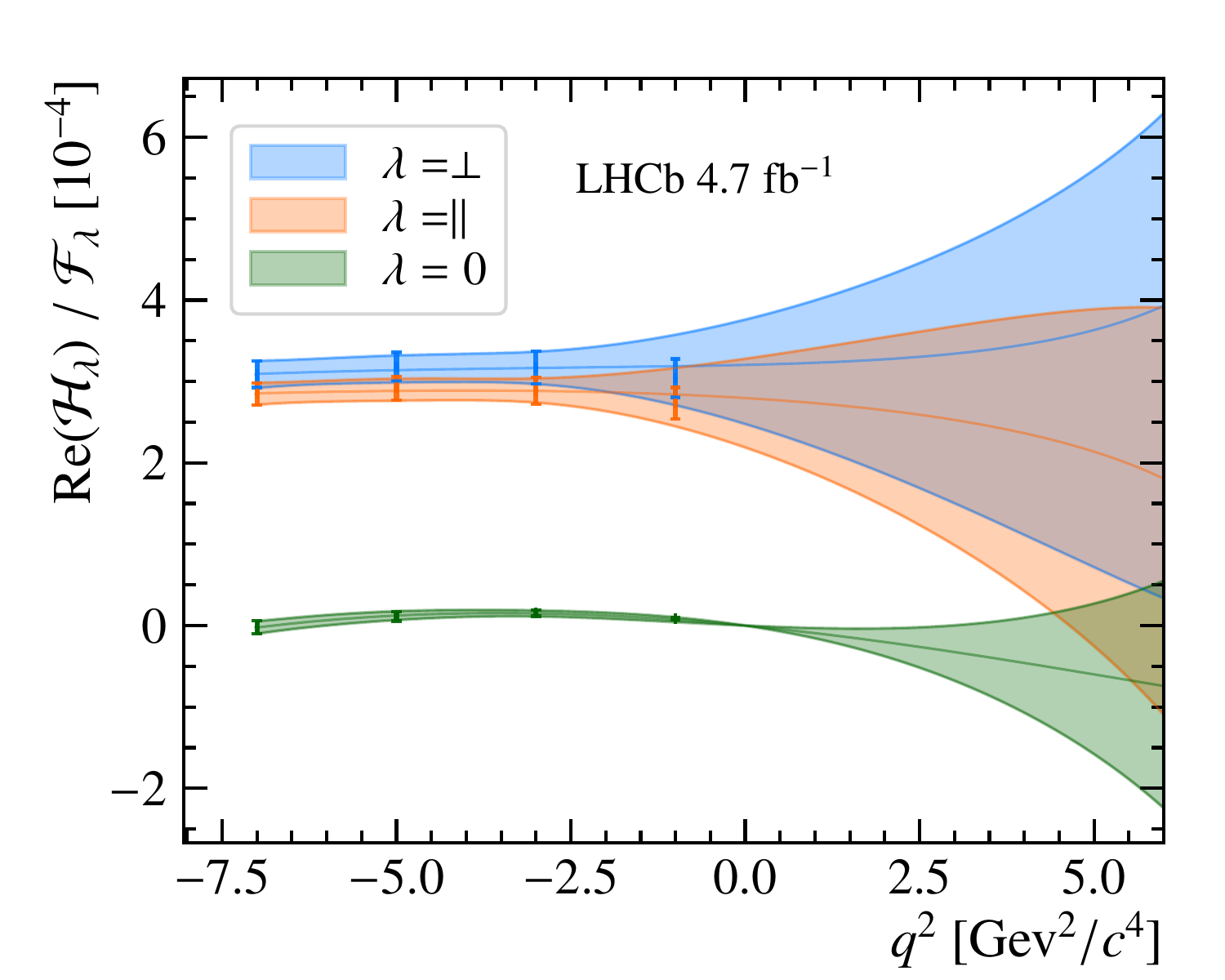} 
\includegraphics[width=0.49\textwidth]{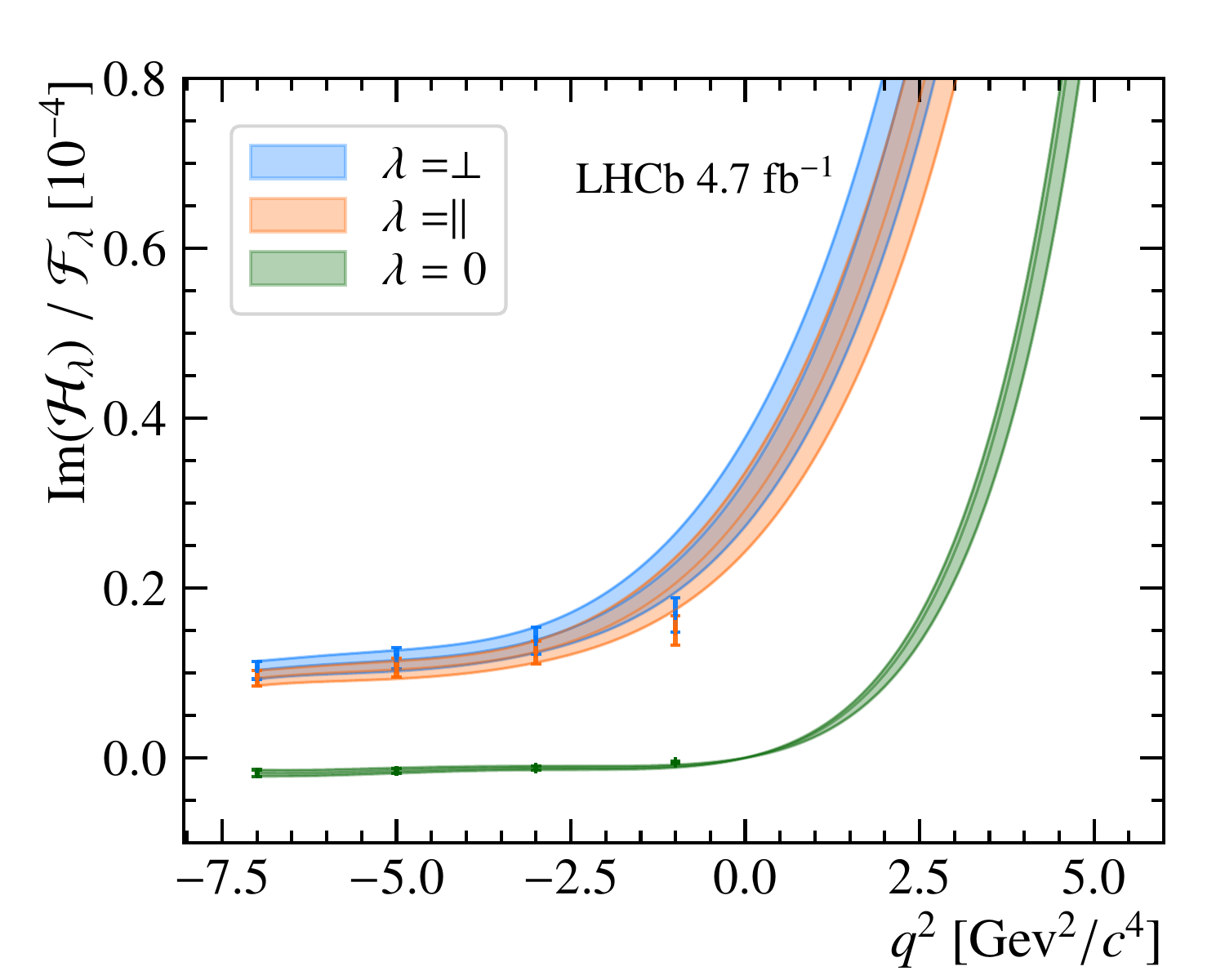} 
\caption{Results for the ratio $\mathcal{H}_\lambda (q^2) / \mathcal{F}_\lambda (q^2)$ obtained from the 
amplitude fit model with theory.
The central values are shown with solid lines, while shaded bands indicate the 68\% confidence intervals.
The theoretical prediction at $q^2<0$ are overlaid for comparison.
 \label{fig:H_neg_q2}}
\end{figure}

The baseline fit with $q^2<0$ constraints includes only three of the four $q^2$ points where SM predictions on the non-local contributions are available~\cite{Gubernari:2020eft}. 
Hence, the remaining point at \mbox{$q^2 = -1 \gevgevcccc$} can be used to test the compatibility of the fit result with the theoretical prediction.
Figure~\ref{fig:H_neg_q2} shows the obtained $\mathcal{H}_\lambda(q^2)$ function at low $q^2$.
While the real part of $\mathcal{H}_\lambda / \mathcal{F}_\lambda$ 
is consistent with the theory prediction at $q^2 = -1\gevgevcccc$,
the imaginary part tends to rise more rapidly than the theoretical predictions.
Note that all theory points are strongly correlated, hence the 
compatibility with the point at $q^2 = -1\gevgevcccc$ is poor.
In fact, to include the theory point at $q^2 = -1\gevgevcccc$ as part of the constraints to the amplitude fit,
it is necessary to further increase the truncation of the polynomial expansion by one additional order.
However, this additional degree of freedom  uniquely modifies the behaviour of the functions $\mathcal{H}_\lambda$  around that point, without providing any significant change to the quality of the fit to data. 
Hence, no additional information is provided by the inclusion of the point at $q^2 = -1\gevgevcccc$ and all conclusions remain unchanged.

\section*{Supplemental Material -- Bootstrapped sets of fit parameters}

In order to favour possible reinterpretation of the analysis, bootstrapped samples of  signal  parameters are uploaded as Supplemental Material~\cite{Suppl}.
These are obtained by sampling the data and repeating the fit for each bootstrapped dataset, where the main sources of systematic uncertainty are included in the fit.
The published set of parameters includes the Wilson coefficients as well as P-wave local and non-local hadronic parameters, for the two studied fit configurations.
These sets of parameters can be employed to derive confidence intervals on any physics quantity of interest, 
\textit{i.e.} checking the compatibility of the result of this analysis with alternative signal parametrisations.


\addcontentsline{toc}{section}{References}
\bibliographystyle{LHCb}
\bibliography{main,standard,LHCb-PAPER,LHCb-CONF,LHCb-DP,LHCb-TDR}

\newpage
\centerline
{\large\bf LHCb collaboration}
\begin
{flushleft}
\small
R.~Aaij$^{35}$\lhcborcid{0000-0003-0533-1952},
A.S.W.~Abdelmotteleb$^{54}$\lhcborcid{0000-0001-7905-0542},
C.~Abellan~Beteta$^{48}$,
F.~Abudin{\'e}n$^{54}$\lhcborcid{0000-0002-6737-3528},
T.~Ackernley$^{58}$\lhcborcid{0000-0002-5951-3498},
B.~Adeva$^{44}$\lhcborcid{0000-0001-9756-3712},
M.~Adinolfi$^{52}$\lhcborcid{0000-0002-1326-1264},
P.~Adlarson$^{78}$\lhcborcid{0000-0001-6280-3851},
C.~Agapopoulou$^{46}$\lhcborcid{0000-0002-2368-0147},
C.A.~Aidala$^{79}$\lhcborcid{0000-0001-9540-4988}
% ...

{\footnotesize \it

% ...
$^{35}$Nikhef National Institute for Subatomic Physics, Amsterdam, Netherlands\\
% ...
$^{44}$Instituto Galego de F{\'\i}sica de Altas Enerx{\'\i}as (IGFAE), Universidade de Santiago de Compostela, Santiago de Compostela, Spain\\
% ...
$^{46}$European Organization for Nuclear Research (CERN), Geneva, Switzerland\\
% ...
$^{48}$Physik-Institut, Universit{\"a}t Z{\"u}rich, Z{\"u}rich, Switzerland\\
% ...
$^{52}$H.H. Wills Physics Laboratory, University of Bristol, Bristol, United Kingdom\\
% ...
$^{54}$Department of Physics, University of Warwick, Coventry, United Kingdom\\
% ...
$^{57}$School of Physics and Astronomy, University of Glasgow, Glasgow, United Kingdom\\
$^{58}$Oliver Lodge Laboratory, University of Liverpool, Liverpool, United Kingdom\\
% ...
$^{66}$Syracuse University, Syracuse, NY, United States\\
% ...
$^{78}$Department of Physics and Astronomy, Uppsala University, Uppsala, Sweden, associated to $^{57}$\\
$^{79}$University of Michigan, Ann Arbor, MI, United States, associated to $^{66}$\\
% ...
}
\end{flushleft}

\end{document}